\documentclass[%
reprint,
superscriptaddress,
amsmath,amssymb,
aps,
pre,
]{revtex4-2}

\usepackage{graphicx}
\usepackage{dcolumn}
\usepackage{bm}
\usepackage{graphicx}
\usepackage{subcaption}
\usepackage{soul}

\usepackage[utf8]{inputenc}
\usepackage{amssymb,amsmath,bm}
\usepackage{textcomp}
\usepackage{amssymb}
\usepackage{xcolor}
\usepackage{listings}
\usepackage{booktabs}
\usepackage{graphicx}
\usepackage{caption}
\usepackage{subcaption}
\usepackage{float}
\usepackage{comment}
\usepackage{etoolbox}

\newcommand{\norm}[1]{\|#1\|}
\newcommand{\normm}[1]{\left\|#1\right\|}
\DeclareMathOperator*{\argmin}{arg\,min}
\DeclareMathOperator*{\argmax}{arg\,max}
\selectcolormodel{cmyk}

\usepackage{subfiles} 

\graphicspath{{figures/}} 

\title{EC12219}

\newcommand{\cgreen}[1]{{#1}}
\newcommand{\cred}[1]{\textcolor{red}{#1}}
\newcommand{\cpurple}[1]{\textcolor{purple}{#1}}

\begin{document}
	
\title{Applications of optimal nonlinear control to a whole-brain network of FitzHugh-Nagumo oscillators}

\author{Teresa Chouzouris}
\thanks{The first two authors contributed equally to this work}
\affiliation{Institut für Softwaretechnik und Theoretische Informatik, Technische Universität Berlin, Marchstraße 23, 10587 Berlin, Germany}
\author{Nicolas Roth$^*$} 
\email{ Correspondence to roth@tu-berlin.de}
\affiliation{Institut für Softwaretechnik und Theoretische Informatik, Technische Universität Berlin, Marchstraße 23, 10587 Berlin, Germany}
\author{Caglar Cakan}
\affiliation{Institut für Softwaretechnik und Theoretische Informatik, Technische Universität Berlin, Marchstraße 23, 10587 Berlin, Germany}
\affiliation{Bernstein Center for Computational Neuroscience Berlin, Philippstraße 13, 10115 Berlin, Germany}
\author{Klaus Obermayer}
\affiliation{Institut für Softwaretechnik und Theoretische Informatik, Technische Universität Berlin, Marchstraße 23, 10587 Berlin, Germany}
\affiliation{Bernstein Center for Computational Neuroscience Berlin, Philippstraße 13, 10115 Berlin, Germany}

\date{August 19, 2021, DOI: \href{http://doi.org/10.1103/physreve.104.024213}{10.1103/PhysRevE.104.024213}}

\begin{abstract}

	We apply the framework of optimal nonlinear control to steer the dynamics of a whole-brain network of FitzHugh-Nagumo oscillators. Its nodes correspond to the cortical areas of an atlas-based segmentation of the human cerebral cortex, and the inter-node coupling strengths are derived from Diffusion Tensor Imaging data of the connectome of the human brain. Nodes are coupled using an additive scheme without delays and are driven by background inputs with fixed mean and additive Gaussian noise. Optimal control inputs to nodes are determined by minimizing a cost functional that penalizes the deviations from a desired network dynamic, the control energy, and spatially non-sparse control inputs. Using the strength of the background input and the overall coupling strength as order parameters, the network’s state-space decomposes into regions of low and high activity fixed points separated by a high amplitude limit cycle all of which qualitatively correspond to the states of an isolated network node. Along the borders, however, additional limit cycles, asynchronous states and multistability can be observed. Optimal control is applied to several state-switching and network synchronization tasks, and the results are compared to controllability measures from linear control theory for the same connectome. We find that intuitions from the latter about the roles of nodes in steering the network dynamics, which are solely based on connectome features, do not generally carry over to nonlinear systems, as had been previously implied. Instead, the role of nodes under optimal nonlinear control critically depends on the specified task and the system's location in state space. Our results shed new light on the controllability of brain network states and may serve as an inspiration for the design of new paradigms for non-invasive brain stimulation.
	
\end{abstract}

\maketitle

\section{Introduction}

The widespread use of noninvasive electrical brain stimulation in clinical applications has sparked ongoing interest in studying the effects of external inputs on brain activity. Stimulation with electric fields in the range of $1-2$ V/m can already modulate oscillatory brain activity \cite{Zaehle2010, Neuling2013}, affect cross-regional synchronization \cite{Polania2012, Strueber2014}, and modulate cognitive performance \cite{Ladenbauer2017}. 
Clinical studies have successfully demonstrated the efficacy of targeted transcranial stimulation in the treatment of neurological and psychiatric disorders and diseases such as epilepsy \cite{Sun2011, Fregni2006}, schizophrenia \cite{hasan2013transcranial, Cole2015}, Alzheimer's disease~\cite{Fried2016} and depression~\cite{George2012}. 

Brain network models offer a way to simulate and understand the human brain as a nonlinear dynamical system, in which each brain region is represented by a node, and the node dynamics is defined by a model of the average neural activity in that region \cite{bassett2017network}. 
Nodes interact with each other according to empirically measured human structural neural connectivity, which quantifies how neural activity in one brain region is coupled to the activity of connected regions.

Parcellation of human brains has yielded various brain atlases \cite{Amunts2014}, which provide information on spatial and functional segregation, dividing the brain into distinct areas \cite{rolls2015aal2, Eickhoff2018}. The relative connectivity strength between these areas is defined by the number of axonal fibers, which can be estimated using structural neuroimaging scans of individual subjects. 
This results in a network model of the human brain where the edges reflect the relative strengths of axonal fiber bundles and the nodes represent individual brain regions. These nodes are then equipped with a dynamical system modeling the average neuronal activity in that region \cite{breakspear2017dynamic}. 
It has been repeatedly shown that when the parameters of brain network models are fitted to functional brain data, optimal operating points were close to the bifurcation lines of these models. This ensures that the model is in a state in which noise fluctuations can be amplified and produce realistic spatial correlation structures which are similar to empirical functional connectivity measurements \cite{deco2012ongoing, deco2013resting, demirtacs2019hierarchical, cakan2020deep}.

Previous theoretical investigations into the impact of external perturbations mostly relied on the assumption of a linear node dynamics, allowing the application of methods from linear control theory \cite{tang2018colloquium}. 
Thus, one can draw conclusions on the effects of external inputs to the system, based on the network topology and independent of its dynamical state \cite{gu2017optimal, gu2015controllability}. 
On the other hand, linear node \cgreen{dynamics cannot} reproduce the dynamics of neural processes close to bifurcations \cite{deco2012ongoing, golos2015multistability}.
In order to describe the transitions from one dynamical regime to another, Muldoon et al.~\cite{muldoon2016stimulation} consider nonlinear node dynamics. They conclude that the effects of stimulation-based control can be predicted by diagnostics from linear control theory based only on the structural connectivity of the network. 

In this work we go beyond linear control theory and explore the framework of optimal nonlinear control \cite{berkovitz2012nonlinear} for the assessing the impact of perturbations on networks of coupled nonlinear systems.
Optimal control is an optimization method that derives control policies based on the minimization of a cost functional which depends on the state and control variables.
Here, we define a cost functional that penalizes the deviation to the desired network dynamics, the control energy, and the spatial sparsity. The latter allows us to find optimal control signals that apply to only a few control sites, as introduced by \cite{herzog2012directional} and further studied by \cite{casas2017analysis}. 
We used methods presented by Tr{\"o}ltzsch in \cite{troltzsch2010optimal} for their calculation, while the handling of the stochastic term is based on the work of Stannat et al.~\cite{stannat2020deterministic}. 
The resulting mathematical formulation is analogous to the formulation presented by Casas et al.~\cite{casas2013sparse} for partial differential equations. 

We evaluate the framework of optimal control for a brain network of FitzHugh-Nagumo (FHN) \cite{FitzHugh1961, Nagumo1962} oscillators with additive coupling and white Gaussian noise, where each oscillator represents a brain region and where the connections between them were chosen according to the human connectome derived from diffusion tensor imaging (DTI) measurements. 
FHN oscillators are well studied models for neural dynamics and detailed analyses of the dynamical states exist for single oscillators \cite{kostova2004fitzhugh} and various network configurations, like two coupled units \cite{scholl2009time, eydam2019FHN}, lattices \cite{shepelev2019FHN}, rings and hierarchical architectures \cite{plotnikov2016synchronization}, as well as random and small-world topologies \cite{lehnert2011loss, cakan2014heterogeneous} and multiplex networks \cite{nikitin2019complex, omelchenko2019control, ruzzene2020remote}. 
Similar empirical DTI-measured brain connectivities were used with FHN dynamics to model highly synchronized epileptic-seizure-like states \cite{chouzouris2018chimera, gerster2020fitzhugh}, unihemispheric sleep \cite{ramlow2019partial}, and the functional organization of the resting brain \cite{ghosh2008cortical, vuksanovic2015dynamic, messe2015closer}. 

We consider two different classes of control problems, targeted attractor switching between multistable network states and increasing network synchronization.
The solutions obtained for given energy, precision, and sparseness constraints are well interpretable and result in intuitively sensible optimal control inputs for all network nodes over time. 

We then correlate the control energies to controllability measures from linear control theory. 
We confirm the findings of Muldoon et al.~\cite{muldoon2016stimulation} for the investigated state transition (from the low fixed-point state to the oscillatory regime).
For other control tasks, however, we show, that diagnostics from linear control theory do not correlate with results from optimal nonlinear control. Conclusions drawn from the structural connectivity alone lead to contradictions, which can only be resolved if the dynamical state of the network and the nonlinear interactions between nodes are taken into account.
Applications of nonlinear control theory to whole-brain models 
enables us to investigate control mechanisms also close to bifurcations. 
It thus may help in the search for more effective paradigms for realistic transcranial brain stimulation protocols.

\section{Nonlinear optimal control}
\label{sec:nonlinear_control}

\subsection{Network model and control inputs}

We consider networks of $N$ equivalent $d$-dimensional and  ``noisy'' nodes with additive and zero-delay internode coupling. Internode coupling strength is described by a $N\times N$ dimensional adjacency matrix $\bm{A}$ which is scaled by a global coupling strength $\sigma$. 
\cgreen{We allow for an instantaneous and additive control input to the network, which is described by $N \times d$ independent control variables $\bm{u}=(\bm{u_1},\ldots,\bm{u_N})$ with $\bm{u_i}=(u_{i1},\dots,u_{id})$. }
The equations describing the \cgreen{controlled} network dynamics thus read:
\begin{eqnarray}
\label{eqn:network_system}
\begin{aligned}
\frac{d }{d t}\bm{x}(t) =\bm{h}(\bm{x}(t) ) + \sigma (\bm{A} \otimes \bm{G}) \bm{x} (t) \\
+(\bm{B} \otimes \bm{K}) \bm{u}(t) + (\bm{I_N} \otimes \bm{D}) \bm{\xi}(t).
\end{aligned}
\end{eqnarray}
where $\otimes$ denotes the Kronecker product. The state of the network is described by a vector $\bm{x}=(\bm{x_1},\ldots,\bm{x_N})$, where $\bm{x_i}=(x_{i1},\dots,x_{id})$ characterize the individual states of the nodes. The local node dynamics is given by $\bm{h}(\bm{x})=(\bm{h}(\bm{x_1}),\ldots,\bm{h}(\bm{x_N}))$ with $\bm{h}(\bm{x_i})=(h_1(\bm{x_i}),\dots,h_d(\bm{x_i}))$.  
All nodes additionally receive Gaussian white noise of similar intensity $\eta$, which may be correlated within but is uncorrelated across the nodes of the network. The stochastic variables $\bm{\xi}=(\bm{\xi_1},\dots,\bm{\xi_N})$ with $\bm{\xi_i}=(\xi_{i1},\dots,\xi_{id})$ are independently drawn from a Gaussian distribution, $\xi_{ij} \in \mathcal{N}(0,1)$, with zero mean and unit variance.
Within node correlations are quantified by the $d\times d$ dimensional local noise scheme $\bm{D}$, while the across-node statistical independence is assured by a N-dimensional identity matrix $\bm{I_N}$. The internode coupling term consists of the Kronecker product of the adjacency matrix $\bm{A}$ and the $d\times d$ dimensional local coupling scheme $\bm{G}$, where the purpose of the former is to describe the relative interaction strength between nodes while the purpose of the latter is to distribute the between-node interactions across the $d$ variables which describe the local node dynamics.
The $N\times N$ matrix $\bm{B}$ \cgreen{in the control term} allows for the control of multiple nodes with different strength
\cgreen{. The} $d\times d$ dimensional matrix $\bm{K}$ implements the local control scheme, which is similar for every node in the network.
Initial conditions are denoted by $\bm{x}(t=0)=\bm{x_0}$.

\subsection{The cost functional and the optimality condition}

The control $\bm{u}$ is considered to be optimal ($\bm{u} = \overline{\bm{u}}$), if it minimizes an appropriate cost functional. To this end, we construct a cost functional $F(\bm{x}(\bm{u}),\bm{u})$ for a state switching task, where the control input drives the network model from one stable state to another (see Section \ref{sec:state_switching}), and for a node synchronization task, where the control input increases the degree of synchronization among its nodes (see Section \ref{sec:synchronize}).  For finite noise, i.e.\ for finite values of $\eta$ in Eq.~\eqref{eqn:network_system}, the overall cost functional $F$ is defined as a mean over noise realizations~\cite{stannat2020deterministic},
\begin{eqnarray}
\label{eqn:cost_functional_general}
F(\bm{x}(\bm{u}),\bm{u})=\langle F_n(\bm{x}(\bm{u}),\bm{u}) \rangle = \langle  F_n^x(\bm{x}) \rangle + F^u(\bm{u}),
\end{eqnarray}
where $F_n$ denotes the cost functional for one noise realization $n$ and the angle brackets $\langle \cdot \rangle$ denote the mean. 
\cgreen{In the noise-free case no averaging is performed.}
The state dependent term $F_n^x(\bm{x})$ only implicitly depends on the control and penalizes the deviation from the desired output. 
It will be different for the switching, $F_{n,\textrm{\cgreen{sw}}}^x$, and synchronization, $F_{n,\textrm{\cgreen{syn}}}^x$, tasks (see below).
The control dependent term $F^u(\bm{u})$ accounts for the cost of the control itself.

For the state switching task, we consider a noise-free system ($\eta=0$), and the state dependent cost functional $F^x_{n,sw}$ is defined in terms of the deviation of the controlled state to a predefined target state $\bm{x}_T(t)$:
\begin{align}
\begin{split}
\label{eqn:cost_functional_f_x1}
F^x_{n,\textrm{\cgreen{sw}}}(\bm{x},t)&= F^x_{\textrm{\cgreen{sw}}}(\bm{x},t)\\ &= \int_{0}^{T}\frac{I_p(t)}{2} \cgreen{\bigl(} \bm{x}(t)-\bm{x}_{T}(t)\cgreen{\bigr)} ^2 dt,
\end{split}
\end{align}
where we consider the control being active within the time period $0\le t\le T$.
In order to penalize the precision only towards the end of the controlled time interval rather than during the transition between the initial and target states, the precision-penalizing variable $I_p$ can be time-dependent (see Section \ref{sec:state_switching}). 

The state dependent cost functional for the synchronization task $F^x_{n,syn}$ is defined in terms of the deviations of the normalized pairwise cross-correlations $R_{ij}$ for all nodes $i$ and $j$ from $R_T$, the desired mean cross-correlation in the synchronized target state:
\begin{eqnarray}
\label{eqn:cost_functional_f_x2}
F^x_{n,\textrm{\cgreen{syn}}}(\bm{x},t) = \frac{I_p}{4N^2} \sum_{i,j=1}^N (R_{ij}-R_T)^2.
\end{eqnarray}

The cross-correlation in component $m$ is defined as
\begin{eqnarray}
\label{eqn:cross_correlation}
R_{ij}^m=\int_0^T \frac{\big(x_{im}(t)- \langle x_{im}\rangle_t\big) \big(x_{jm}(t)- \langle x_{jm}\rangle_t\big)}{\sigma_{ \langle x_{im}\rangle_t} \sigma_{ \langle x_{jm}\rangle_t} } dt,
\end{eqnarray}
with $i, j \in [0,\dots ,N]$ and where $\langle .\rangle_t$ and $\sigma_{\langle .\rangle_t}$ denotes the temporal mean and the standard deviation. The mean $R^m$ over all values $R_{ij}^m$ 
\begin{eqnarray}
\label{eqn:nw_cross_correlation}
R^m = \frac{1}{N^2} \sum_{i=1}^{N} \sum_{j=1}^N R_{ij}^m,
\end{eqnarray}
is the component-wise network cross-correlation. Later we specialize to $m=1$ and suppress the index $m$. 

The input dependent cost functional $F^u(\bm{u})$ penalizes the energy of the control signal and enforces its directional sparsity \cite{herzog2012directional, casas2017analysis}. It is given by
\begin{eqnarray}
\begin{split}
\label{eqn:cost_functional_f_u}
F^u(\bm{u}) = \ &\frac{{I_e}}{2}    \int_{0}^{T}  \bm{u}^2(t) dt  \\
&+ {I_s} 
\cgreen{\sum_{k=1}^N} \Big( \int_{0}^{T} {u}_k^2(t)dt \Big)^{\frac{1}{2}} ,
\end{split}
\end{eqnarray}
where $I_e$ and $I_s$ are the weights for the energy and sparsity terms. \cgreen{The first term corresponds to the L2-regulation and the second term to the L1-regulation of the cost-functional.} Typically, increasing the sparsity $I_s$ reduces the number of controlled nodes, i.e.\ the nodes for which the control input is non-zero for at least part of the time period $0 \leq t \leq T$, while a higher penalty on the energy term typically leads to an overall reduction of control strengths.

Our goal is to find the optimal control $\overline{\bm{u}}$ that minimizes the cost functional for chosen $I_p$, $I_e$, and $I_s$, leading to the minimization problem
\begin{eqnarray}
\label{eqn:minimization_problem}
\overline{\bm{u}} = \argmin_{\bm{u}} \langle F_n(\bm{x}(\bm{u}),\bm{u}) \rangle.
\end{eqnarray}

Similarly to Troeltzsch et al.\ and Casas et al.\ \cite{casas2013sparse, troltzsch2010optimal, buchholz2013optimal}, we analyze the gradient $g=\bm{\nabla_u} F$ of the cost functional, which has to vanish when evaluated at the optimal control $\overline{\bm{u}}$. 
By applying the method of Lagrange multipliers, we obtain an expression for the optimality condition that depends on the adjoint states $\bm{\phi}(\bm{x},{\bm{u}},t)$ corresponding to the Lagrange multipliers. The control $\overline{\bm{u}}$ is optimal, if
\begin{eqnarray}
\label{eqn:zerogradient} 
\resizebox{0.85\hsize}{!}{%
	$\bm{g}(t)= (\bm{B} \otimes \bm{K})^T \langle \bm{\phi}(\bm{x},\overline{\bm{u}},t) \rangle +  I_e \overline{\bm{u}}(t) + I_s \overline{\bm{\lambda}}(t) \overset{!}{=} \bm{0}$%
}
\end{eqnarray}
for $0\leq t \le T$, where $\overline{\lambda}(t)$ is the derivative of the sparsity term, Eq.~\eqref{eqn:cost_functional_f_u}, with respect to the control inputs $\bm{u}$. The adjoint states are governed by a set of linear differential equations:
\begin{eqnarray}
\label{eqn:adjoint_state} 
\resizebox{0.85\hsize}{!}{%
	$-\cgreen{\frac{d }{d t}}\bm{\phi}(t)=  \big[ D_{\bm{x}}(\bm{h}) + \sigma (\bm{A} \otimes \bm{G}) \big]^T \bm{\phi}(t) + \bm{\nabla_x} f^x_n(\bm{x},t),$%
}
\end{eqnarray}
where $D_{\bm{x}}(\bm{h})$ is the Jacobian matrix of the state equations of the dynamical system
, and $f^x_n(\bm{x})$ is the integrand of the cost functional, $F^x_n(\bm{x})=\int_0^T f^x_n(\bm{x}) dt$. The adjoint state satisfies the boundary condition $\bm{\phi}(T)=\bm{0}$, and the differential equation is solved backwards in time.

Following Ref.~\cite{casas2017analysis}, $\overline{\lambda}(t)$ is given by
\begin{equation}
\label{eqn:lambda}
\overline{\lambda_k}(t) =
\begin{cases} 
\frac{\overline{\bm{u}}_k(t)}{\sqrt{E_k}} & \text{if $E_k \neq 0$}, \\
-\frac{1}{I_s} \big[(\bm{B} \otimes \bm{K})^T \langle \bm{\phi}(\bm{x},\overline{\bm{u}},t) \rangle \big]_k & \text{otherwise}.
\end{cases}
\end{equation}
Here, $k \in [1,N]$, and $E_k$ is the nodewise control energy of the resulting optimal control $\overline{\bm{u}}$, which is defined as
\begin{eqnarray}
\label{eqn:control_energy} 
E_k=\int_0^T \overline{\bm{u}}_k^2(t) dt,
\end{eqnarray}
resulting in the total control energy $E = \sum_{k} E_k$.
A detailed derivation of Eqs. \eqref{eqn:zerogradient} and \eqref{eqn:adjoint_state} is provided in Appendix~\ref{sec:appendixMath}.

\subsection{Numerical solution of the minimization problem}
\label{sec:min_problem}

The optimization problem is numerically solved using the conjugate gradient method. We integrate the equations for the network state and the adjoint given in Eqs.~\eqref{eqn:network_system} and~\eqref{eqn:adjoint_state} 
\cgreen{(see Appendix~\ref{sec:appendixNumerics} for details)}. The direction \cgreen{$\bm{d}_l$} for each step of the conjugate gradient algorithm is defined by the Polak-Ribiere method~\cite{polak1969note}, while its step size \cgreen{$s_l$} is derived using simple bisection \cite{nocedal2006numerical}. 
We apply the Fletcher and Reeves algorithm~\cite{fletcher1964function} as presented in the following. 

We initialize at iteration  \cgreen{$l=0$} by choosing an initial control $\bm{u}_0$ and drawing $20$ noise realizations $\bm{\xi}_n(t)$. The corresponding states $\bm{x}_{0}(t)$, Eq.~\eqref{eqn:network_system}, and adjoint states $\bm{\phi}_{0}(t)$, Eq.~\eqref{eqn:adjoint_state},  
are calculated for every noise realization. Then $\bm{g}(t)$ as given in Eq.~\eqref{eqn:zerogradient} is evaluated. 
The descent direction is initialized with $\bm{d}_0(t) =-\bm{g}(t)$.
We then iterate until convergence:
\begin{itemize}
	\item[] while $\norm{\bm{g}(t)}_{\infty} > \epsilon$: 
	\begin{enumerate}
		\item compute the step size $s_l$ using bisection 
		\item set $\bm{u}_{l+1}(t)= \bm{u}_l(t) + s_l \bm{d}_l(t)$ 
		\item calculate the states $\bm{x}_{l+1}(t)$ and the adjoint states $\bm{\phi}_{l+1}(t)$ for every noise realization and compute the mean $\langle \bm{\phi}_{l+1}(t) \rangle$  
		\item evaluate the gradient $\bm{g}_{l+1}(t)$ 
		\item compute $\beta_l$ using the Polak-Ribiere method: $\beta_l=\frac{\bm{g}_{l+1}(\bm{g}_{l+1}-\bm{g}_{l})}{\norm{\bm{g}_{l+1}}^2}$
		\item set the direction $\bm{d}_{l+1}(t)= - \bm{g}_{l+1} (t)+ \beta_l \bm{d}_l(t)$
		\item if $\bm{d}_{l+1}(t)$ is not a descent direction, set $\bm{d}_{l+1}(t)= - \bm{g}_{l+1}(t) $
		\item set $l=l+1$\\
	\end{enumerate}
\end{itemize}

\cgreen{Gradient-based optimization is only guaranteed to converge to local optima of the cost function. In general, however, multiple initial conditions $\bm{u}_0$ converged to the same optimum. The only exception occurred for the state-switching task for high values of the sparseness parameter $I_s$, for which a solution with finite values of $\bm{u}$ coexists with the solution $\bm{u}(t)=0$ (see Appendix \ref{sec:appendixNumerics} for details).}

\section{The whole-brain network}
\label{sec:State_space}


\subsection{\cgreen{The local node dynamics}}
\label{sec:population_dynamics}
\begin{figure}
	\centering
	\includegraphics[width=0.45\textwidth]{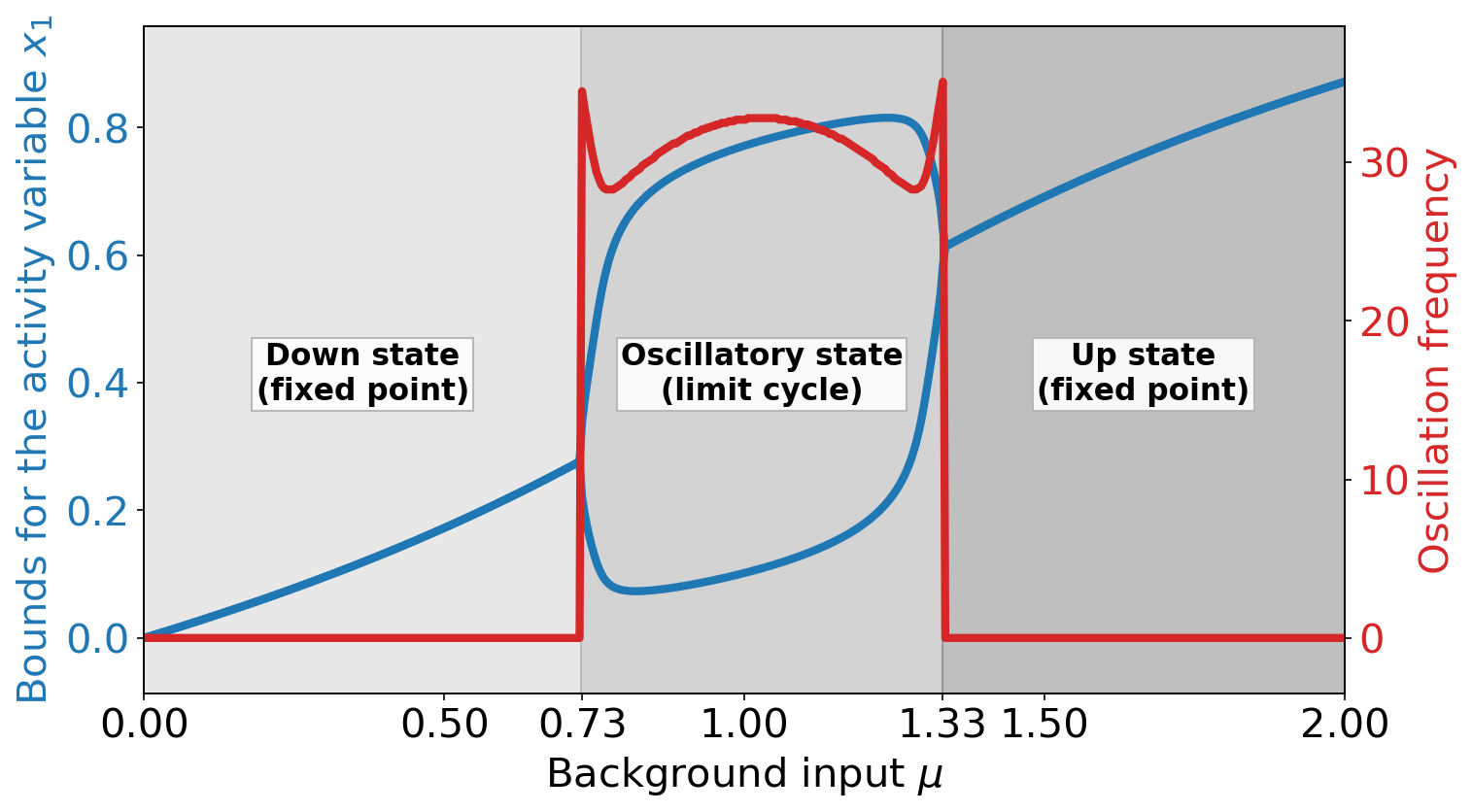}
	\caption{Bifurcation diagram of an uncoupled FitzHugh-Nagumo oscillator, as given in Eq.~\eqref{eqn:FHN}, with the parameters $\alpha=3$, $\beta=4$, $\gamma=1.5$, $\delta=0.5$, and $\tau=20$. The blue line shows the minimal and maximal values of the activity variable $x_1$ (identical in fixed points) for the respective background input $\mu$ to the node. The red line shows the frequency of the oscillation (time and therefore also frequency are measured in arbitrary units).}
	\label{fig:sgl_bifur_diag}
\end{figure}

We consider a single FitzHugh-Nagumo (FHN) oscillator, with the activity variable $x_1$ and the linear recovery variable $x_{2}$:
\begin{align}
\label{eqn:FHN}
\bm{h}(\bm{x})= \begin{pmatrix}h_{1}(\bm{x})\\h_{2}(\bm{x})\end{pmatrix} = \begin{pmatrix}\cgreen{\frac{d}{d t}}x_{1}\\\cgreen{\frac{d}{d t}}x_{2}\end{pmatrix}=
\begin{pmatrix} R(x_{1}) - x_{2} + \mu \\\frac{1}{\tau} (x_{1}  - \delta x_{2})\end{pmatrix},
\end{align}
where $\mu$ is a node-independent, constant background input and $R(x)=- \alpha x^3 + \beta x^2 - \gamma x$. The parameters in $R$ 
are chosen to obtain the bifurcation diagram shown in Fig.~\ref{fig:sgl_bifur_diag}. 
This bifurcation diagram, depending on the background input $\mu$, shows three distinct states. In the down state, the node is in a stable fixed point and has a low constant value of the activity variable $x_1$ (in blue). In the oscillatory state, the activity variable $x_1$ oscillates at an input-dependent frequency (in red). In the up state, the node is again in a stable fixed point with a high constant value of the activity variable $x_1$. 

The succession of these states in the single node dynamics -- a supercritical Andronov-Hopf bifurcation from the down state to the oscillatory state and another supercritical Andronov-Hopf bifurcation from the oscillatory state to the up state \cite{kostova2004fitzhugh, izhikevich2007dynamical} -- \cgreen{closely} resembles the states found in large random networks of excitatory and inhibitory spiking neuron models and their corresponding mean-field description \cite{cakan2019}.  
\cgreen{The mean firing rate of these models changes as a function of background input in a way which is qualitatively similar to the value of the activity variable shown in Fig.~\ref{fig:sgl_bifur_diag}. For this reason one can interpret the value of the activity variable $x_1$ as the difference between the output firing rate of a cortical node to a baseline value.} 

\subsection{The brain network model}

\begin{figure}
	\centering
	\includegraphics[width=0.325\textwidth]{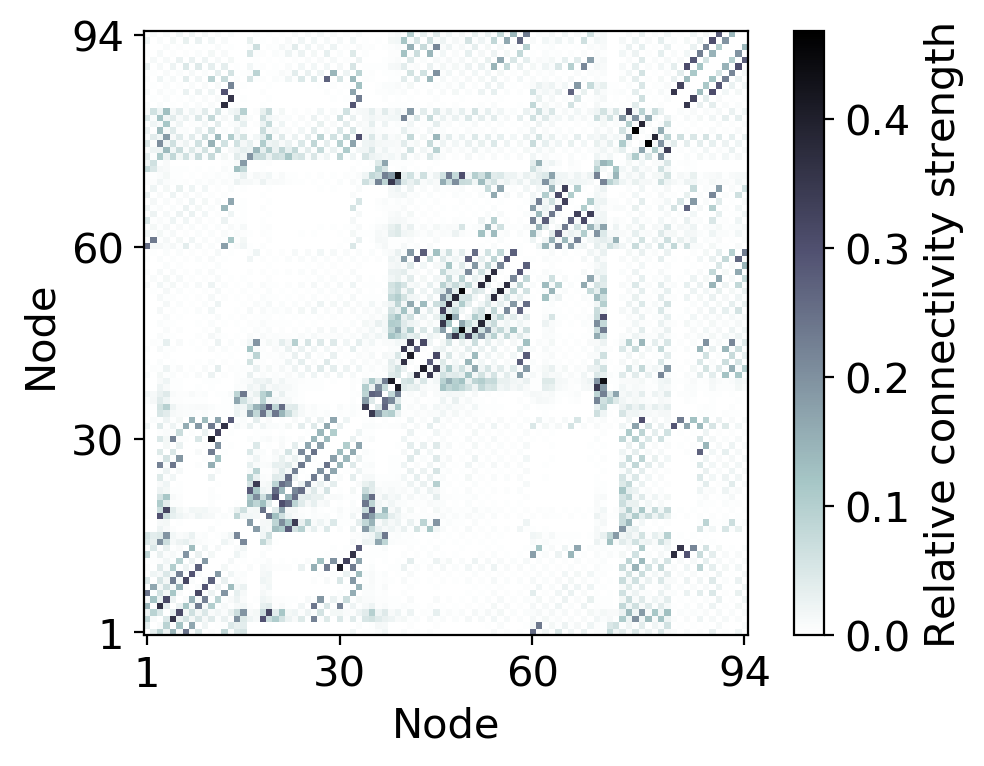}
	\caption{Weighted adjacency matrix $\bm{A}$ describing the topology of the whole-brain network. Color denotes the relative connection strength of the structural connectivity for every pair of the $N=94$ nodes.}
	\label{fig:Adjacency_matrix}
\end{figure}

The topology of the whole-brain network model is derived from diffusion tensor imaging (DTI) data of 12 human subjects (all male and 26-30 years old) from the Human Connectome Project \cite{van2013hcp} (see \cgreen{Appendix~\ref{sec:appendixDTI}} for subject IDs). 
The $N=94$ nodes in the network correspond to the cortical and subcortical regions defined by the AAL2 atlas-based segmentation \cite{rolls2015aal2} (cerebellum excluded). 
We performed probabilistic fiber tracking (using \cgreen{the Functional Magnetic Resonance Imaging of the Brain Software Library (FSL)} \cite{jenkinson2012fsl}, details on the processing pipeline \cgreen{in Appendix~\ref{sec:appendixDTI}}) to determine the relative connection strength (edges) between these brain regions (nodes). 
The pairwise structural connectivity is summarized in the weighted adjacency matrix~$\bm{A}$ shown in Fig.~\ref{fig:Adjacency_matrix}.
\cgreen{Since DTI data carries no directional information, $\bm{A}$ must be symmetric. For the following, we define the weighted degree $d_k$ of a node $k$ as the sum over all afferent connection strengths, i.e.\ $d_k = \sum_{i=1}^{N}\bm{A}_{ik}$.}

As motivated in Section \ref{sec:population_dynamics}, the dynamics of each node in the network is described by a FHN oscillator [Eq.~\eqref{eqn:FHN}].
The general network dynamics, Eq.~\eqref{eqn:network_system}, then simplifies to
\begin{eqnarray}
\label{eqn:network_simple}
\begin{aligned}
\frac{d }{d t}{x_{k1}}(t) &= 
{h_1}(\bm{x}_k(t) ) 
+ \sigma \sum_{i=1}^N \bm{A}_{ki} x_{i1} (t)
+ 
\xi_k(t)
+{u}_k(t), \\
\frac{d }{d t}{x_{k2}}(t) &= 
{h_2}(\bm{x}_k(t) ), 
\end{aligned}
\end{eqnarray}
for all nodes $k\in [1,N]$. This implies setting the control matrix $\bm{B}=\bm{I_N} $, which means that all individual nodes can potentially receive independent control inputs. 
The local coupling, control, and noise schemes are set to  $\bm{G}=\bm{K}=\bm{D}=[[1,0],[0,0]]$\cgreen{, i.e.\ the indvidual FHN nodes receive node-external inputs only through their activity variables $x_{k1}$}. 

\cgreen{We do not consider finite delays for reasons of simplicity. Finite delays induce additional dynamical states. Although of high interest in terms of whole-brain modeling, these would not add to the conclusions drawn in Sections \ref{sec:applications} and \ref{sec:linear_vs_nonlinear} about the impact of non-linear optimal control and its comparison with diagnostics \cite{gu2015controllability} derived from linear control and based on connectome properties only. Please refer to Section~\ref{sec:summary} for a discussion of the biological plausibility of this assumption.}

\subsection{State space exploration}

In this section, we describe and characterize the different dynamical states that can emerge in a whole-brain network of coupled FHN oscillators for a large range of parameter configurations. Having such an overview of the dynamical landscape of the system will allow us to formulate well defined control tasks in Section~\ref{sec:applications}.
Throughout the state space exploration we do not apply any control input [${u}_k(t) = 0 \,\, \forall \,k, t$]. 

Initially we consider the noise free case ($\eta = 0$). 
Here, only two free model parameters remain: the global coupling strength $\sigma$ and the time independent background input $\mu$ which is the same for each FHN oscillator.
The state space of the FHN network is explored by simulating the network dynamics for wide ranges of these parameters \cgreen{(see Appendix~\ref{sec:appendixNumerics} for details)}.
The initial conditions $\bm{x_0}$ are drawn randomly from a uniform distribution in the interval $\mathopen[0,1\mathclose)$ or taken as the state vector $\bm{x'}(t_{end})$ of the last time step of the previous simulation with same $\sigma$ and slightly smaller $\mu$ (\textit{continuation}).
In order to avoid analyzing transient effects, we show and evaluate the network states only after a sufficiently long transient time~$\bar{t}$. 

If oscillations are present, the dynamical state is characterized by the strength of synchronization quantified by the total cross-correlation, Eq.~\eqref{eqn:nw_cross_correlation}, and by the dominant frequency $f_{dom}$. The latter is given by the frequency of the highest peak of the combined power spectrum of all nodes,
\begin{align}
\label{eqn:dom_freq}
f_{dom} = \argmax_{f}  \sum_{i=1}^{N} S_{xx,i1}(f),  
\end{align}
where $S_{xx,i1}=\left| (\mathcal{F}x_{i1})(f)\right|^{2},$ with Fourier transform $\mathcal{F}$, denotes the power spectral density of an individual node.

\begin{figure*}
	\centering
	\includegraphics[width=\textwidth]{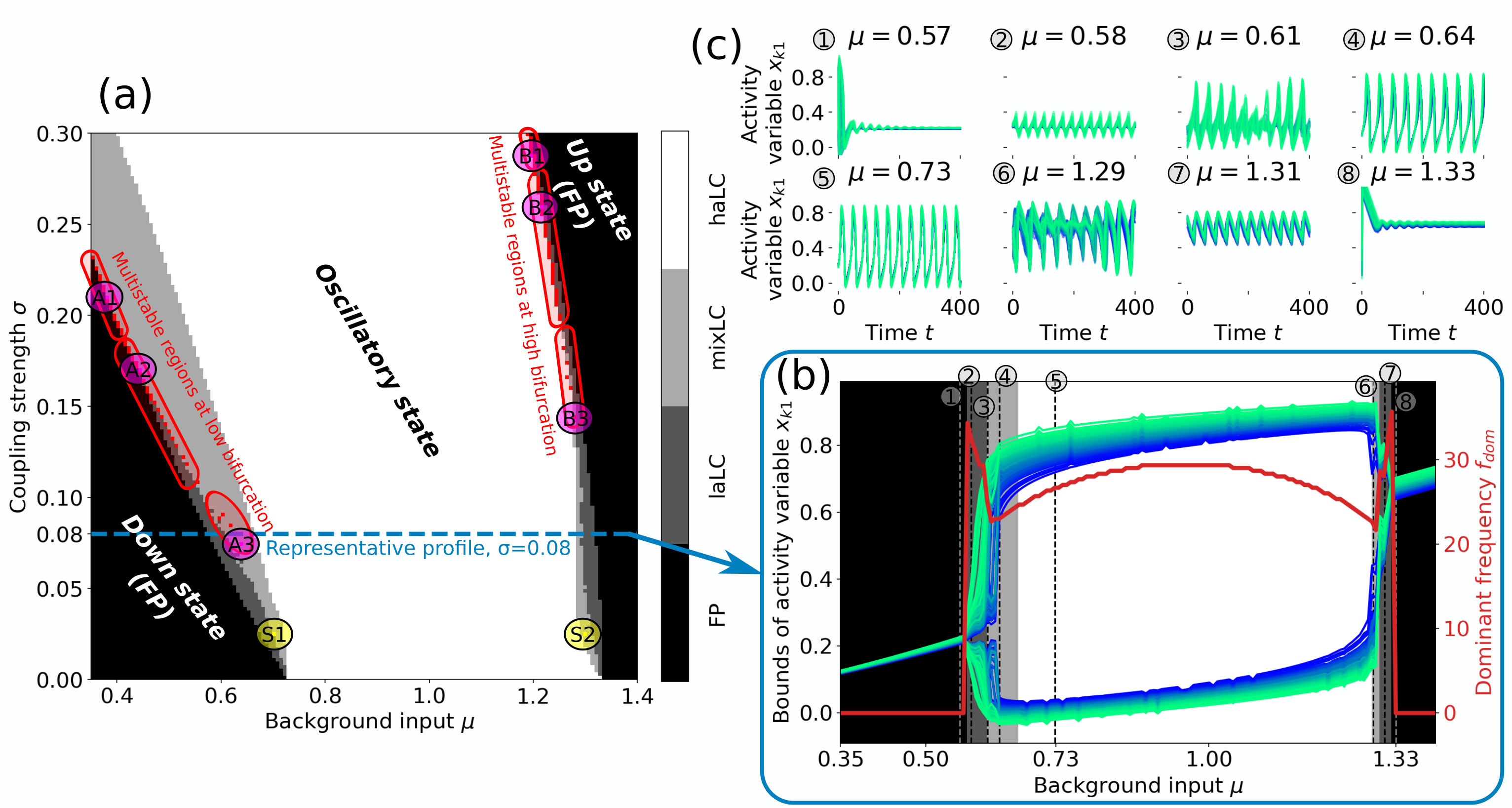}
	\caption{Overview of the state space of the brain network model, Eqs.~\eqref{eqn:FHN} and \eqref{eqn:network_simple}, for the noise free case ($\eta = 0$). Time $t$ measured in arb.\ units and simulations of each parameter configuration are started with initial conditions $\bm{x_0}$ based on continuation and evaluated [except for (c1) and (c8)] after $\bar{t}=5000$ (see text for details). 
		(a) Each pixel corresponds to a network with a particular parameter configuration for $\mu \in [0.35,1.4]$ and $\sigma  \in [0,0.3]$. States are classified based on the network oscillations, and oscillatory states are distinguished depending on whether no (laLC, light gray), some (mixLC, dark gray), or all (haLC, white) nodes fulfill the criterion of Eq.~\eqref{eqn:LC_crit}. Parameter configurations for which multiple stable solutions were detected (see text for details) and visually confirmed are marked with red pixels. Red lines enclose regions where we observe bistability between FP and mixLC (around point A1), laLC and mixLC (around points A2, B2), and mixLC and haLC (around point B1), as well as multistability in mixLC (around points A3, B3). Examples for bi- and multistable states are shown in Fig.~\ref{fig:State_space_traces}. Points S1 and S2 indicate states with a low mean network cross-correlation $R$ that are explored in Section~\ref{sec:synchronize}.
		(b) One dimensional bifurcation diagram for a network as a function of $\mu$  for a fixed coupling strength of $\sigma = 0.08$. The dominant network frequency (red line) is calculated according to Eq.~\eqref{eqn:dom_freq}. Minimal and maximal values of the activity variable $x_{k1}$ are plotted for each node individually, where line color indicates the weighted degree (green to blue indicates high to low values).  (c) Example traces of the activity of all nodes in the network for different values of $\mu$ [see (b)] and for $\sigma=0.08$, showing different dynamical states. Line color indicates the weighted degree of the nodes [see (b)]. Traces at the FPs (1, 8) are shown after $\bar{t}=0$ (see text for details) to show the stability of the respective fixed point.} 
	\label{fig:State_space_overview}
\end{figure*}

Figure \ref{fig:State_space_overview} provides an overview of the state space of the FHN network. 
Similar to the case of a single FHN oscillator (cf.\ Fig.~\ref{fig:sgl_bifur_diag}), the network shows two regions in parameter space, for which a stable fixed point (FP) exists. 
For small values of $\mu$ and $\sigma$ the activity variable $x_{k1}$ has a low value for all nodes [\textit{down state}, cf.\ Fig.~\ref{fig:State_space_overview}c(1)]; for large values of $\mu$ and $\sigma$ the activity value is high [\textit{up state}, cf.\ Fig.~\ref{fig:State_space_overview}c(8)]. 
Figure~\ref{fig:State_space_overview}a shows that the dynamics can transition from an down state to an oscillatory state, as well as from an oscillatory state to an up state by either increasing $\mu$ or $\sigma$. For convenience, we call these transitions \textit{low} and \textit{high bifurcation}, respectively. 

Within the oscillatory regime, we observe network states that are qualitatively different in their appearance, which we explain in the following by their different underlying mechanisms. 
A single, uncoupled FHN node oscillates in its limit cycle (LC) if it receives a background input in the range of $\mu \in [0.73,1.33]$ (cf.\ Fig.~\ref{fig:sgl_bifur_diag}). We therefore expect a node in the network to show sustained oscillation (indiv.\ LC) -- and consequently affecting the dynamics of the network -- if its combined input is in this interval.
We thus call a node ``to be in its individual LC regime'' if
\begin{align}
\label{eqn:LC_crit}
0.73 \lesssim \mu + \sigma \sum_{i=1}^N \bm{A}_{ki} x_{i1} (t)\lesssim 1.33, 
\end{align}
for at least one point in time within the considered time interval. 
This criterion is further motivated by observations of the dynamics  (cf.\ Fig.~S1 in \cgreen{\cite{supp}}) which show qualitative changes in behavior depending on whether it is met by all, some, or none of the nodes in the network.

If all nodes fulfill Eq.~\eqref{eqn:LC_crit}, we observe the network to be in a synchronous, high amplitude limit cycle [haLC in Fig.~\ref{fig:State_space_overview}a, cf.\ Fig.~\ref{fig:State_space_overview}c(5)].
If, however, no node fulfills Eq.~\eqref{eqn:LC_crit}, the network either inherits the fixed point solution from the individual nodes, or a new state emerges due to network effects. We observe both cases: The first one is true in both down and up state of the network in Fig.~\ref{fig:State_space_overview}a. The second case 
appears for low and intermediate coupling strength $\sigma$ at the transition from down or up state to oscillatory state.
In these states 
the FP of the network dynamics is unstable due to coupling effects and a self-sustained low amplitude limit cycle [laLC in Fig.~\ref{fig:State_space_overview}a, cf.\ Fig.~\ref{fig:State_space_overview}c(2,7)] emerges. 
The geometric interpretation of this laLC is a rotation around the FP that would be transient for an isolated node, but is prevented from converging to a fixed-point by the network couplings. 

It is also possible that only a fraction of the network nodes fulfill the criterion in Eq.~\eqref{eqn:LC_crit}, which can lead -- even in the absence of noise and without network delays -- to asynchronous and apparent aperiodic behavior [mixLC in Fig.~\ref{fig:State_space_overview}a, cf.\ Fig.~\ref{fig:State_space_overview}c(3,6), further indications for aperiodicity are provided in \cgreen{\cite{supp}}]. 
On the network level, this can be seen as the result of the different frequencies of the two coexisting network limit cycles, laLC and haLC, interacting. 
Close to the bifurcation the frequency of the individual nodes (cf.\ Fig.~\ref{fig:sgl_bifur_diag}), and their trajectories (cf.\ Fig.~S3 in \cgreen{\cite{supp}}), are particularly sensitive to their respective input. If the additive coupling input between the nodes is not sufficient to entrain a common frequency, this may result in asynchronous and potentially aperiodic network oscillations.
States that are classified as mixLC are, however, not necessarily asynchronous. 
If the driving force of nodes that fulfill the criterion in Eq.~\eqref{eqn:LC_crit} is high enough, it will lead to frequency entrainment. Therefore, the oscillations of the driven nodes [where Eq.~\eqref{eqn:LC_crit} is not fulfilled] are similar to the ones of the driver nodes, resulting in dynamics that are similar to the haLC state [cf.\ Fig.~\ref{fig:State_space_overview}c(4,5)].

\begin{figure*}
	\centering
	\includegraphics[width=\textwidth]{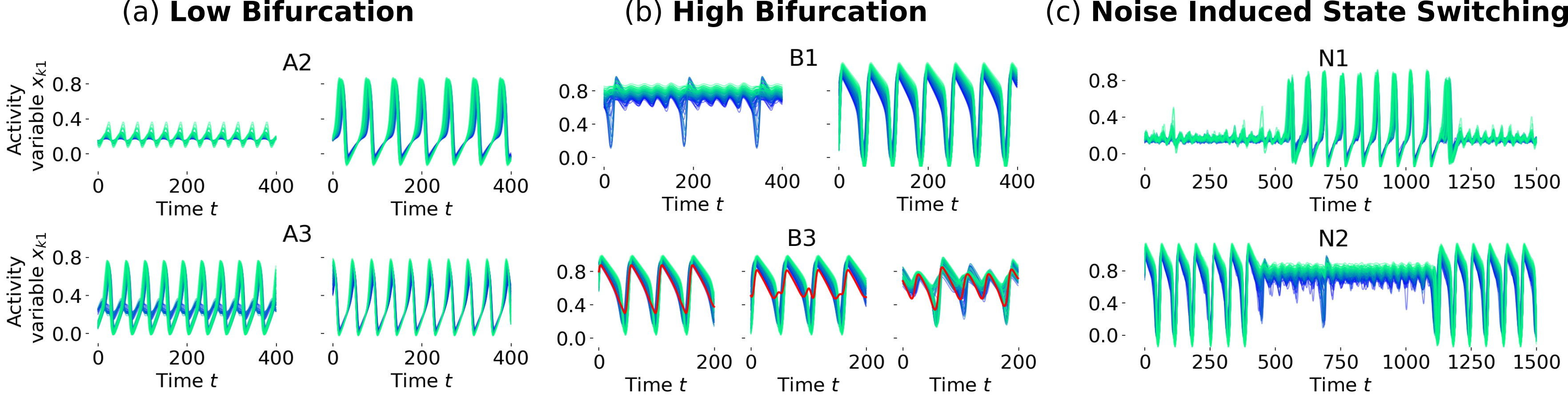}
	\caption{Example traces of the activity variable $x_{k1}$ of each node from the regions of bi- and multistability indicated by the labeled points in Fig.~\ref{fig:State_space_overview}a. Plots show the dynamics for random initial conditions after a sufficiently long transient time ($\bar{t}=5000$ arb.\ units). Line color indicates the weighted degree (green to blue indicates high to low values). (a) Traces at low bifurcation, top: bistability between laLC (left) and mixLC (right) in point A2 ($\mu=0.44, \sigma=0.17, \eta=0$); bottom: bistability between states in mixLC in point A3 ($\mu=0.64, \sigma=0.07, \eta=0$). (b) Traces at high bifurcation, top: bistability between mixLC (left) and haLC (right) in point B1 ($\mu=1.2, \sigma=0.29, \eta=0$); bottom: multistability between states in mixLC in point B3 ($\mu=1.28, \sigma=0.17, \eta=0$). The main difference between left and middle panels is the trajectory of the node with the smallest in-degree (highlighted in red). 
		(c) Traces showing noise induced state switching. Locations N1 and N2 are also close to the bifurcations but are not shown in Fig.~\ref{fig:State_space_overview}a since the state space changes under the influence of noise.  Top: Example at the low bifurcation in point N1 ($\mu=0.39, \sigma=0.2, \eta=0.024$). Bottom: Example at the high bifurcation in point N2 ($\mu=1.2, \sigma=0.3, \eta=0.024$). 
	} 
	\label{fig:State_space_traces}
\end{figure*}

The distinction between different dynamical states and types of network oscillations (laLC, mixLC, haLC) is particularly interesting since the transitions between these are the regions in state space, where we find multistable network states. 
Multistability is detected by simulating the network dynamics for 21 different initial conditions $\bm{x_0}$ and comparing the resulting time series by calculating their node-wise correlation in the activity variable.
We ignore the first $\bar{t}=5000$ arb.\ units of each time series to avoid analyzing transient effects and then compare the interval $t\in[5000, 6000]$ (equiv.\ to 10--35 periods) of one initial condition with $t\in[5000, 6200]$ of all other traces (second interval is sufficiently longer to account for all phase shifts).
If the auto-correlation of at least one initial condition is close to one, this indicates the existence of a stable (periodic) state, and if the cross-correlation between two (or more) initial conditions is different from one, this indicates bi-(or multi-)stability. Red pixels in Fig.~\ref{fig:State_space_overview}a show states with more than one stable solution, which are observed along both, the low and high bifurcation. 

Numerically we found regions of bistability between FP and mixLC (A1), laLC and mixLC (A2, B2), and mixLC and haLC (B1), as well as multistability in mixLC with different numbers of nodes fulfilling the criterion in Eq.~\eqref{eqn:LC_crit} (A3, B3) (cf.\ Fig.~\ref{fig:State_space_overview}a). 
The corresponding time series of the activity variables $x_{k1}$ are shown in Fig.~\ref{fig:State_space_traces}.
The automatic detection of multistability might not capture all multistable states, especially because some of the states may be very similar (cf.\ Fig.~\ref{fig:State_space_traces}b, bottom panel). More detailed analyses could thus also uncover bistabilities between up state and mixLC, as well as between mixLC and haLC at the low bifurcation.
We inspected the detected multistable states visually to assure that their difference are not due to transient effects and confirm their assignment to the specified multistable regions in Fig.~\ref{fig:State_space_overview}a.
\cgreen{We do not observe state switches in this noise-free case even for very long simulation times ($200,000$ arb.\ units simulated for multiple initial conditions in each multistable region, A1--3, B1--3, of Fig.~\ref{fig:State_space_overview}a), meaning that the states are stable and no temporal intermittency is found.}

Adding noise to the network most strongly affects the dynamics of the states close to the bifurcations.
The influence of noise on synchronous, mono-stable oscillatory states is, as expected, comparably small (cf.\ Figs.~S4d,~e in \cgreen{\cite{supp}}).
If the network dynamics is in a FP far away from the bifurcation line, the additional Gaussian white noise results in uncorrelated fluctuations in the activity variable (cf.\ Fig.~S4i in \cgreen{\cite{supp}}). 
Parametrizations that without noise would lead to a stable FP close to the bifurcation, show oscillations when sufficient noise is introduced (cf.\ Figs.~S4a,~h in \cgreen{\cite{supp}}). 
The clear distinction between the FP and laLC network states in the noise-free case (cf.\ Fig.~\ref{fig:State_space_overview}a) is therefore blurred for noisy dynamics.
For asynchronous mixLC states, the additional noise can have a synchronizing effect  (cf.\ Fig.~S4c in \cgreen{\cite{supp}}). This can be explained by more nodes fulfilling the criterion in Eq.~\eqref{eqn:LC_crit}, resulting in a stronger drive for the remaining nodes and therefore more synchronous oscillation.
These effects lead to a shift of the bifurcation lines compared to the noise-free case and consequently also affects the location of multistable network states. 

In the case of multistability, we observe that this stochastic additional input can lead to sufficient perturbations that drive the dynamics from one attractor to the other.
This results in noise-induced state switching, as shown in Fig.~\ref{fig:State_space_traces}c, at the bifurcation lines of the noisy state space. Such noise-induced transitions are a known phenomenon in systems of coupled oscillators \cite{horsthemke1984noise} and may be exploited for the control of the neural dynamics in practice \cite{schmidt2013micha}.

\section{Optimal control of the brain network dynamics}
\label{sec:applications}

\subsection{Switching between bistable network states}
\label{sec:state_switching}

In this section, we present optimal control inputs that induce a switch between previously identified multistable states. 
All control inputs optimize the cost functional given in Eq.~\eqref{eqn:cost_functional_general} with the state dependent terms given in  Eq.~\eqref{eqn:cost_functional_f_x1} and the energy and sparsity terms given in Eq.~\eqref{eqn:cost_functional_f_u}. 
Energy and sparsity terms are evaluated over the whole time interval during which the control is active. The cost functional itself, however, considers the deviation from the target state only the end of the control period.  
For this we set 
$I_p(t)=I_p^*$ for $(T - \tau) \leq t \leq T$ and $I_p=0$ else. 
The convergence criterion of the numerical solution of the minimization problem, as described in Section \ref{sec:min_problem}, is set to $\epsilon =  10^{-5}$ for all our applications. We ensured for all presented examples that the method actually converges and that results do not change for lower values for $\epsilon$.
We computed the optimal control for different phase shifts of the initial and target states with respect to the control onset and present the results for the phase shift which leads to the smallest total control energy $E$.

\begin{figure}
	\centering
	\includegraphics[width=0.5\textwidth]{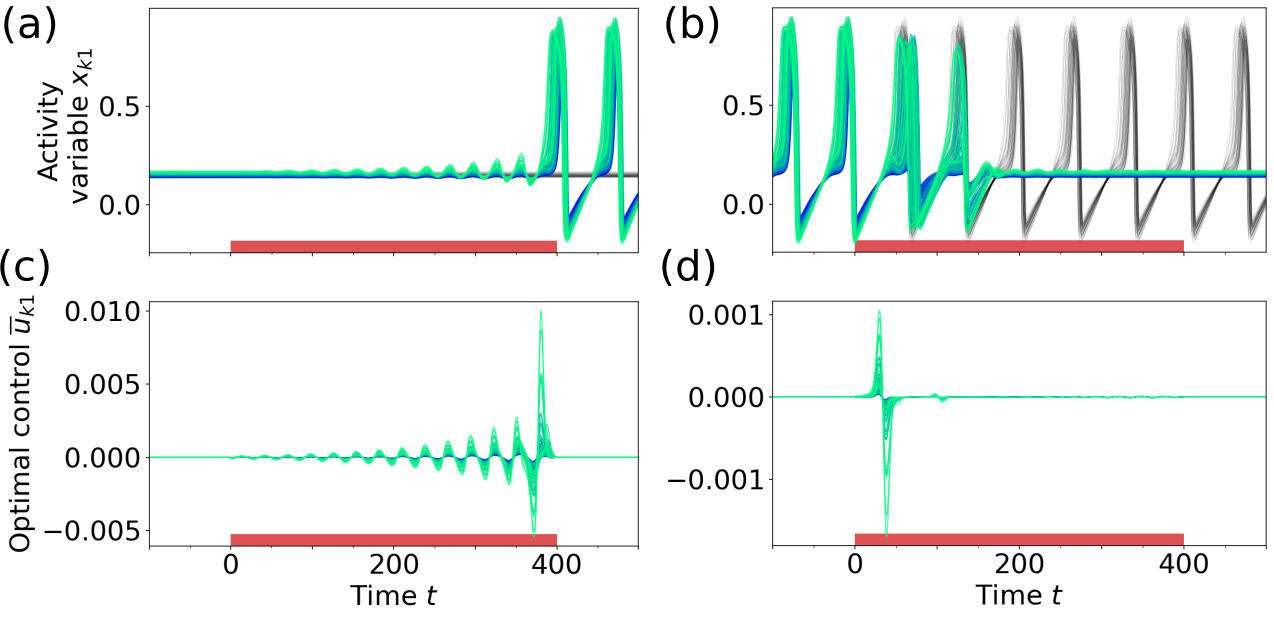}
	\caption{State switching with optimal control between bistable states at the low bifurcation (point A1 in Fig.~\ref{fig:State_space_overview}a). 
		(a) Switch from down-state to oscillatory state and (b) vice versa. Activities $x_{k1}$ of each node $k$ are shown over time. The colored lines show the controlled activities with \cgreen{green to blue indicating nodes with high to low weighted degree}. The \cgreen{ black} lines correspond to the uncontrolled activities. (c, d) Corresponding optimal control inputs $\overline{u}_{k1}$ to each node. The red \cgreen{bar indicates the time} interval during which the control is active (from $t=0$ to $t=400$). The deviation from the target state is penalized in the last $\tau=25$ time units of the control. Parameters were: $\mu=0.378$, $\sigma=0.21$, $\eta=0.0$, $I_p^*=0.0005$, $I_e=1.0$
		, $I_s=0$.}
	\label{fig:state_switching_fig1}
\end{figure}

Figure~\ref{fig:state_switching_fig1} shows the result of applying optimal control at point A1 in Fig.~\ref{fig:State_space_overview}a, where a low activity fixed point coexists with an oscillatory mixed state. 
The weight $I_s$ for the sparseness term in Eq.~\eqref{eqn:cost_functional_f_u} was set to zero.
Since the target state in this task is always stable, the network remains in this state once it is reached also after the control is turned off.
If the initial state is the down state FP, 
the obtained optimal control input 
oscillates synchronously for all nodes with increasing amplitude. The frequency of the control input [$f_{dom}(\overline{u}_1)=35.0$] corresponds to the frequency of the fixed point's focus [$f_{dom}(x^{FP}_1)=35.0$, measured by perturbing the FP], inducing resonance effects in the node activity. This strategy can be well observed in Video~1 (in \cgreen{\cite{supp}}), showing the node oscillations and the respective optimal control inputs in state space. 

When switching from the oscillatory to the down-state, as in Fig.~\ref{fig:state_switching_fig1}d, the optimal control input consists of one short biphasic pulse applied to all of the nodes followed by minor corrections. 
It is a known result from control theory, that applying a biphasic control pulse around an extreme point is an efficient way to drive an oscillating system to a stop. 
Interestingly, the optimal control strategy in this example is not to apply the pulse at an extreme point of the activity variables $x_{k1}$, but rather in a way that the gradient of the control is highest when the phase velocity of the limit cycle oscillation is the lowest (best observed in Video~2 in \cgreen{\cite{supp}}).
Although the deviation from the target state is only penalized at the end of the control interval, the pulse is applied early in order to give the system enough time to come to rest. 

\begin{figure}
	\centering
	\includegraphics[width=0.5\textwidth]{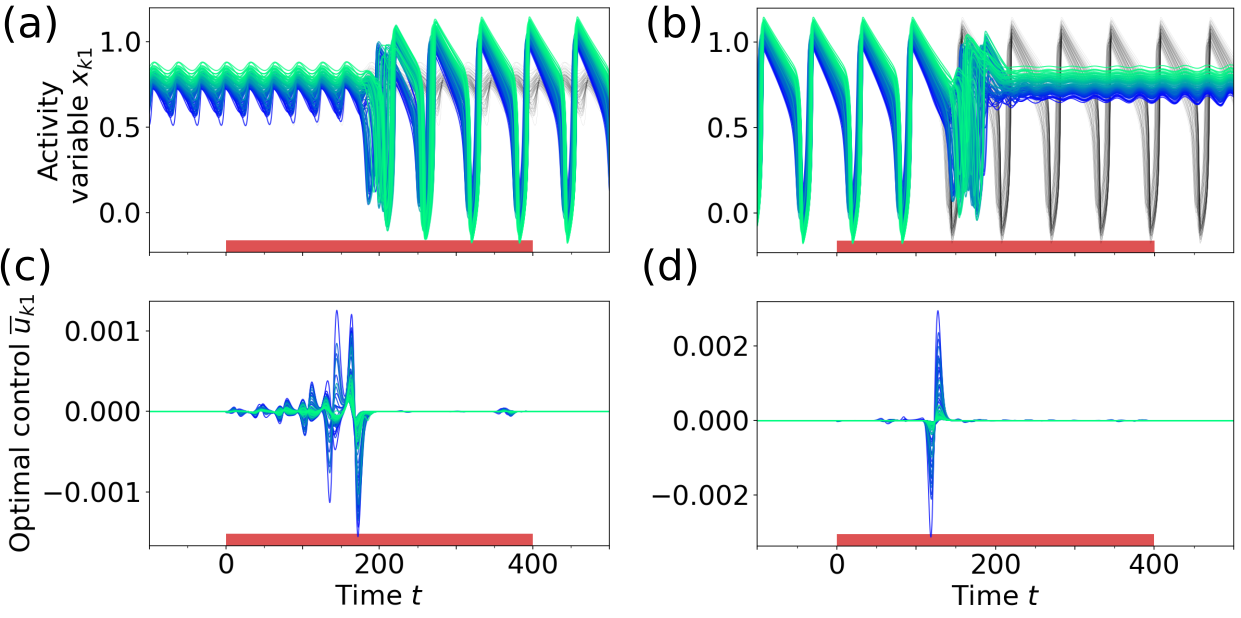}
	\caption{State switching with optimal control between bistable states at the high bifurcation (point B2 in Fig.~\ref{fig:State_space_overview}a). (a) Switch from low-amplitude oscillation to high-amplitude oscillation and (b) vice versa. Activities $x_{k1}$ of each node $k$ are shown over time. The colored lines show the controlled activities with \cgreen{green to blue indicating nodes with high to low weighted degree}. The \cgreen{ black} lines correspond to the uncontrolled activities. (c, d) Corresponding optimal control inputs $\overline{u}_{k1}$ to each node. The red \cgreen{bar indicates the time} interval during which the control is active (from $t=0$ to $t=400$). The deviation from the target state is penalized in the last $\tau=25$ time units of the control. Parameters were $\mu=1.22$, $\sigma=0.26$, $\eta=0.0$,  $I_p^*=7\times 10^{-5}$, $I_e=1.0$
		, $I_s=0$.}
	\label{fig:state_switching_fig2}
\end{figure}

Figure~\ref{fig:state_switching_fig2} shows the result of applying optimal control at point B1 in Fig.~\ref{fig:State_space_overview}a, where a low amplitude limit cycle (laLC) around the high activity up-state 
coexists with an oscillatory mixed state (mixLC, cf.\ Section \ref{sec:State_space}).
When switching from the laLC to the mixLC state (Figs.~\ref{fig:state_switching_fig2}a, c), the frequency of the control input [$f_{dom}(\overline{u}_1)=30.0$] is again adapted to the frequency of the initial state [$f_{dom}(x^{laLC}_1)=30.0$], utilizing resonance effects. While the nodes in the laLC state all oscillate with the same frequency ($f_k=f_{dom}=30.0 \ \forall \ k$), their phase speed is not the same at all times [reflected by lower synchrony $R(x^{laLC}_1)=0.861$, best observed in Video~3 in \cgreen{\cite{supp}}].
This results in a less synchronous oscillation of the control signals in Fig.~\ref{fig:state_switching_fig2}c ($R(\overline{u}_1)=0.714$) compared to the control inputs at the low bifurcation [Fig.~\ref{fig:state_switching_fig1}c, $R(\overline{u}_1)=0.999$]. 
In the other direction (Figs.~\ref{fig:state_switching_fig2}b,~d, Video~4 in \cgreen{\cite{supp}}), the optimal control finds a short, off-phase biphasic pulse analogous to Fig.~\ref{fig:state_switching_fig1}d. As a result, the amplitude of the network oscillation decreases to almost zero (for $t\in [200,250]$ in Fig.~\ref{fig:state_switching_fig2}b). Since there is no stable FP, the oscillation amplitude increases again over time 
and converges to the laLC target state approx.\ 550 time units after the control pulse (see also Fig.~S5 in \cgreen{\cite{supp}}).

\begin{figure}
	\centering
	\includegraphics[width=0.5\textwidth]{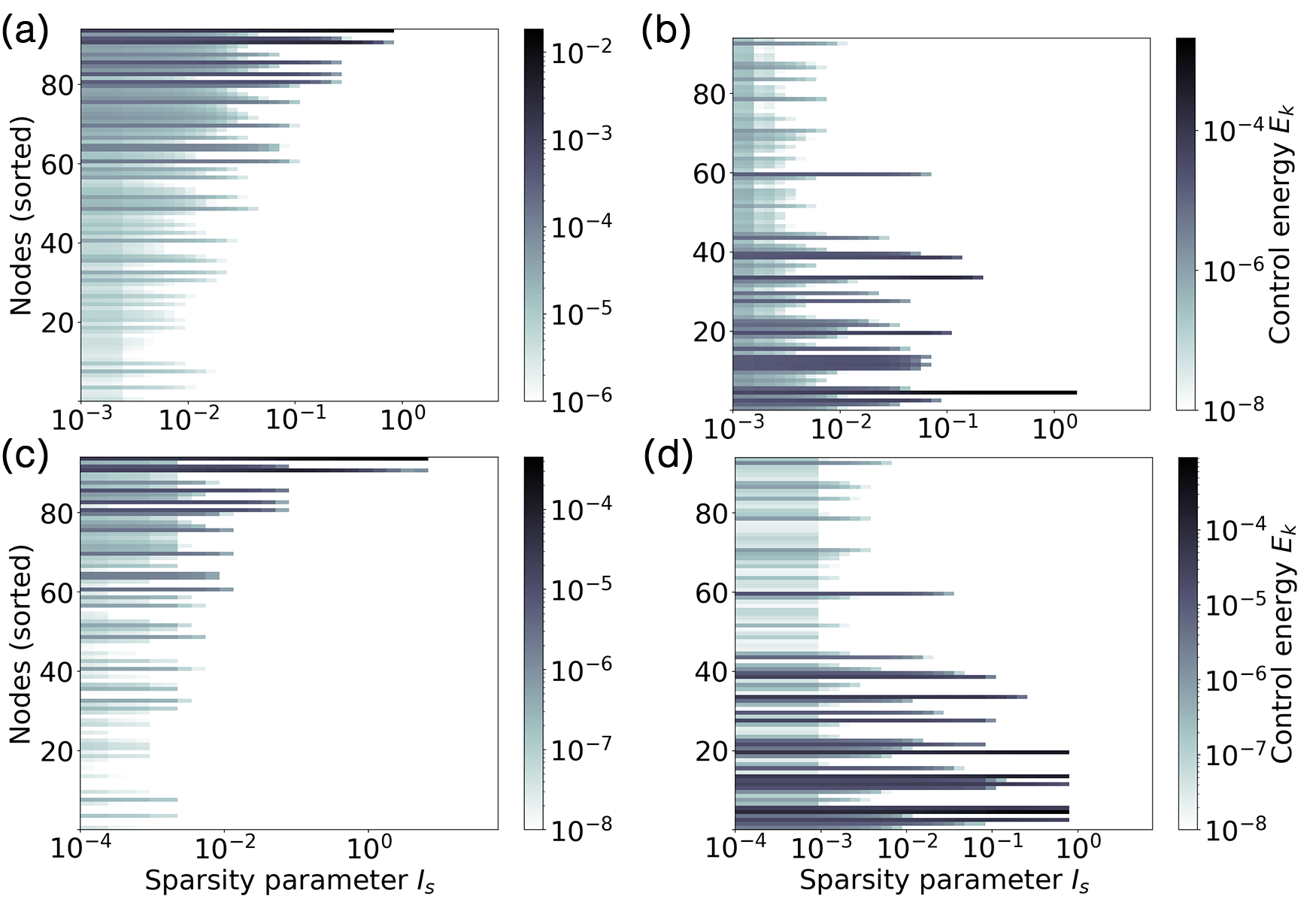}
	\caption{State switching with sparse optimal control. The nodes are sorted from lowest (bottom) to highest (top) weighted degree. The length of the bars indicate up to which value of the spatial sparsity parameter $I_s$ the node still receives finite control input. Their color denote the corresponding optimal control energies $E_k$ [Eq.~\eqref{eqn:control_energy}] for each node. (a) Switching from the down state (FP) to the oscillatory state (mixLC) with parameters close to the low bifurcation as in Fig.~\ref{fig:state_switching_fig1} and (b) Switching from low-amplitude (laLC) to high-amplitude oscillatory state (mixLC) with parameters close to the high bifurcation as in Fig.~\ref{fig:state_switching_fig2}. 
		(c) Same as (a) but switching from mixLC to FP. (d) Same as (b) but switching from mixLC to laLC.
	}
	\label{fig:state_switching_fig3}
\end{figure}

By increasing the sparsity parameter $I_s$ of the cost functional, we can tune the number of controlled nodes. Figure~\ref{fig:state_switching_fig3} shows the control energy $E_k$ [Eq.~\eqref{eqn:control_energy}] of the optimal control input $\overline{u}_{k1}$ to each node $k$ as a function of the sparsity parameter $I_s$. 
For low values of $I_s$, all nodes receive a finite control signal. 
When $I_s$ is increased, less nodes are controlled, until $I_s$ becomes so large, that the target state can no longer be reached.
As expected, decreasing the number of controlled nodes needs to be compensated for with a higher control energy. 

The results also show that at the low bifurcation (Figs.~\ref{fig:state_switching_fig3}a,~c), nodes with a high weighted degree receive, independent of the switching direction, a stronger control signal and remain controlled even for high values of $I_s$. 
This shows that at the low bifurcation it is most important to control the network hubs since they act as driver nodes for attractor switching. 
At the high bifurcation (Figs.~\ref{fig:state_switching_fig3}b,~d), on the other hand, nodes with low weighted degree receive the strongest control inputs. 

This different behavior can be explained based on the additive coupling scheme between the nodes 
(cf.\ Section~\ref{sec:State_space}), which causes nodes with higher degree to typically receive stronger inputs. 
When choosing parameters close to the low bifurcation, these high degree nodes therefore have higher oscillation amplitudes (cf.\ Fig.~\ref{fig:sgl_bifur_diag}) and are consequently driving the oscillation of the remaining nodes in the network. 
Close to the high bifurcation the nodes with lower degree, who receive less inputs, are more likely to remain in the limit cycle regime. Consequently -- and in contrast to the dynamics close to the low bifurcation -- the low degree nodes drive the oscillation of the network hubs. Videos 1 and 3 (in \cgreen{\cite{supp}}) illustrate how the control inputs on the respective driver nodes force the network dynamics to the high-amplitude oscillation target states.

The optimal control inputs to the control sites have well interpretable shapes. 
When the objective is to transition from a low-amplitude state to one with a higher amplitude (mixLC, Figs.~\ref{fig:state_switching_fig1}a and~\ref{fig:state_switching_fig2}a), the optimal control strategy utilizes the characteristics of the flow field in state space and synchronously drives the network with its resonant frequency from one attractor towards the other. Switching in the opposite direction, and therefore leaving this basin of attraction, is achieved most efficiently with one strong, biphasic pulse. This deflects the system from its initial mixLC trajectory and causes it to drop to the attractor with a lower amplitude.

\subsection{Synchronizing the network dynamics}
\label{sec:synchronize}

In this section, we apply the nonlinear control method to find the optimal inputs to synchronize the dynamics of the individual nodes. For this application we use the state dependent cost functionals given in Eq.~\eqref{eqn:cost_functional_f_x2} to penalize the deviation from the target cross correlation ($R_T=1$, fully synchronous state) and in Eq.~\eqref{eqn:cost_functional_f_u} to penalize the energy of the control and enforce its sparsity in space. 

\begin{figure}
	\centering
	\includegraphics[width=0.5\textwidth]{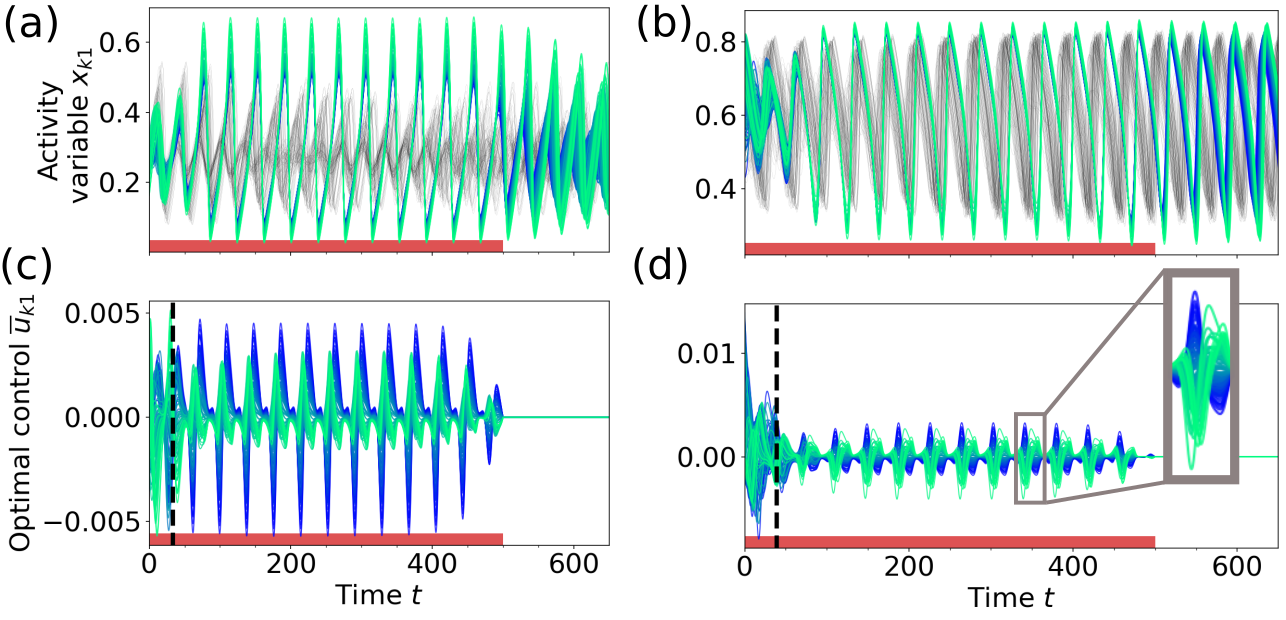}
	\caption{
		Synchronizing the noise-free network with optimal control inputs. 
		(a) Synchronization task at point S1 (low bifurcation) in Fig.~\ref{fig:State_space_overview}a. Activities $x_{k1}$ of each node $k$ are shown over time. The colored lines show the controlled activities (average cross-correlation $R = 0.995$ [Eq.~\eqref{eqn:nw_cross_correlation}]) with \cgreen{green to blue indicating nodes with high to low weighted degree}. The \cgreen{ black} lines show the uncontrolled activities ($R = 0.289$). 
		(b) Synchronization task at point S2 (high bifurcation) in Fig.~\ref{fig:State_space_overview}a (controlled: $R = 0.996$, uncontrolled: $R = 0.368$). 
		(c, d) Corresponding optimal control input $\overline{u}_{k1}$ to each node. The red  \cgreen{bar indicates the time} interval during which the control is active (from $t=0$ to $t=500$). The vertical black dashed line indicates the critical time $t_c$ (see text).
		Parameters were: Point S1 $\mu=0.7$ and $\sigma=0.025$, point S2 $\mu=1.3$ and $\sigma=0.025$, other parameters: $\eta=0$, $I_p=0.1$, $I_e=1.0$
		, $I_s=0$.
	}
	\label{fig:synchronize_trace}
\end{figure}

\begin{figure}
	\centering
	\includegraphics[width=0.5\textwidth]{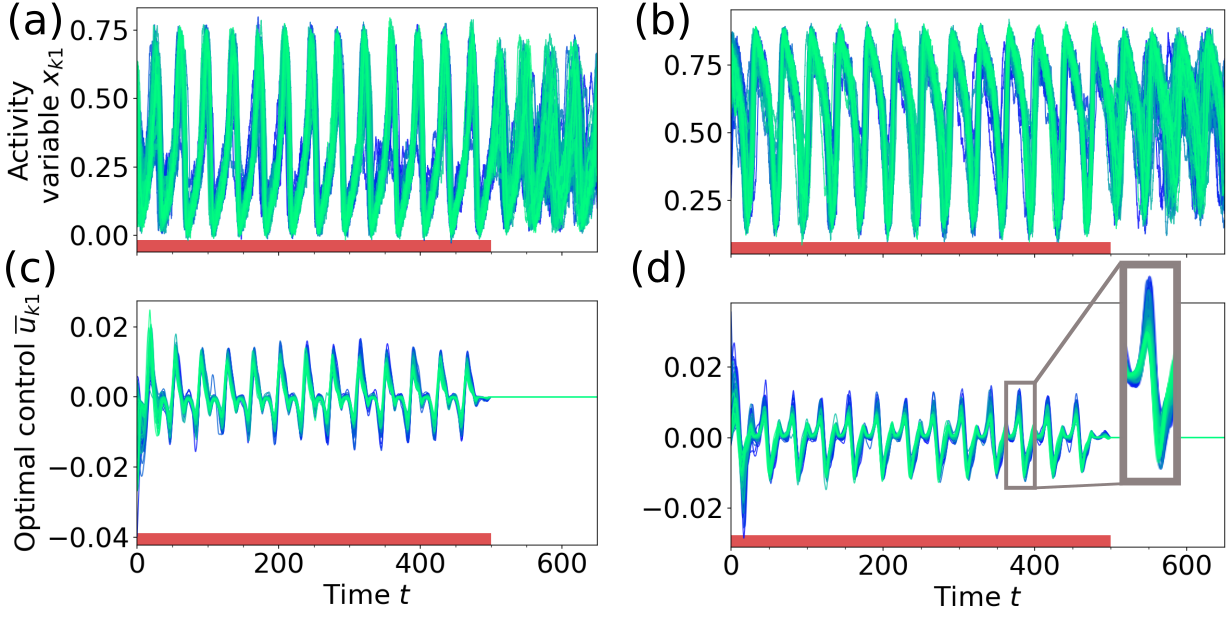}
	\caption{
		Synchronizing the noisy network with optimal control inputs. 
		(a) Synchronization task at point S1 (low bifurcation) in Fig.~\ref{fig:State_space_overview}a. Activities $x_{k1}$ of each node $k$ are shown over time with \cgreen{green to blue indicating nodes with high to low weighted degree} (average cross-correlation $R = 0.845$ [Eq.~\eqref{eqn:nw_cross_correlation}], uncontrolled: $R = 0.210$). 
		(b) Synchronization task at point S2 (high bifurcation) in Fig.~\ref{fig:State_space_overview}a (controlled: $R = 0.853$, uncontrolled: $R = 0.321$).
		(c, d) Corresponding optimal control input $\overline{u}_{k1}$ to each node. The red \cgreen{bar indicates the time} interval during which the control is active (from $t=0$ to $t=500$). 
		Parameters were $\eta=0.024$, $I_p=0.1$, $I_e=1.0$
		, $I_s=0$, all other parameters are as in Fig.~\ref{fig:synchronize_trace}.
	}
	\label{fig:synchronize_trace-noise}
\end{figure}

We parameterize the system to be in the two asynchronous regions marked by the labels S1 and S2 in Fig.~\ref{fig:State_space_overview}a, close to the low and high bifurcation lines. The optimal control method is then applied to synchronize the dynamics 
for the noise-free (Fig.~\ref{fig:synchronize_trace}) and noisy (Fig.~\ref{fig:synchronize_trace-noise}) case. Both figures show the controlled (synchronous) and uncontrolled (asynchronous) time series and the corresponding optimal control inputs. 
Since the dynamics in points S1 and S2 are monostable, the synchronous state is not a stable solution and the system returns to its original state as soon as the control is switched off. The network dynamics and the optimal control strategies are best seen in Videos 5-8 (in \cgreen{\cite{supp}}).

Figure~\ref{fig:synchronize_trace} shows how the optimal control inputs synchronize the oscillation of all nodes in the network. 
In the uncontrolled state (in gray, Figs.~\ref{fig:synchronize_trace}a, b) the nodes at both bifurcations oscillate with similar frequencies (mean and standard deviation of oscillation frequencies across nodes at S1: $\langle f\rangle_N = 32.2, \, \sigma_{ \langle f\rangle_N}=1.1,$ and at S2: $\langle f\rangle_N = 27.0, \, \sigma_{ \langle f\rangle_N}=0.9$).
Close inspection of the control inputs in Figs.~\ref{fig:synchronize_trace}c and d shows that the optimal control acts in two phases. First, the control aligns the phases of all oscillators until all nodes are synchronized for the first time, which we defined as the critical time $t_c$ (\cgreen{please refer to} \cgreen{\cite{supp}} for details about the computation of $t_c$). Second, a periodic input maintains the synchronous oscillation during the period the control is active.

\begin{figure}
	\centering
	\includegraphics[width=0.5\textwidth]{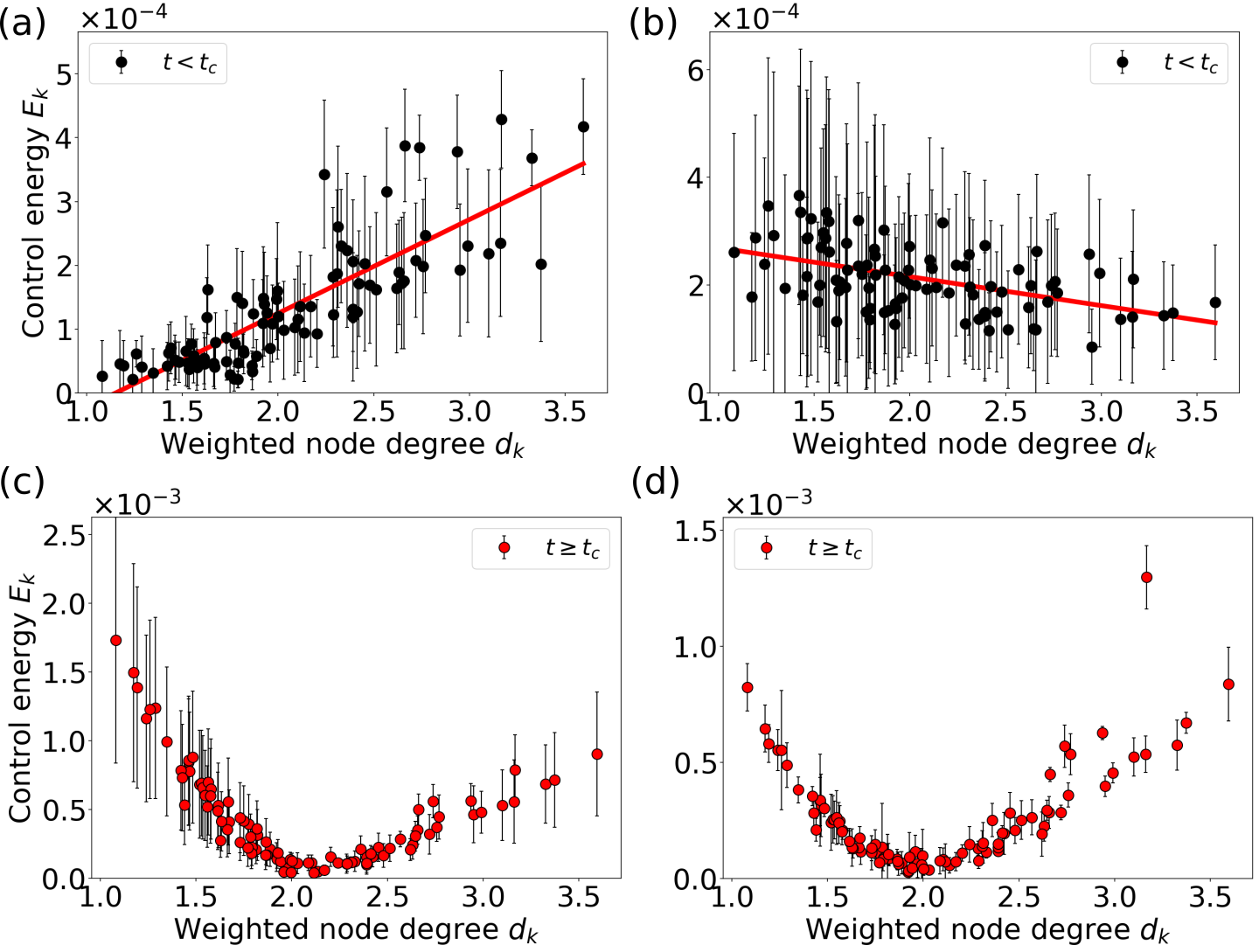}
	\caption{
		Mean energies $E_k$ of the control inputs [Eq.~\eqref{eqn:control_energy}] as a function of the weighted node degree $d_k$ for the noise-free synchronization task.
		The error bars show the standard deviation over 10  different initial conditions of the network dynamics. (a, b) The control signal for time $t \in [0,t_c)$ is considered. Linear regression (red line) coefficients are $r=0.83, p<10^{-24}$ and $r=-0.49, p<10^{-6}$. (c, d) The control signal for time $t \in [t_c, T]$ is considered. (a, c) Point S1 (low bifurcation) in Fig.~\ref{fig:State_space_overview}a. (b, d) Point S2 (high bifurcation) in Fig.~\ref{fig:State_space_overview}a. All parameters are as in Fig. \ref{fig:synchronize_trace}.
	}
	\label{fig:synchronize_energy}
\end{figure}

For times $t<t_c$, Fig.~\ref{fig:synchronize_energy}a shows that at the low bifurcation the energies $E_k$ of the control inputs to the nodes are positively correlated with the weighted node degrees.  
Thus, it is beneficial to focus the control input on the network hubs for alignment. 
A possible interpretation for this strategy is that the optimal control utilizes the influence of the network hubs on the remaining nodes to force them on the synchronous limit cycle trajectory. 
However, we again observe a different behavior at the high bifurcation, as shown in Fig.~\ref{fig:synchronize_energy}b. Here, $E_k$ is negatively correlated with the weighted node degree for times $t<t_c$. 
In this case, the network coupling is much smaller compared to the background input~$\mu$ and the node's phase space oscillations are similar to the trajectory of the limit cycle of an uncoupled node (cf.\ Video~6 in \cgreen{\cite{supp}}).
The negative correlation in Fig.~\ref{fig:synchronize_energy}b suggests that if the dynamics of the nodes are similar in the first place, it is beneficial to focus the control on more weakly coupled nodes and align their phase space trajectory  with the trajectory of the network hubs. Similar to the attractor switching at the high bifurcation (cf.\ Figs.~\ref{fig:state_switching_fig3}b,~d), the control of the low degree nodes drives the network oscillation towards the target state.

In the second phase, for $t\geq t_c$, the  high degree nodes without control would have a higher (or lower) phase velocity than the nodes with intermediate degree, while the opposite is true for nodes with low degree. 
The optimal control strategy is thus to act on the low and high degree nodes in opposite directions, which can be observed as antiphase control inputs shown in Figs.~\ref{fig:synchronize_trace}c and~d, where nodes with intermediate degree receive only small control inputs. 
(Videos 5 and 6 in \cgreen{\cite{supp}} show how the control inputs keep the nodes together in phase space.)
This is reflected in the arch-like form of the mean energies $E_k$ of the control inputs at both bifurcations, as seen in Figs.~\ref{fig:synchronize_energy}c and~d, with high control energies for nodes with low or high degree.
As a result, the optimal control achieves a mean network cross-correlation of $R = 0.995$  in the time interval $t\in [t_c, T]$, compared to the uncontrolled scenario with $R = 0.289$ at point S1 (Fig.~\ref{fig:synchronize_trace}a), and $R = 0.996$ compared to $R=0.368$ at point S2 (Fig.~\ref{fig:synchronize_trace}b).

Figure \ref{fig:synchronize_trace-noise} shows results for the application of optimal control to the synchronization task for the case of finite noise.
We obtain a mean network cross-correlation of $R = 0.83$ (analysis of the deviation from $R_T$ in \cgreen{\cite{supp}}, uncontrolled: $R=0.25$) at point S1 (Fig.~\ref{fig:synchronize_trace-noise}a) \cgreen{and $R = 0.85$} (uncontrolled: $R=0.35$) at point S2 (Fig.~\ref{fig:synchronize_trace-noise}b). 
We conclude that the optimal control input successfully synchronizes the network dynamics, just as in the noise-free case. 

The control strategy that leads to the respective target state, however, changes substantially when noise is added to the system.
The control input in the noise-free case is tailored individually for each node, slowing the phase space velocity of some nodes down and speeding others up at the same time (cf.\ blue inset in Fig.~\ref{fig:synchronize_trace}d, mean cross-correlation of the control inputs during this control period is 0.03).
In contrast to that, the control input for the noisy network dynamics has a similar shape for all nodes (cf.\ blue inset in Fig.~\ref{fig:synchronize_trace-noise}d, mean cross-correlation of the control inputs during control period is 0.88) with varying amplitude depending on the weighted node degree $d_k$.

\begin{figure}
	\centering
	\includegraphics[width=0.5\textwidth]{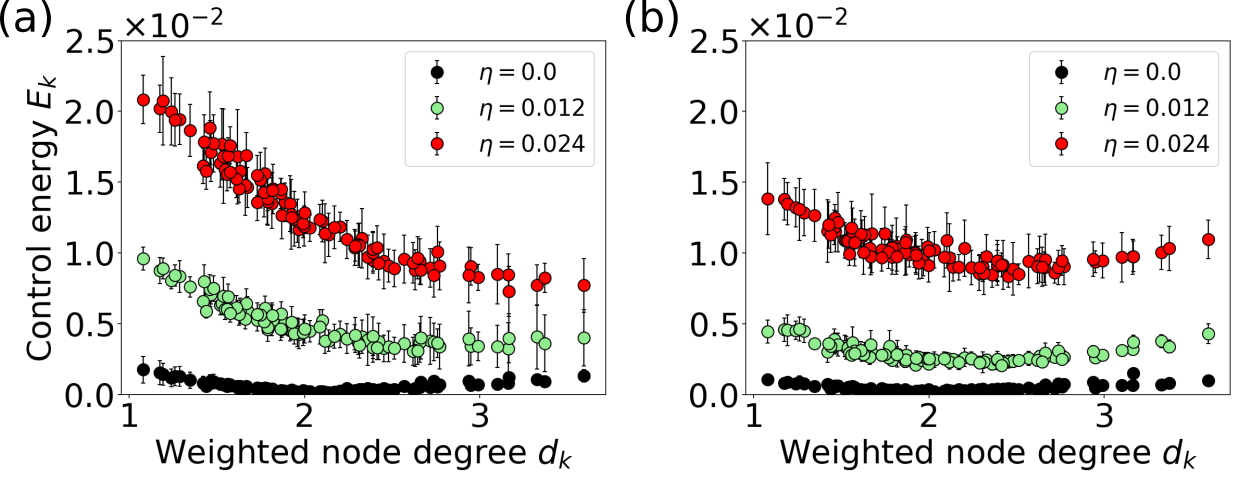}
	\caption{Mean energies $E_k$ of the optimal control inputs [Eq.~\eqref{eqn:control_energy}] as a function of the weighted node degree $d_k$ for different noise levels $\eta$ for the synchronization task. The error bars show the standard deviation over 10 different initial conditions of the network dynamics with 20 independent noise realizations of the optimization each. \cgreen{$E_k$ is, in contrast to Fig.~\ref{fig:synchronize_energy}, calculated over the whole interval during which the control is active.} 
	(a) Point S1 (low bifurcation) in Fig.~\ref{fig:State_space_overview}a. (b) Point S2 (high bifurcation) in Fig.~\ref{fig:State_space_overview}a. Parameters were $I_p=0.1$, $I_e=1.0$
	, $I_s=0$, all other parameters are as in Fig.~\ref{fig:synchronize_trace}.
	}
	\label{fig:synchronize_energy-noise}
\end{figure}

This change in control strategy is also reflected in Fig.~\ref{fig:synchronize_energy-noise}, which shows the control energy $E_k$ per node for different noise levels.
The control energy increases for all nodes with increasing noise strength, both at the low (Fig.~\ref{fig:synchronize_energy-noise}a) and the high bifurcation (Fig.~\ref{fig:synchronize_energy-noise}b).
With increasing noise level $\eta$, we observe a gradual transition of the optimal control strategy. 
Instead of balancing out the phase velocity differences between nodes towards the phase of nodes with intermediate degree (cf.\ Videos 5 and~7 in \cgreen{\cite{supp}}), the control in the noisy case forces the system on a limit cycle with a slightly higher oscillation amplitude (cf.\ Videos 6 and~8 in \cgreen{\cite{supp}}).
This strategy requires a higher control energy but is independent of the specific realization of the noise. 

\begin{figure}
	\centering
	\includegraphics[width=0.5\textwidth]{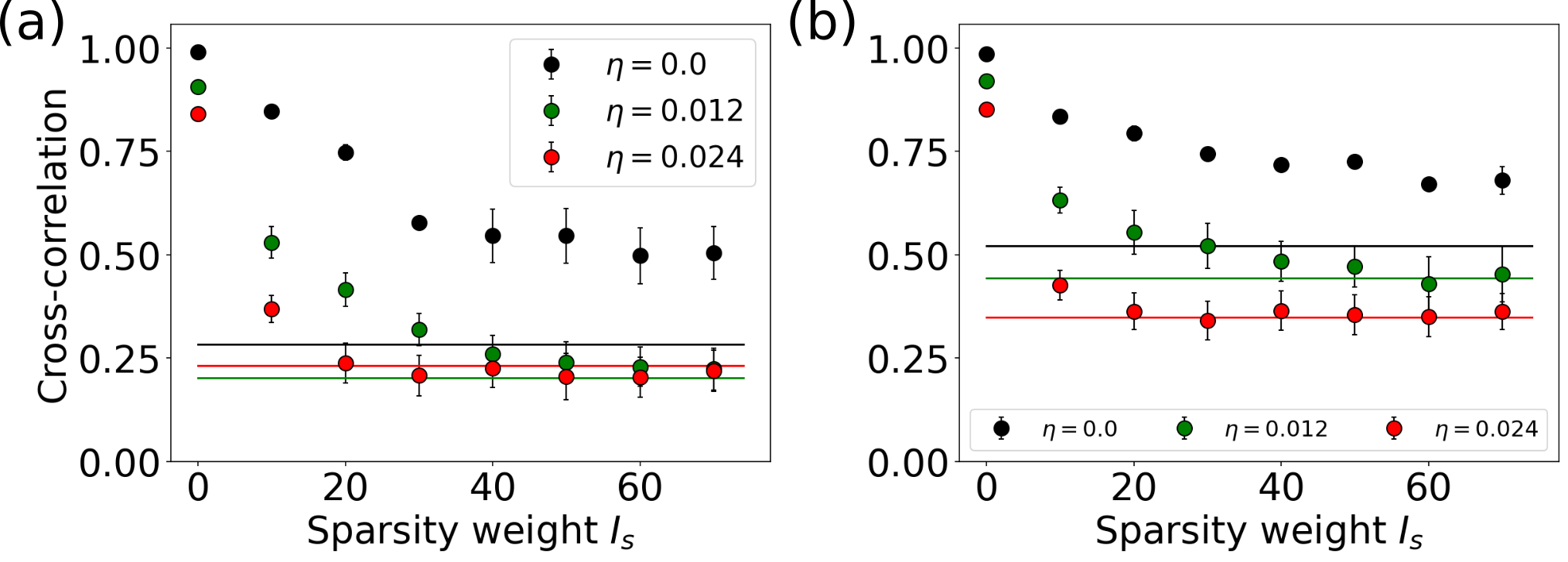}
	\caption{Average network cross-correlation $R$ [Eq.~\eqref{eqn:nw_cross_correlation}] with sparse optimal control as a function of the sparsity parameter $I_s$ [Eq.~\eqref{eqn:cost_functional_f_u}] for different noise levels $\eta$. 
		The mean and standard deviation are shown for 5 different initial conditions of the network dynamics, each with 20 independent noise realizations. Horizontal lines indicate the cross correlation $R$ for the uncontrolled case. (a)  Point S1 (low bifurcation) in Fig.~\ref{fig:State_space_overview}a. (b)  Point S2 (high bifurcation) in Fig.~\ref{fig:State_space_overview}a. Parameters were $I_p=0.1$, $I_e=1.0$
		, all other parameters are as in Fig.~\ref{fig:synchronize_trace}.}
	\label{fig:synchronize_sparsity}
\end{figure}

By imposing sparsity constraints on the system, we can investigate to what extent synchronization can be achieved with fewer control sites.   
Figure \ref{fig:synchronize_sparsity} shows the relation of the synchrony in the network, measured by the average cross-correlation~$R$ [Eq.~\eqref{eqn:nw_cross_correlation}], to the sparsity parameter~$I_s$. 
The average cross-correlation $R$ decreases with increasing sparsity parameter $I_s$ at both locations in state space and for all noise levels.
This shows that, under the imposed constraints, it is not possible to fully synchronize the network with sparse control. 
Optimally synchronizing aperiodic states is a collective effort. Unlike state switching it cannot be achieved by controlling only a few sites, because the reduced number of control sites cannot sufficiently be compensated for by a higher control energy. 
Especially in the case of noisy network dynamics, where the control must drive all nodes, the network's synchrony quickly drops to the value of the uncontrolled network when $I_s$ is increased. 
For fixed $I_s$ the number of controlled nodes and the control sites optimal for synchronizing the network changes for each initial condition $\bm{x_0}$ (cf.\ Fig. S1 in \cgreen{\cite{supp}}), as it depends on details of the exact network dynamics. 

\cgreen{\section{Comparison with controllability measures derived from linear control theory}}
\label{sec:linear_vs_nonlinear}
When relating functional properties of a neural system to properties of the underlying connectome, neural activity is often approximated by linear dynamical systems \cite{honey2009predicting, galan2008network, gu2015controllability}. This has obvious benefits for analyzing the effects of perturbations\cgreen{. C}alculating optimal control inputs for linear systems, as in Refs. \cite{bassett2017network, gu2017optimal}, can be done analytically and with little computational effort\cgreen{, and conclusions about the effects of external inputs can be drawn from controllability measures, which depend on network topology only \cite{tang2018colloquium}}. 

\cgreen{Two of these measures have previously been applied to quantify the impact of perturbations in a whole-brain network setting \cite{gu2015controllability, muldoon2016stimulation, gu2017optimal}.}
\textit{Modal controllability} refers to the ability of a node to control each evolutionary mode of a dynamical network \cite{hamdan1989measures}, and the \textit{average controllability} is given by the average control input energy to the respective node over all possible target states \cite{muldoon2016stimulation, gu2015controllability} (mathematical description in Appendix~\ref{sec:appendixControllability}). 
Nodes with high average controllability require only low energy input to move a linear system into ``easy-to-reach'' states, 
while nodes with high modal controllability require a high control energy input to have an effect on the dynamics but are crucial when the target state of the system is ``difficult-to-reach''~\cite{gu2015controllability, tang2020control, tang2018colloquium}.
It has also been shown that the average controllability of a node is strongly correlated with its weighted degree~$d_k$ ($r=0.85, p<10^{-26}$ for our structural connectivity matrix) while the modal controllability is known to have a strong inverse correlation ($r=-0.82, p<10^{-23}$, see Fig.~S7 in \cgreen{\cite{supp}}).

\begin{figure}
	\centering
	\includegraphics[width=0.5\textwidth]{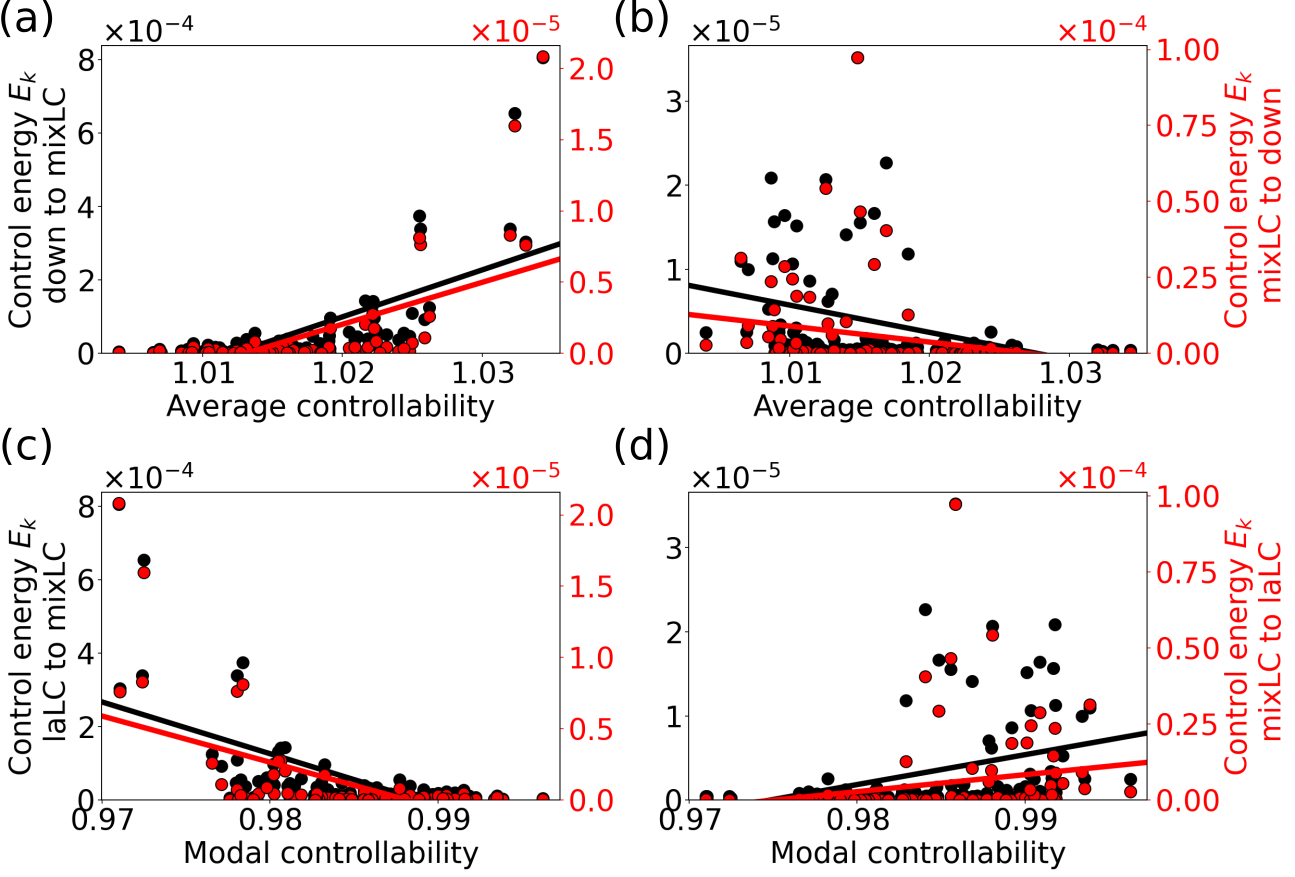}
	\caption{Mean energies $E_k$ of the optimal control inputs [Eq.~\eqref{eqn:control_energy}] plotted against measures from linear control theory for state switching tasks from down state / low amplitude oscillation towards high amplitude oscillation (black) and the opposite direction (red).
		(a) $E_k$ vs. average controllability when switching at the low bifurcation (Point A1 in Fig.~\ref{fig:State_space_overview}a).
		(b) Same as (a) but at the high bifurcation (Point B2 in Fig.~\ref{fig:State_space_overview}a).
		(c) $E_k$ vs. modal controllability at A1 and
		(d) at B2.
		Solid lines denote the results of a linear regression with the following obtained coefficients: (a) $r=0.66, p<10^{-12}$ for the down-to-up switch (black) and $r=0.61, p<10^{-10}$ for the up-to-down switch (red). (b) $r=-0.33, p=0.001$ (black) and $r=-0.24, p=0.02$ (red). (c) $r=-0.62, p<10^{-10}$ (black) and $r=-0.57, p<10^{-8}$ (red). (d) $r=0.31, p=0.003$ (black) and $r=0.21, p=0.04$ (red). Same parameters as in Figs.~\ref{fig:state_switching_fig1} and \ref{fig:state_switching_fig2}, for low and high bifurcation, respectively.}
	\label{fig:ln_switch}
\end{figure}

\cgreen{We now apply the diagnostics from linear control theory to the brain network model, Eq.~\eqref{eqn:network_simple}.} Figure \ref{fig:ln_switch} shows the correlations of the energies $E_k$ of the optimal control inputs with the average and modal controllability for the attractor switching tasks. 
At the low bifurcation A1 (Figs. \ref{fig:ln_switch}a, c) we find, that the energy of the control input for a node is positively correlated with its average controllability and negatively correlated with its modal controllability, both irrespective of the switching direction.
Therefore, the most efficient strategy for switching between the attractors is to control the network hubs with high average controllability which then force the other nodes towards the target state.
These results can be considered consistent with predictions from linear control theory \cite{gu2015controllability} and previous results on the global impact of stimulation \cite{muldoon2016stimulation}.  

At the high bifurcation at location B2 (Figs.~\ref{fig:ln_switch}b, d), however, we observe the opposite trend, with the control energy being negatively correlated with the average and positively correlated with the modal controllability. 
Predictions based on linear control theory alone are indifferent to the actual dynamics and therefore not able to distinguish between the two cases. 
Yet, nodes with high modal controllability play a much more important role in affecting the global dynamics at this location in state space.

\begin{figure}
	\centering
	\includegraphics[width=0.5\textwidth]{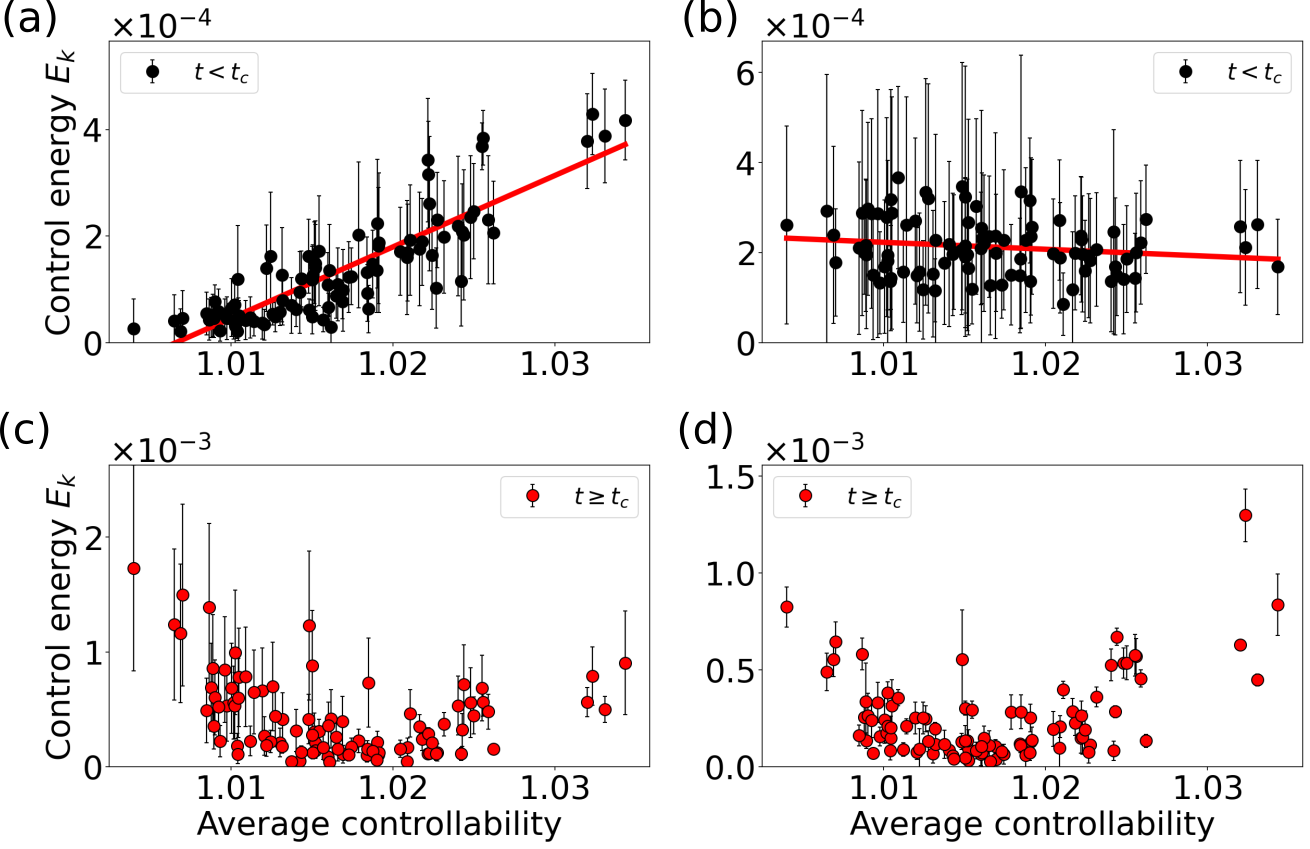}
	\caption{Mean energies $E_k$ of the optimal control inputs [Eq.~\eqref{eqn:control_energy}] plotted against the average controllability for the synchronization task (cf.\ Fig.~\ref{fig:synchronize_energy}). 	
		(a)~$E_k$ vs. average controllability, with control signal considered during the initial time interval $t \in [0,t_c)$, at the low bifurcation (Point S1 in Fig.~\ref{fig:State_space_overview}a). 
		(b)~Same as (a) but at the high bifurcation (Point S2 in Fig.~\ref{fig:State_space_overview}a).
		Solid red lines denote the linear regression results with the following obtained coefficients:
		(a)~$r=0.87, p<10^{-29}$, (b)~$r=-0.16, p=0.12$ (not significant).
		(c)~$E_k$ vs. average controllability, with control signal considered for time $t \in [t_c, T]$, at S1.
		(d)~Same as (c) but at S2.
		All parameters as in Fig.~\ref{fig:synchronize_energy}.}
	\label{fig:ln_syn}
\end{figure}

For the synchronization task, we observe a similar pattern. While there is a clear correlation between the optimal node-wise control energies $E_k$ with the average controllability when aligning the phases (i.e.\ $t<t_c$) at the low bifurcation (Fig.~\ref{fig:ln_syn}a), no correlation is observed at the high bifurcation (Fig.~\ref{fig:ln_syn}b). (The corresponding results involving modal controllability are shown in Fig.~S8 in \cgreen{\cite{supp}}.)
In the case of maintaining the synchronous network dynamics ($t\ge t_c$, Figs.~\ref{fig:ln_syn}c,~d), we observe -- at both bifurcations -- a similar but less obvious arclike shape as in Figs.~\ref{fig:synchronize_energy}c,~d. 
For noisy network dynamics (cf.\ Fig.~S9 in \cgreen{\cite{supp}}), however,  $E_k$ is negatively correlated with the average controllability at both bifurcations (cf.\ Fig.~S9 in~\cgreen{\cite{supp}}).

This shows that in our nonlinear setting, linear controllability measures do not provide additional insights compared to the weighted node degree, and that intuitions based on these measures can be misleading.
\cgreen{Diagnostics from linear control theory have previously been claimed to be predictive for a node's role in driving brain state transitions \cite{gu2015controllability, gu2017optimal} or for the global impact of local brain stimulation \cite{muldoon2016stimulation}. Our results however show, that the  optimal control inputs and sites in nonlinear systems not only depend }
on the structural network connectivity, but also on the location in state space, the control task, and other factors like the amount of noise.\\

\section{Discussion}
\label{sec:summary}

In this contribution we apply techniques from the optimal control of nonlinear dynamical systems to the dynamics of brain network models. 
Nodes were equipped with FitzHugh-Nagumo oscillators, since they are simple and well studied nonlinear models for neural dynamics. 
Changing the background input for the FHN nodes or the global coupling strength of the network can both lead to transitions between two different stable fixed points and two different limit cycle attractors (laLC and haLC). The interaction between nodes in different oscillation states (mixLC) can lead to an asynchronous network dynamics.
At the bifurcations, we also find different coexisting stable states. 

The general mathematical framework of the optimal control of partial differential equations \cite{troltzsch2010optimal} is adapted for noisy dynamical systems on graphs,  where the local network dynamics, the noise level, the network connectivity, and the local coupling schemes [Eq.~\eqref{eqn:network_system}] can be freely chosen.
The state dependent part of the cost functional is averaged over multiple noise realizations and penalizes the deviation of the network dynamics from a task dependent control target [Eq.~\eqref{eqn:cost_functional_f_x1} for attractor switching, Eq.~\eqref{eqn:cost_functional_f_x2} for synchronization].
The part of the cost functional which depends on the control input [Eq.~\eqref{eqn:cost_functional_f_u}] penalizes the control energy and non-sparse solutions. 
The presented method is applicable to any network that can be described in the form of Eq.~\eqref{eqn:network_system}, including models of power grids, social networks, or climate dynamics.

A common problem for gradient based methods are potential local optima of the cost functional which may prevent the convergence to a globally optimal solution. To alleviate this problem we performed the minimization as described in Section~\ref{sec:min_problem} with different initial conditions $\bm{u}_0$ for the control time-series and choose the result with minimal cost. The set of initial conditions included $\bm{u}_0=0$ as well as valid control time-series taken from different parametrizations. 

We use the optimal control method to cause targeted attractor switching between previously identified coexisting stable states.
When no sparsity is enforced, we show that it is optimal to resonantly drive all nodes to transition to the high amplitude oscillatory state. 
When the task is to switch to a state with lower amplitude or no oscillation, the optimal strategy is to apply a precisely timed biphasic pulse.
The nodes that receive the largest control energy are the same for both of these switching directions and are also the ones that are still controlled when we enforce sparsity in space.
Depending on the location in state space, either nodes with high degree (at the low bifurcation) or low degree (at the high bifurcation) are the ones that most efficiently drive the network dynamics from the initial to the target state. 
When sparsity is enforced, controlling only a small number of these driving nodes with increased control energies $E_k$ is sufficient to switch from one attractor to another in an optimal way.

In the second application, we show that our method can also be used to control global properties of the network dynamics, such as the average cross-correlation between node activities in the oscillating regime.
Immediately with control onset, the control signal acts on all nodes to align their phases. As soon as this is achieved, the control maintains the synchronous oscillation with periodic control signals.
Which nodes receive larger control input for the initial alignment again depends on the location in the state space. 
Both at the low and high bifurcation, individually adapted control inputs lead to a successful synchronization of the network dynamics.
While the average cross-correlation $R$ is increased also with sparse control, the synchronous target state is only achieved in an optimal way, when all nodes in the network receive a finite control input.
This suggests that synchronizing all nodes needs collective intervention, while attractor switching can be caused by controlling a few selected nodes only.
The introduction of noise to the system makes the dynamics of each node less predictable, resulting in a loss in specificity and more similar control inputs to the nodes.
Consequently, the optimal control of noisy network dynamics requires higher total control energy $E$ (cf.\ Fig.~\ref{fig:synchronize_energy-noise}) while resulting in a lower precision (cf.\ Fig.~\ref{fig:synchronize_sparsity} for $I_s = 0$).

The information on the different states and bifurcations, which we show to be a decisive factor for choosing the optimal control sites in our applications, is lost when techniques from linear control theory are applied. While predictions do qualitatively agree for certain dynamical systems, control tasks, and locations in state space (cf.\ \cite{muldoon2016stimulation}), this agreement does not hold in a general setting. It would, however, be worthwhile investigating for what classes of dynamical systems and control tasks  ``controllability measures'' can be defined, which only depend on the properties of the connectome.

In this contribution, the techniques of nonlinear optimal control were applied to a simplified model of the global brain dynamics. The bifurcations in our FHN model (cf.\ Fig.~\ref{fig:sgl_bifur_diag}) phenomenologically capture the state transitions found in more complex, biophysically motivated network models \cite{cakan2019} as we have adapted the FHN parameters to resemble their dynamics. 
\cgreen{Consequently, the activity variables of the FHN nodes can be interpreted as relative output firing rates of cortical nodes, and their total input $\mu + \sigma \sum_{i=1}^N \bm{A}_{ki} x_{i1} (t)$ can -- at least qualitatively -- be interpreted as a proxy of the local field potential \cite{mazzoni2015computing}. 
Given this interpretation, model down-(up-)states are related to down-(up-)states in cortical physiology, while an oscillatory regime of the whole-brain model corresponds to a brain state with oscillations present.}
Locations at the lower bifurcation of the network model were implicated as proper ``operating points'' for modeling the brain's resting state activity \cite{deco2012ongoing, deco2013resting, demirtacs2019hierarchical}, while locations at the high bifurcation can be considered a cartoon model for the global brain dynamics \cgreen{during non-Rapid Eye Movement (non-REM) sleep (see \cite{cakan2020deep} for a biophysically more detailed model)}. 

\cgreen{In this work we consider a network with instantaneous couplings, although coupling delays in human white matter are finite. The interpretation of results regarding the model’s state space should thus be limited to cortical states whose dynamics is slow compared to these delays. Given that delays are of the order of approx.\ 5-15~ms \cite{caminiti2013diameter} brain oscillations with periods of approx.\ 1~s or less would qualify. These include the so-called slow oscillations, which are the prominent brain rhythm during non-REM sleep and which have already been subject to external perturbation experiments in human neuroscience and clinical settings \cite{marshall2004transcranial, Ladenbauer2016}.}

\cgreen{In lieue of a biologically more detailed model of the effects of control inputs on whole-brain dynamics, delays must be included as soon as brain oscillations with frequencies above a few Hz are of interest. Relative delays can be computed from DTI-based estimates of the length of white matter fiber tracts and included in the coupling terms of Eqs.~\eqref{eqn:network_system} and \eqref{eqn:network_simple}.}

\cgreen{When relative delays are scaled by a global delay constant, the ratio of the coupling delays to the FHN oscillation period can be adapted to match the physiologically observed ratios for any given brain rhythm of interest. The formalism of nonlinear optimal control summarized in Section~\ref{sec:nonlinear_control} must extended, though, to cover finite delays by modifying the coupling term in Eq.~\eqref{eqn:network_system} to, e.g., $\sigma (\bm{A} \otimes \bm{G}) \bm{x}(t-d)$ for the case of constant delays. The adapted coupling term reappears when computing the adjoint state with the differential equation [Eq.~\eqref{eqn:adjoint_state}], and thus the iterative algorithm for the numerical optimization (cf.\ Section~\ref{sec:min_problem}) has to be adapted accordingly in the calculation of step~3. }

Here we used a \cgreen{highly} simplified model of the global brain dynamics to showcase the applicability of nonlinear optimal control and its added value beyond \cgreen{connectome-based diagnostics derived} from linear control theory. When applied to biophysically more realistic models of whole-brain activity (cf.\ \cite{cakan2020deep}), nonlinear optimal control may serve as a tool to evaluate the impact of external stimulation and to facilitate the design of new brain stimulation protocols. 
Non-invasive brain stimulation like transcranial current stimulation \cite{nitsche2011transcranial, antal2008comparatively, terney2008increasing} is a highly promising technique to perturb the global brain dynamics with the goal to improve sensory \cite{behrens2017long}, motor \cite{moisa2016brain}, and cognitive abilities \cite{marshall2004transcranial} of human subjects. Using a cost functional which penalizes spatial sparseness and control energy may -- for example -- help reducing the required current applied to a subject's brain as well as focusing the electrical perturbation on the relevant brain areas only. While exact target trajectories are unlikely to be available in a practical setting, state-dependent cost functionals which refer to global quantities [cf.\ Eq.~\eqref{eqn:cost_functional_f_x2} for the synchronization task] may well be. Particularly, reformulating the control formalism in frequency space will be beneficial here. It would for example allow for computing optimal interventions for changing the power of certain brain rhythms, which can be monitored by electroencephalography and which are common control targets in clinical settings (cf.\ \cite{Ladenbauer2016}). As shown in other computational and physiological studies \cite{cakan2019, alagapan2016modulation} and emphasized again by our results, the timing of the control inputs may also be crucial for successful interventions. Here we see a high potential of nonlinear optimal control in guiding electrical brain stimulation under electro- or magnetoencephalography \cite{thut2017guiding, bergmann2018brain}, a setting which allows for such precisely timed control inputs. 

\cgreen{For quantitative predictions, an application of non-linear optimal control to biophysically grounded network models (cf.\ \cite{cakan2020deep}) is desirable, where the coupling of neurons to externally applied electric fields is included in a biophysically realistic way. Still, results obtained with simplified brain-network models will strongly facilitate biologically detailed but computationally expensive in-silico experiments 
and may already} inspire physiological studies on brain-stimulation to explore new control paradigms for clinical stimulation protocols with potentially better control results, reduced energy expenditure, and higher spatial sparseness.

\section{Acknowledgments}
We thank Dr.\ Michael Scholz, Cristiana Dimulescu, and Lena Salfenmoser for scientific discussions, and Prof.\ Dr.\ Eckehard Schöll for his comments on the manuscript.\\
This work was funded by the German Research Foundation (163436311  --  SFB  910, 327654276 -- SFB 1315, and under Germany's Excellence Strategy -- EXC 2002/1 “Science of Intelligence” -- 390523135).

Human connectivity data were provided by the Human Connectome Project, WU-Minn Consortium (Principal Investigators: David Van Essen and Kamil Ugurbil; 1U54MH091657) funded by the 16 NIH Institutes and Centers that support the NIH Blueprint for Neuroscience Research; and by the McDonnell Center for Systems Neuroscience at Washington University.

\appendix

\section{Mathematical formulation of the optimal control problem}
\label{sec:appendixMath}
\subsection{Optimization problem: general formulation}
In this section the mathematical foundations for controlling the network dynamics are laid. We use nonlinear sparse optimal control. All derivations below are similar to the ones in Refs. \cite{casas2013sparse, troltzsch2010optimal}, 
where sparse optimal control was applied to a system of partial differential equations with diffusive coupling.

We consider a network of $N$ nodes with $d$-dimensional dynamics each. Its dynamics is defined by a system of $N$ stochastic differential equations,
\begin{eqnarray}
\label{eqn:GSystem}
\begin{aligned}
\frac{d }{d t}\bm{x}(t) =\bm{h}(\bm{x}(t) ) + \sigma (\bm{A} \otimes \bm{G}) \bm{x} (t) \\
+(\bm{B} \otimes \bm{K}) \bm{u}(t) + \eta (\bm{I_N} \otimes \bm{D}) \bm{\xi}(t).
\end{aligned}
\end{eqnarray}
with the state vector $\bm{x}=(\bm{x_1},...,\bm{x_N})$ and $\bm{x_i}=(x_{i1},\dots,x_{id})$. $\otimes$ denotes the Kronecker product. The local node dynamics is governed by $\bm{h}(\bm{x})=(\bm{h}(\bm{x_1}),...,\bm{h}(\bm{x_N}))$ with $\bm{h}(\bm{x_i})=(h_1(\bm{x_i}),\dots,h_d(\bm{x_i}))$. 
The coupling term consists of the $N\times N$-dimensional adjacency matrix $\bm{A}$ and the  $d\times d$-dimensional local coupling scheme $\bm{G}$. $\bm{u}=(\bm{u_1},...,\bm{u_N})$ with $\bm{u_i}=(u_{i1},\dots,u_{id})$ denotes the vector of control inputs, $\bm{B}$ is a diagonal $N \times N$ dimensional control matrix, and $\bm{K}$ the $d\times d$-dimensional local control scheme.
The noise term consists of the Kronecker product of an N-dimensional identity matrix $\bm{A}$ and the  $d\times d$-dimensional local noise scheme $\bm{G}$. The independent white Gaussian stochastic process is given by $\xi$ \cgreen{with standard deviation $\eta$}. Boundary conditions are given by
\begin{align}
\label{eqn:GBC}
\bm{x}(t=0)=\bm{x_0}.
\end{align}


Since our system is stochastic while the control is deterministic, the optimal control must minimize a cost functional which is an expectation over many realizations of the noise. It is defined as
\begin{align}
\label{eqn:Gfunctional}
\langle F(\bm{x}(\bm{u}),\bm{u}) \rangle=\int_{0}^{T} \langle f(\bm{x}(\bm{u}),\bm{u}) \rangle dt,
\end{align}
where the angle brackets denote the expectation.

Our goal is to find the optimal control $\overline{\bm{u}}$ that minimizes the above mean cost functional. Hence, the variational inequality
\begin{align}
\label{eqn:Gvarineq}
\begin{aligned}
\langle F(\bm{x}(u_i),u_i)\rangle- \langle F(\bm{x}(\overline{u}_i),\overline{u}_i)\rangle \geq 0
\end{aligned}
\end{align}
holds when regarding every dimension $i$ separately. 
We linearize around the optimal solution and write $ \delta u_i = (u_i -\overline{u}_i)$ to obtain
\begin{eqnarray}
\label{eqn:Gvarineq1}
\begin{aligned}
\langle F(\bm{x}(u_i),u_i)\rangle- \langle F(\bm{x}(\overline{u}_i),\overline{u}_i)\rangle& \\ \approx \frac{\partial}{\partial u_i} \langle F(\bm{x}(u_i),u_i)\rangle &\bigg|_{\overline{u}_i} \delta u_i \geq 0.
\end{aligned}
\end{eqnarray}
Since above inequality (\ref{eqn:Gvarineq1}) has to hold for all $\delta u_i$ and  $-\delta u_i$, we find the stronger necessary condition
\begin{align}
\label{eqn:Gvarineq2}
\begin{aligned}
\langle F(\bm{x}(u_i),u_i)\rangle- \langle F(\bm{x}(\overline{u}_i),\overline{u}_i)\rangle = 0
\end{aligned}
\end{align}
for the optimal solution.

Inserting the expression from the right side of Eq.~(\ref{eqn:Gfunctional}) yields
\begin{eqnarray}
\label{eqn:Gvarineq3}
\begin{aligned}
\int_{0}^{T} \Big[ \langle f(\bm{x}(u_i),u_i) \rangle - \langle f(\bm{x}(\overline{u}_i),\overline{u}_i) \rangle \Big] dt& \\
\approx \int_{0}^{T} \frac{\partial}{\partial u_i} \langle f(\bm{x}(u_i),u_i)\rangle &\bigg|_{\overline{u}_i} (u_i-\overline{u}_i) dt = 0.
\end{aligned}
\end{eqnarray}
Since Eq.~(\ref{eqn:Gvarineq3}) must hold for all components $i$, we write
\begin{align}
\label{eqn:Gvarineq4}
\int_{0}^{T} \bm{\nabla_u} \langle f(\bm{x}(\bm{\overline{u}}),\bm{\overline{u}})\rangle \circ (\bm{u}-\overline{\bm{u}}) dt = 0,
\end{align}
where $\circ$ denotes the element-wise multiplication (Schur product). Here we assume that the cost functional can be separated into a part $F^u(\bm{u})$, which depends on control inputs only and which is deterministic, and a part $\langle F^x(\bm{x}(\bm{u})) \rangle$, which is a function of the network state $\bm{x}$ and only indirectly depends on the applied control, i.e.:
\begin{align}
\begin{aligned}
\label{eqn:cf2}
\langle F(\bm{x}(\bm{u}),\bm{u}) \rangle = F^u(\bm{u})+\langle F^x(\bm{x}(\bm{u})) \rangle.
\end{aligned}
\end{align}
Equivalently, one can write
\begin{align}
\begin{aligned}
\label{eqn:cf3}
\int_{0}^{T} \langle f(\bm{x}(\bm{u}),\bm{u}) \rangle dt=\int_{0}^{T} \Big[ f^u(\bm{u})+\langle f^x(\bm{x}(\bm{u})) \rangle  \Big]dt.
\end{aligned}
\end{align}

Applying the chain rule in Eq.~(\ref{eqn:Gvarineq4}) yields
\begin{align}
\label{eqn:Gder_f} 
\begin{aligned}
&\int_{0}^{T} \bm{\nabla_u} \langle f(\bm{x}(\bm{\overline{u}}),\bm{\overline{u}})\rangle \circ (\bm{u}-\overline{\bm{u}}) dt \\
&= \int_{0}^T \Big[ \bm{\nabla_u}  f^u(\bm{\overline{u}}) + \langle D_{\bm{u}}^T(\bm{x}(\overline{\bm{u}} ))\bm{\nabla_x}  f^x(\bm{x}(\overline{\bm{u}}))\rangle \Big] \circ (\bm{u} -\overline{\bm{u}})  dt \\ &= 0,
\end{aligned}
\end{align}
where $ D_{\bm{u}}(\bm{x}(\overline{\bm{u}} ))$ is the Jacobian matrix of the state $\bm{x}$ with respect to the control vector $\bm{u}$, evaluated at the optimal control $\overline{\bm{u}}$.

To calculate the above derivative we must find an expression for the Jacobian matrix $ D_{\bm{u}}(\bm{x}(\overline{\bm{u}} ))$ . However, the dependence of the network state on the control inputs cannot be provided in closed form. The Lagrange Formalism 
enables us to derive an alternative expression for the second term on the right-hand side of Eq.~(\ref{eqn:Gder_f}), which can be evaluated to find the optimal control.

\subsection{The method of Lagrange multipliers}
We can use the formalism of Lagrange multipliers to formulate conditions for the optimization of the cost functional, Eq.~(\ref{eqn:Gfunctional}), that is subject to the constraint given in Eq.~(\ref{eqn:GSystem}) and the boundary conditions given in Eq.~(\ref{eqn:GBC}). Our approach is similar to the method used in [68], 
and we will obtain a result that resembles what is found in~\cite{casas2013sparse} 
for a system reaction-diffusion equations.

We define the mean Lagrange function $\langle \mathcal{L} (\bm{x}(t) ,\bm{u}(t)) \rangle$ as
\begin{align}
\label{eqn:GLagrangian}
\begin{aligned}
\langle \mathcal{L} (\bm{x}(t) &,\bm{u}(t)) \rangle \\
=& \langle F(\bm{x} ,\bm{u}) - \int_{0}^T  \Big[\frac{d}{d t} \bm{x} -\bm{h}(\bm{x} ) - \sigma ( \bm{A} \otimes \bm{G}) \bm{x} \\
&-(\bm{B} \otimes \bm{K}) \bm{u} - (\bm{I_N} \otimes \bm{G}) \bm{\xi}  \Big]^T\bm{\phi}(t) ~dt  \rangle \\
=& \langle \int_{0}^T \Big[ f(\bm{x} ,\bm{u}) -  \Big( \frac{d}{d t} \bm{x} -\bm{h}(\bm{x} ) - \sigma (\bm{A} \otimes \bm{G}) \bm{x} \\
&-(\bm{B} \otimes \bm{K}) \bm{u} - (\bm{I_N} \otimes \bm{G}) \bm{\xi}  \Big)^T \bm{\phi}(t)\Big] dt  \rangle
\end{aligned}
\end{align}
Equation (\ref{eqn:GSystem}) for the dynamics of the network state $\bm{x}$ has to be satisfied for all times $0\le t\le T$, which is ensured by the time integration of the constraint. $\bm{\phi}(t)=(\bm{\phi_1}(t),...,\bm{\phi_N}(t))$ with $\bm{\phi_i}(t)=(\phi_{i1}(t),\dots,\phi_{id}(t))$ is the $d \times N$-dimensional vector of time-dependent Lagrange multipliers.

Linearizing around the optimal solution [cf.\ Eqs. (\ref{eqn:Gvarineq}), (\ref{eqn:Gvarineq4})], we obtain the optimality conditions
\begin{align}
\label{eqn:Gvariational ineq}
\begin{aligned}
\langle \int_{0}^{T} \bm{\nabla_u} \mathcal{L}(\bm{x}(\bm{\overline{u}}),\bm{\overline{u}})\circ \bm{\delta u}~ dt \rangle = 0
\end{aligned}
\end{align}
and 
\begin{align}
\label{eqn:Gvariational ineq2}
\begin{aligned}
\langle \int_{0}^{T} \bm{\nabla_x}  \mathcal{L}(\bm{x}(\bm{\overline{u}}),\bm{\overline{u}})\circ \bm{\delta x}~ dt \rangle = 0.
\end{aligned}
\end{align}
with $\bm{\delta u} = \bm{u} -\overline{\bm{u}}$ and $\bm{\delta x} = \bm{x}(\bm{u}) - \bm{x}(\overline{\bm{u}})$.

After inserting the Lagrange function, Eq.~(\ref{eqn:GLagrangian}), into Eq.~(\ref{eqn:Gvariational ineq2}), we apply partial integration and the chain rule, and obtain
\begin{align}
\label{eqn:Gder_x}
\begin{aligned}
\langle \bm{\nabla_x} &\mathcal{L}(\bm{x}(\bm{\overline{u}}),\bm{\overline{u}})\circ \bm{\delta x} \rangle  \\
=&\langle \int_{0}^T \bm{\nabla_x}  f(\bm{x}(\overline{\bm{u}}),\bm{\overline{u}}) \circ \bm{\delta x}~ dt \\
&- \int_0^T \bm{\nabla_x} \big(\frac{d}{d t} \bm{x} ({\bm{\overline{u}}}) \big)^T \bm{\phi} \circ \bm{\delta x} ~dt \\
&-\int_0^T \bm{\nabla_x} \Big[ \Big( -\bm{h}(\bm{x}(\overline{\bm{u}}) )  - \sigma (\bm{A} \otimes \bm{G}) \bm{x}(\overline{\bm{u}}) \\
&-(\bm{B} \otimes \bm{K}) \bm{\overline{u}} - (\bm{I_N} \otimes \bm{G}) \bm{\xi} \Big)^T\bm{\phi} \Big] \circ \bm{\delta x}~ dt \rangle \\
=&\langle  \int_{0}^T \Big[\Big( \bm{\nabla_x} f^x(\bm{x})) + \frac{\partial}{\partial t} \bm{\phi} \\
&+\big(D_{\bm{x}}(\bm{h}) + \sigma (\bm{A} \otimes \bm{G}) \Big)^T \bm{\phi} \Big] \circ \bm{\delta x}~dt - \bm{\phi}(T) \circ \bm{\delta x}(T) \rangle \\
=& \, \bm{0}.
\end{aligned}
\end{align}

The boundary term
\begin{align}
\label{eqn:GBAd0}
\bm{\phi}(T) \circ \bm{\delta x}(T)  = \bm{0}
\end{align}
must hold for every realization. Since $\bm{\delta x}(T)$ could be finite, the boundary condition $\bm{\phi}(T)$ for the vector of Lagrange multipliers is
\begin{align}
\label{eqn:GBAd}
\bm{\phi}(T)=\bm{0}.
\end{align}

Equation (\ref{eqn:Gder_x}) is fulfilled if the integrand vanishes. We thus obtain a $d \times N$-dimensional system of linear ordinary differential equations for the so-called adjoint states $\bm{\phi}$,
\begin{align}
\label{eqn:GAd} 
-  \frac{d}{d t} \bm{\phi}(t)=  \big(D_{\bm{x}}(\bm{h}) + \sigma (\bm{A} \otimes \bm{G}) \big)^T \bm{\phi}(t) + \bm{\nabla_x} f^x(\bm{x})),
\end{align}
where $D_{\bm{x}}(\bm{h})$ denotes the Jacobian matrix of the local dynamics with respect to the network state $\bm{x}$. The adjoint states are obtained for every realization by solving Eq.~(\ref{eqn:GAd}) backwards in time obeying the boundary conditions (\ref{eqn:GBAd}). 

Inserting the Lagrange function [Eq.~(\ref{eqn:GLagrangian})] into Eq.~(\ref{eqn:Gvariational ineq}), we obtain
\begin{align}
\label{eqn:Gder_u0}
\begin{aligned}
\langle &\bm{\nabla_u} \mathcal{L}(\bm{x}(\bm{\overline{u}}),\bm{\overline{u}}) \circ \bm{\delta u}\rangle \\
=& \langle \int_{0}^T \bm{\nabla_u}  f(\bm{x}(\overline{\bm{u}}),\overline{\bm{u}}) \circ \bm{\delta u}~ dt \\
&- \int_0^T  \frac{d}{d t} [D_{\bm{u}}(\bm{x}(\overline{\bm{u}}))]^T \bm{\phi} \circ \bm{\delta u} ~dt \\
&- \int_0^T \bm{\nabla_u} \Big[\Big( \bm{h}(\bm{x}(\overline{\bm{u}}) ) - \sigma (\bm{A} \otimes \bm{G} )\bm{x}(\overline{\bm{u}}) \\
&-(\bm{B} \otimes \bm{K}) \bm{\overline{u}} \Big)^T\bm{\phi} \Big] \circ \bm{\delta u}~dt \rangle\\
=& \, \bm{0}.
\end{aligned}
\end{align}

The first term in Eq.~(\ref{eqn:Gder_u0}) vanishes [see Eq.~(\ref{eqn:Gvarineq4})]. Furthermore, after applying partial integration and the chain rule, above Eq.~(\ref{eqn:Gder_u0}) simplifies to
\begin{align}
\label{eqn:Gder_u1}
\begin{aligned}
\langle \bm{\nabla_u}& \mathcal{L} \circ \bm{\delta u}\rangle = \langle  \int_0^T  \big(D_{\bm{u}}(\bm{x}(\overline{\bm{u}}))\big)^T \frac{d}{d t} \bm{\phi} \circ  \bm{\delta u} ~dt \\
&+ \int_0^T \Big[ \big(D_{\bm{u}}(\bm{x}(\overline{\bm{u}}))\big)^T \Big( \bm{\nabla_x} \big( \bm{h}(\bm{x}(\overline{\bm{u}})) \\
&+  \sigma (\bm{A} \otimes \bm{G}) \big)^T \phi \Big) + (\bm{B} \otimes \bm{K})^T \phi \Big] \circ \bm{\delta u} ~dt \rangle\\
=& \langle \int_0^T \Big[ \big( D_{\bm{u}}(\bm{x}(\overline{\bm{u}}))\big)^T \big( \frac{d}{d t} \bm{\phi}
+  D_{\bm{x}}(\bm{h}) +  \sigma (\bm{A} \otimes \bm{G}) \big)^T \phi \\
& + (\bm{B} \otimes \bm{K})^T \phi \Big] \circ  \bm{\delta u} ~dt \rangle \\
=& \bm{0},
\end{aligned}
\end{align}
where the boundary term of the partial integration was set to zero because of the boundary conditions imposed on $\bm{\phi}$ and $\bm{x}$.

Inserting the differential equations (\ref{eqn:GAd}) that govern the dynamics of the adjoint states into Eq.~(\ref{eqn:Gder_u1}), we obtain
\begin{align}
\label{eqn:Gder_u3} 
\begin{split}
\langle \int_0^T \Big(& - D_{\bm{u}}^T(\bm{x}(\overline{\bm{u}}))  \bm{\nabla_x}  f^x(\bm{x}(\overline{\bm{u}})) +(\bm{B} \otimes \bm{K})^T \bm{\phi}\Big) \circ  \bm{\delta u} ~dt \rangle \\ &=\bm{0}.
\end{split}
\end{align}
The first term corresponds to the expression found in Eq.~(\ref{eqn:Gder_f}). We can solve Eq.~(\ref{eqn:Gder_u3}) for this term and insert it into Eq.~(\ref{eqn:Gder_f}) to obtain the optimality condition
\begin{align}
\label{eqn:Gder_f21} 
\begin{split}
\int_0^T \Big(&\bm{\nabla_u} f^u(\overline{\bm{u}}) + (\bm{B} \otimes \bm{K})^T \langle \bm{\phi} \rangle \Big) \circ \bm{\delta u} ~dt = \int_0^T \bm{g}(t) ~dt \\ &=\bm{0}.
\end{split}
\end{align}

Equation (\ref{eqn:Gder_f21}) is fulfilled if the stronger optimality condition
\begin{align}
\label{eqn:Gder_f2} 
\bm{g}(t) = \bm{\nabla_u} f^u(\overline{\bm{u}}) + (\bm{B} \otimes \bm{K})^T \langle \bm{\phi} \rangle =\bm{0}
\end{align}
holds for all times $t \in [0,T]$. This equation is similar to the equation derived by Casas et al. \cite{casas2013sparse} 
for a system of reaction-diffusion equations. 


\section{\cgreen{Numerical integration method and local optima}}
\label{sec:appendixNumerics}
\cgreen{We integrate the equations for the network dynamics [Eq.~\eqref{eqn:network_system}] and the adjoint state [Eq.~\eqref{eqn:adjoint_state}] using the fourth order Runge-Kutta (RK4) method. A small step size of $\Delta t=0.1$ arb.\ units is used to ensure numeric stability and small truncation errors. The noise term is scaled with the factor $\frac{1}{\sqrt{\Delta t}}$, such that the effective noise strength does not depend on the integration step size.}
	

\cgreen{When solving the minimization problem for the optimal control numerically (see Section \ref{sec:min_problem}) using gradient based optimization, we may converge to a local minimum of the cost functional. In order to alleviate this problem, we choose multiple initial conditions $\bm{u_0}$ for the conjugate gradient algorithm. In the case of synchronizing the network dynamics, the control scheme was robust to changes in $\bm{u_0}$.
The coexistence of multiple local minima of the cost functional was observed only in the case of switching between network states using sparse control (cf.\ Fig.~\ref{fig:state_switching_fig3}). For larger values for $I_s$ [cf.~Eq.~\eqref{eqn:cost_functional_f_u}] we find one minimum at $\bm{u}(t)=\bm{0}$ (smaller cost for the sparsity term) coexisting with another minimum corresponding to a control input $\bm{u}(t)$ affecting a small number of nodes (smaller cost for the precision term). Which of these local optima has minimal cost depends on the value of $I_s$.}

\cgreen{To ensure that the global optimum was found in the latter case for a given value of $I_s$, we systematically varied the initial conditions. Starting with $I_s = 0$, we first calculated the optimal control signal, then increased the value of $I_s$, and used the previous optimal control as the new initial condition (\textit{continuation}). Likewise, starting with a high enough value for $I_s$, this process is repeated, but for decreasing values of $I_s$. For a given value of $I_s$ the control input with the lower cost was chosen for further analysis.}



\section{Definitions of average and modal controllability}
\label{sec:appendixControllability}
We consider a simplified linear network dynamics of the form
\begin{align}
\bm{x}(t+1) = \bm{A} \bm{x}(t) + \bm{B} \bm{u}(t),
\end{align}
with activity vector $\bm{x}$, adjacency matrix $\bm{A}$, control matrix $\bm{B}$ and control inputs $\bm{u}$. 
In order to calculate the average controllability of node $k$ in the network, we choose the input matrix $\bm{B}^{(k)}$ to target a single node, i.e.\ ${B}^{(k)}_{ij}=1$ for $i,j=k$ and ${B}^{(k)}_{ij}=0$, otherwise.
The controllability Gramian of this system is defined as
\begin{align}
\bm{W}^{(k)} = \sum^\infty_{\tau = 0}\bm{A}^\tau \bm{B}^{(k)}  {\bm{B}^{(k) T}} (\bm{A}^\tau)^{T}
\end{align}
and the average controllability as its trace (cf.\ \cite{gu2015controllability})
\begin{align}
c_k^{av} = \mathrm{Tr}\hspace{1pt} \bm{W}^{(k)}.
\end{align}
Network nodes with high average controllability have a large impact on the network dynamics. This makes them important control sites since by controlling these nodes a large number of possible activity vectors $\bm{x}$ can be reached with little control energy. They are consequently said to be able to move the system to many \textit{easy-to-reach states}~\cite{gu2015controllability}. 

The modal controllability is defined based on on an eigenvalue decomposition of the network adjacency matrix $\bm{A}$,
\begin{align}
c^{mod}_k = \sum^N_{j = 1}\left(1-\lambda^2_j\right) v_{kj}^2,
\end{align}
where $v_{kj}$ are the elements of the eigenvector matrix $\bm{V} = [v_{kj}]$ and $\lambda_1\dots \lambda_N$ are the corresponding eigenvalues~\cite{gu2015controllability}. 
Nodes with high modal controllability tend to be sparsely connected. They are capable of pushing the network dynamics towards activity vectors $\bm{x}$ which require substantial control energy to be reached. Thus, nodes with high modal controllability are said to be capable of steering the network into \textit{difficult-to-reach states}~\cite{gu2015controllability}. 

For a detailed description and derivation of these regional controllability measures please refer to Refs. \cite{gu2015controllability, muldoon2016stimulation, tang2020control}. 
We use the matlab code provided with the publication of Ref.~\cite{gu2015controllability} 
to compute the controllability measures \cite{bassetcode}. For their calculation, only the adjaciency matrix $\bm{A}$ (cf.\ Fig.~\ref{fig:Adjacency_matrix}) is required.

\section{Diffusion Tensor Imaging}
\label{sec:appendixDTI}

\subsection{Subjects}
We use the Diffusion Tensor Imaging (DTI) data of 12 human subjects from the Human Connectome Project (Young Adults HCP) \cite{van2013hcp} 
with the following IDs: 101309, 121416, 211215, 211619, 212116, 213522, 219231, 220721, 268749, 284646, 303119, 329844. All subjects are male, healthy, and 26-30 years old. 

\subsection{Data processing pipeline}
Data acquisition and preprocessing is described in Ref.~\cite{van2013hcp}. 
We used BEDPOSTX (Bayesian Estimation of Diffusion Parameters Obtained using Sampling Techniques) from the FSL toolbox \cite{jenkinson2012fsl}
for building up distributions on diffusion parameters, which automatically determines the number of crossing fibres per voxel. Based on these diffusion parameters at each voxel, we then applied the PROBTRACKX (probabilistic tractography) algorithm from the same toolbox with 5000 samples per voxel.
The resulting connectivity matrix $A^s$ for each subject $s$ was normalized by dividing the connections between any two regions by the number of voxels in the source region multiplied by 5000. 
Self connections were deleted, $A^s_{ii} \overset{!}{=}0, \, \forall \, i$, and we averaged the structural connectivity matrices of the individual subjects to obtain the adjacency matrix $A = \frac{1}{12}\sum^{N=12}_s A^s$. 
Probabilistic fiber tracking does not provide information about directionality; hence the \cgreen{resulting graph must be undirected. Following standard procedures (cf.\ \cite{deco2017single}), the} structural connectivity matrix was symmetrized $\left(A \leftarrow \frac{A+A^T}{2}\right)$, \cgreen{as any deviations from symmetry are artifacts of the processing method}. 
Since application of the PROBTRACKX algorithm typically results in a certain number of false positive fibers and thus assigns non-zero connection strengths to all pairs of brain regions, we enforced a sparsity of 20\% \cite{gong2009age}
by discarding all connections with a relative connection strength smaller than $0.00071$ (0.15\% of strongest relative connection strength).

\AtEndEnvironment{thebibliography}{
	\bibitem{supp} See Supplemental Material (attached to this document) for additional information on the state space exploration of the whole-brain network (limit cycle criterion, aperiodic behavior in mixed states, influence of noise 	on the network dynamics), the optimal control of the brain network dynamics (recovery of laLC target state, computation	of critical times $t_c$, controlled nodes in case of sparse control, videos of the node dynamics in phase space with optimal control), and linear network control theory (correlation of average	and modal controllability with weighted degree, modal controllability evaluated for the synchronization task in the noise free case, controllability measures for the synchronization task for noisy network dynamics).
	\bibitem{bassetcode}\href{https://complexsystemsupenn.com/codedata}{https://complexsystemsupenn.com/codedata}.
}
\bibliography{refs}

\begin{thebibliography}{77}%
\makeatletter
\providecommand \@ifxundefined [1]{%
 \@ifx{#1\undefined}
}%
\providecommand \@ifnum [1]{%
 \ifnum #1\expandafter \@firstoftwo
 \else \expandafter \@secondoftwo
 \fi
}%
\providecommand \@ifx [1]{%
 \ifx #1\expandafter \@firstoftwo
 \else \expandafter \@secondoftwo
 \fi
}%
\providecommand \natexlab [1]{#1}%
\providecommand \enquote  [1]{``#1''}%
\providecommand \bibnamefont  [1]{#1}%
\providecommand \bibfnamefont [1]{#1}%
\providecommand \citenamefont [1]{#1}%
\providecommand \href@noop [0]{\@secondoftwo}%
\providecommand \href [0]{\begingroup \@sanitize@url \@href}%
\providecommand \@href[1]{\@@startlink{#1}\@@href}%
\providecommand \@@href[1]{\endgroup#1\@@endlink}%
\providecommand \@sanitize@url [0]{\catcode `\\12\catcode `\$12\catcode
  `\&12\catcode `\#12\catcode `\^12\catcode `\_12\catcode `\%12\relax}%
\providecommand \@@startlink[1]{}%
\providecommand \@@endlink[0]{}%
\providecommand \url  [0]{\begingroup\@sanitize@url \@url }%
\providecommand \@url [1]{\endgroup\@href {#1}{\urlprefix }}%
\providecommand \urlprefix  [0]{URL }%
\providecommand \Eprint [0]{\href }%
\providecommand \doibase [0]{https://doi.org/}%
\providecommand \selectlanguage [0]{\@gobble}%
\providecommand \bibinfo  [0]{\@secondoftwo}%
\providecommand \bibfield  [0]{\@secondoftwo}%
\providecommand \translation [1]{[#1]}%
\providecommand \BibitemOpen [0]{}%
\providecommand \bibitemStop [0]{}%
\providecommand \bibitemNoStop [0]{.\EOS\space}%
\providecommand \EOS [0]{\spacefactor3000\relax}%
\providecommand \BibitemShut  [1]{\csname bibitem#1\endcsname}%
\let\auto@bib@innerbib\@empty
\bibitem [{\citenamefont {Zaehle}\ \emph {et~al.}(2010)\citenamefont {Zaehle},
  \citenamefont {Rach},\ and\ \citenamefont {Herrmann}}]{Zaehle2010}%
  \BibitemOpen
  \bibfield  {author} {\bibinfo {author} {\bibfnamefont {T.}~\bibnamefont
  {Zaehle}}, \bibinfo {author} {\bibfnamefont {S.}~\bibnamefont {Rach}},\ and\
  \bibinfo {author} {\bibfnamefont {C.}~\bibnamefont {Herrmann}},\ }\bibfield
  {title} {\bibinfo {title} {Transcranial alternating current stimulation
  enhances individual alpha activity in human {EEG}},\ }\href
  {https://doi.org/10.1371/journal.pone.0013766} {\bibfield  {journal}
  {\bibinfo  {journal} {PLoS One}\ }\textbf {\bibinfo {volume} {5}},\ \bibinfo
  {pages} {e13766} (\bibinfo {year} {2010})}\BibitemShut {NoStop}%
\bibitem [{\citenamefont {Neuling}\ \emph {et~al.}(2013)\citenamefont
  {Neuling}, \citenamefont {Rach},\ and\ \citenamefont
  {Herrmann}}]{Neuling2013}%
  \BibitemOpen
  \bibfield  {author} {\bibinfo {author} {\bibfnamefont {T.}~\bibnamefont
  {Neuling}}, \bibinfo {author} {\bibfnamefont {S.}~\bibnamefont {Rach}},\ and\
  \bibinfo {author} {\bibfnamefont {C.}~\bibnamefont {Herrmann}},\ }\bibfield
  {title} {\bibinfo {title} {Orchestrating neuronal networks: sustained
  after-effects of transcranial alternating current stimulation depend upon
  brain states},\ }\href {https://doi.org/10.3389/fnhum.2013.00161} {\bibfield
  {journal} {\bibinfo  {journal} {Frontiers in Human Neuroscience}\ }\textbf
  {\bibinfo {volume} {7}},\ \bibinfo {pages} {161} (\bibinfo {year}
  {2013})}\BibitemShut {NoStop}%
\bibitem [{\citenamefont {Polan{\'\i}a}\ \emph {et~al.}(2012)\citenamefont
  {Polan{\'\i}a}, \citenamefont {Nitsche}, \citenamefont {Korman},
  \citenamefont {Batsikadze},\ and\ \citenamefont {Paulus}}]{Polania2012}%
  \BibitemOpen
  \bibfield  {author} {\bibinfo {author} {\bibfnamefont {R.}~\bibnamefont
  {Polan{\'\i}a}}, \bibinfo {author} {\bibfnamefont {M.~A.}\ \bibnamefont
  {Nitsche}}, \bibinfo {author} {\bibfnamefont {C.}~\bibnamefont {Korman}},
  \bibinfo {author} {\bibfnamefont {G.}~\bibnamefont {Batsikadze}},\ and\
  \bibinfo {author} {\bibfnamefont {W.}~\bibnamefont {Paulus}},\ }\bibfield
  {title} {\bibinfo {title} {The importance of timing in segregated theta
  phase-coupling for cognitive performance},\ }\href@noop {} {\bibfield
  {journal} {\bibinfo  {journal} {Current Biology}\ }\textbf {\bibinfo {volume}
  {22}},\ \bibinfo {pages} {1314} (\bibinfo {year} {2012})}\BibitemShut
  {NoStop}%
\bibitem [{\citenamefont {Strüber}\ \emph {et~al.}(2014)\citenamefont
  {Strüber}, \citenamefont {Rach}, \citenamefont {Trautmann-Lengsfeld},
  \citenamefont {Engel},\ and\ \citenamefont {Herrmann}}]{Strueber2014}%
  \BibitemOpen
  \bibfield  {author} {\bibinfo {author} {\bibfnamefont {D.}~\bibnamefont
  {Strüber}}, \bibinfo {author} {\bibfnamefont {S.}~\bibnamefont {Rach}},
  \bibinfo {author} {\bibfnamefont {S.}~\bibnamefont {Trautmann-Lengsfeld}},
  \bibinfo {author} {\bibfnamefont {A.}~\bibnamefont {Engel}},\ and\ \bibinfo
  {author} {\bibfnamefont {C.}~\bibnamefont {Herrmann}},\ }\bibfield  {title}
  {\bibinfo {title} {Antiphasic 40 {Hz} oscillatory current stimulation affects
  bistable motion perception},\ }\href
  {https://doi.org/10.1007/s10548-013-0294-x} {\bibfield  {journal} {\bibinfo
  {journal} {Brain Topography}\ }\textbf {\bibinfo {volume} {27}},\ \bibinfo
  {pages} {158} (\bibinfo {year} {2014})}\BibitemShut {NoStop}%
\bibitem [{\citenamefont {Ladenbauer}\ \emph {et~al.}(2017)\citenamefont
  {Ladenbauer}, \citenamefont {Ladenbauer}, \citenamefont {K{\"u}lzow},
  \citenamefont {de~Boor}, \citenamefont {Avramova}, \citenamefont {Grittner},\
  and\ \citenamefont {Fl{\"o}el}}]{Ladenbauer2017}%
  \BibitemOpen
  \bibfield  {author} {\bibinfo {author} {\bibfnamefont {J.}~\bibnamefont
  {Ladenbauer}}, \bibinfo {author} {\bibfnamefont {J.}~\bibnamefont
  {Ladenbauer}}, \bibinfo {author} {\bibfnamefont {N.}~\bibnamefont
  {K{\"u}lzow}}, \bibinfo {author} {\bibfnamefont {R.}~\bibnamefont {de~Boor}},
  \bibinfo {author} {\bibfnamefont {E.}~\bibnamefont {Avramova}}, \bibinfo
  {author} {\bibfnamefont {U.}~\bibnamefont {Grittner}},\ and\ \bibinfo
  {author} {\bibfnamefont {A.}~\bibnamefont {Fl{\"o}el}},\ }\bibfield  {title}
  {\bibinfo {title} {Promoting sleep oscillations and their functional coupling
  by transcranial stimulation enhances memory consolidation in mild cognitive
  impairment},\ }\href@noop {} {\bibfield  {journal} {\bibinfo  {journal}
  {Journal of Neuroscience}\ }\textbf {\bibinfo {volume} {37}},\ \bibinfo
  {pages} {7111} (\bibinfo {year} {2017})}\BibitemShut {NoStop}%
\bibitem [{\citenamefont {Sun}\ \emph {et~al.}(2011)\citenamefont {Sun},
  \citenamefont {Fu}, \citenamefont {Mao}, \citenamefont {Wang},\ and\
  \citenamefont {Wang}}]{Sun2011}%
  \BibitemOpen
  \bibfield  {author} {\bibinfo {author} {\bibfnamefont {W.}~\bibnamefont
  {Sun}}, \bibinfo {author} {\bibfnamefont {W.}~\bibnamefont {Fu}}, \bibinfo
  {author} {\bibfnamefont {W.}~\bibnamefont {Mao}}, \bibinfo {author}
  {\bibfnamefont {D.}~\bibnamefont {Wang}},\ and\ \bibinfo {author}
  {\bibfnamefont {Y.}~\bibnamefont {Wang}},\ }\bibfield  {title} {\bibinfo
  {title} {Low-frequency repetitive transcranial magnetic stimulation for the
  treatment of refractory partial epilepsy},\ }\href
  {https://doi.org/10.1177/155005941104200109} {\bibfield  {journal} {\bibinfo
  {journal} {Clinical EEG and Neuroscience: Official Journal of the EEG and
  Clinical Neuroscience Society (ENCS)}\ }\textbf {\bibinfo {volume} {42}},\
  \bibinfo {pages} {40} (\bibinfo {year} {2011})}\BibitemShut {NoStop}%
\bibitem [{\citenamefont {Fregni}\ \emph {et~al.}(2006)\citenamefont {Fregni},
  \citenamefont {Otachi}, \citenamefont {Valle}, \citenamefont {Boggio},
  \citenamefont {Thut}, \citenamefont {Rigonatti}, \citenamefont
  {Pascual-Leone},\ and\ \citenamefont {Valente}}]{Fregni2006}%
  \BibitemOpen
  \bibfield  {author} {\bibinfo {author} {\bibfnamefont {F.}~\bibnamefont
  {Fregni}}, \bibinfo {author} {\bibfnamefont {P.}~\bibnamefont {Otachi}},
  \bibinfo {author} {\bibfnamefont {A.}~\bibnamefont {Valle}}, \bibinfo
  {author} {\bibfnamefont {P.}~\bibnamefont {Boggio}}, \bibinfo {author}
  {\bibfnamefont {G.}~\bibnamefont {Thut}}, \bibinfo {author} {\bibfnamefont
  {S.}~\bibnamefont {Rigonatti}}, \bibinfo {author} {\bibfnamefont
  {A.}~\bibnamefont {Pascual-Leone}},\ and\ \bibinfo {author} {\bibfnamefont
  {K.}~\bibnamefont {Valente}},\ }\bibfield  {title} {\bibinfo {title} {A
  randomized clinical trial of repetitive transcranial magnetic stimulation in
  patients with refractory epilepsy},\ }\href
  {https://doi.org/10.1002/ana.20950} {\bibfield  {journal} {\bibinfo
  {journal} {Annals of Neurology}\ }\textbf {\bibinfo {volume} {60}},\ \bibinfo
  {pages} {447} (\bibinfo {year} {2006})}\BibitemShut {NoStop}%
\bibitem [{\citenamefont {Hasan}\ \emph {et~al.}(2013)\citenamefont {Hasan},
  \citenamefont {Falkai},\ and\ \citenamefont
  {Wobrock}}]{hasan2013transcranial}%
  \BibitemOpen
  \bibfield  {author} {\bibinfo {author} {\bibfnamefont {A.}~\bibnamefont
  {Hasan}}, \bibinfo {author} {\bibfnamefont {P.}~\bibnamefont {Falkai}},\ and\
  \bibinfo {author} {\bibfnamefont {T.}~\bibnamefont {Wobrock}},\ }\bibfield
  {title} {\bibinfo {title} {Transcranial brain stimulation in schizophrenia:
  targeting cortical excitability, connectivity and plasticity},\ }\href@noop
  {} {\bibfield  {journal} {\bibinfo  {journal} {Current Medicinal Chemistry}\
  }\textbf {\bibinfo {volume} {20}},\ \bibinfo {pages} {405} (\bibinfo {year}
  {2013})}\BibitemShut {NoStop}%
\bibitem [{\citenamefont {Cole}\ \emph {et~al.}(2015)\citenamefont {Cole},
  \citenamefont {Bernacki}, \citenamefont {Helmer}, \citenamefont {Pinninti},\
  and\ \citenamefont {O'reardon}}]{Cole2015}%
  \BibitemOpen
  \bibfield  {author} {\bibinfo {author} {\bibfnamefont {J.}~\bibnamefont
  {Cole}}, \bibinfo {author} {\bibfnamefont {C.}~\bibnamefont {Bernacki}},
  \bibinfo {author} {\bibfnamefont {A.}~\bibnamefont {Helmer}}, \bibinfo
  {author} {\bibfnamefont {N.}~\bibnamefont {Pinninti}},\ and\ \bibinfo
  {author} {\bibfnamefont {J.}~\bibnamefont {O'reardon}},\ }\bibfield  {title}
  {\bibinfo {title} {Efficacy of transcranial magnetic stimulation ({TMS}) in
  the treatment of schizophrenia: A review of the literature to date},\
  }\href@noop {} {\bibfield  {journal} {\bibinfo  {journal} {Innovations in
  Clinical Neuroscience}\ }\textbf {\bibinfo {volume} {12}},\ \bibinfo {pages}
  {12} (\bibinfo {year} {2015})}\BibitemShut {NoStop}%
\bibitem [{\citenamefont {Fried}(2016)}]{Fried2016}%
  \BibitemOpen
  \bibfield  {author} {\bibinfo {author} {\bibfnamefont {I.}~\bibnamefont
  {Fried}},\ }\bibfield  {title} {\bibinfo {title} {Brain stimulation in
  {A}lzheimer's disease},\ }\href@noop {} {\bibfield  {journal} {\bibinfo
  {journal} {Journal of Alzheimer's Disease}\ }\textbf {\bibinfo {volume}
  {54}},\ \bibinfo {pages} {789} (\bibinfo {year} {2016})}\BibitemShut
  {NoStop}%
\bibitem [{\citenamefont {George}\ \emph {et~al.}(2013)\citenamefont {George},
  \citenamefont {Taylor},\ and\ \citenamefont {Short}}]{George2012}%
  \BibitemOpen
  \bibfield  {author} {\bibinfo {author} {\bibfnamefont {M.~S.}\ \bibnamefont
  {George}}, \bibinfo {author} {\bibfnamefont {J.~J.}\ \bibnamefont {Taylor}},\
  and\ \bibinfo {author} {\bibfnamefont {E.~B.}\ \bibnamefont {Short}},\
  }\bibfield  {title} {\bibinfo {title} {The expanding evidence base for rtms
  treatment of depression},\ }\href@noop {} {\bibfield  {journal} {\bibinfo
  {journal} {Current Opinion in Psychiatry}\ }\textbf {\bibinfo {volume}
  {26}},\ \bibinfo {pages} {13} (\bibinfo {year} {2013})}\BibitemShut {NoStop}%
\bibitem [{\citenamefont {Bassett}\ and\ \citenamefont
  {Sporns}(2017)}]{bassett2017network}%
  \BibitemOpen
  \bibfield  {author} {\bibinfo {author} {\bibfnamefont {D.~S.}\ \bibnamefont
  {Bassett}}\ and\ \bibinfo {author} {\bibfnamefont {O.}~\bibnamefont
  {Sporns}},\ }\bibfield  {title} {\bibinfo {title} {Network neuroscience},\
  }\href@noop {} {\bibfield  {journal} {\bibinfo  {journal} {Nature
  Neuroscience}\ }\textbf {\bibinfo {volume} {20}},\ \bibinfo {pages} {353}
  (\bibinfo {year} {2017})}\BibitemShut {NoStop}%
\bibitem [{\citenamefont {Amunts}\ \emph {et~al.}(2014)\citenamefont {Amunts},
  \citenamefont {Hawrylycz}, \citenamefont {{Van Essen}}, \citenamefont {{Van
  Horn}}, \citenamefont {Harel}, \citenamefont {Poline}, \citenamefont {{De
  Martino}}, \citenamefont {Bjaalie}, \citenamefont {Dehaene-Lambertz},
  \citenamefont {Dehaene}, \citenamefont {Valdes-Sosa}, \citenamefont
  {Thirion}, \citenamefont {Zilles}, \citenamefont {Hill}, \citenamefont
  {Abrams}, \citenamefont {Tass}, \citenamefont {Vanduffel}, \citenamefont
  {Evans},\ and\ \citenamefont {Eickhoff}}]{Amunts2014}%
  \BibitemOpen
  \bibfield  {author} {\bibinfo {author} {\bibfnamefont {K.}~\bibnamefont
  {Amunts}}, \bibinfo {author} {\bibfnamefont {M.}~\bibnamefont {Hawrylycz}},
  \bibinfo {author} {\bibfnamefont {D.}~\bibnamefont {{Van Essen}}}, \bibinfo
  {author} {\bibfnamefont {J.}~\bibnamefont {{Van Horn}}}, \bibinfo {author}
  {\bibfnamefont {N.}~\bibnamefont {Harel}}, \bibinfo {author} {\bibfnamefont
  {J.-B.}\ \bibnamefont {Poline}}, \bibinfo {author} {\bibfnamefont
  {F.}~\bibnamefont {{De Martino}}}, \bibinfo {author} {\bibfnamefont
  {J.}~\bibnamefont {Bjaalie}}, \bibinfo {author} {\bibfnamefont
  {G.}~\bibnamefont {Dehaene-Lambertz}}, \bibinfo {author} {\bibfnamefont
  {S.}~\bibnamefont {Dehaene}}, \bibinfo {author} {\bibfnamefont
  {P.}~\bibnamefont {Valdes-Sosa}}, \bibinfo {author} {\bibfnamefont
  {B.}~\bibnamefont {Thirion}}, \bibinfo {author} {\bibfnamefont
  {K.}~\bibnamefont {Zilles}}, \bibinfo {author} {\bibfnamefont
  {S.}~\bibnamefont {Hill}}, \bibinfo {author} {\bibfnamefont {M.}~\bibnamefont
  {Abrams}}, \bibinfo {author} {\bibfnamefont {P.}~\bibnamefont {Tass}},
  \bibinfo {author} {\bibfnamefont {W.}~\bibnamefont {Vanduffel}}, \bibinfo
  {author} {\bibfnamefont {A.}~\bibnamefont {Evans}},\ and\ \bibinfo {author}
  {\bibfnamefont {S.}~\bibnamefont {Eickhoff}},\ }\bibfield  {title} {\bibinfo
  {title} {Interoperable atlases of the human brain},\ }\href
  {https://doi.org/https://doi.org/10.1016/j.neuroimage.2014.06.010} {\bibfield
   {journal} {\bibinfo  {journal} {Neuroimage}\ }\textbf {\bibinfo {volume}
  {99}},\ \bibinfo {pages} {525 } (\bibinfo {year} {2014})}\BibitemShut
  {NoStop}%
\bibitem [{\citenamefont {Rolls}\ \emph {et~al.}(2015)\citenamefont {Rolls},
  \citenamefont {Joliot},\ and\ \citenamefont
  {Tzourio-Mazoyer}}]{rolls2015aal2}%
  \BibitemOpen
  \bibfield  {author} {\bibinfo {author} {\bibfnamefont {E.~T.}\ \bibnamefont
  {Rolls}}, \bibinfo {author} {\bibfnamefont {M.}~\bibnamefont {Joliot}},\ and\
  \bibinfo {author} {\bibfnamefont {N.}~\bibnamefont {Tzourio-Mazoyer}},\
  }\bibfield  {title} {\bibinfo {title} {Implementation of a new parcellation
  of the orbitofrontal cortex in the automated anatomical labeling atlas},\
  }\href@noop {} {\bibfield  {journal} {\bibinfo  {journal} {Neuroimage}\
  }\textbf {\bibinfo {volume} {122}},\ \bibinfo {pages} {1} (\bibinfo {year}
  {2015})}\BibitemShut {NoStop}%
\bibitem [{\citenamefont {Eickhoff}\ \emph {et~al.}(2018)\citenamefont
  {Eickhoff}, \citenamefont {Yeo},\ and\ \citenamefont {Genon}}]{Eickhoff2018}%
  \BibitemOpen
  \bibfield  {author} {\bibinfo {author} {\bibfnamefont {S.~B.}\ \bibnamefont
  {Eickhoff}}, \bibinfo {author} {\bibfnamefont {B.~T.}\ \bibnamefont {Yeo}},\
  and\ \bibinfo {author} {\bibfnamefont {S.}~\bibnamefont {Genon}},\ }\bibfield
   {title} {\bibinfo {title} {Imaging-based parcellations of the human brain},\
  }\href@noop {} {\bibfield  {journal} {\bibinfo  {journal} {Nature Reviews
  Neuroscience}\ }\textbf {\bibinfo {volume} {19}},\ \bibinfo {pages} {672}
  (\bibinfo {year} {2018})}\BibitemShut {NoStop}%
\bibitem [{\citenamefont {Breakspear}(2017)}]{breakspear2017dynamic}%
  \BibitemOpen
  \bibfield  {author} {\bibinfo {author} {\bibfnamefont {M.}~\bibnamefont
  {Breakspear}},\ }\bibfield  {title} {\bibinfo {title} {Dynamic models of
  large-scale brain activity},\ }\href@noop {} {\bibfield  {journal} {\bibinfo
  {journal} {Nature Neuroscience}\ }\textbf {\bibinfo {volume} {20}},\ \bibinfo
  {pages} {340} (\bibinfo {year} {2017})}\BibitemShut {NoStop}%
\bibitem [{\citenamefont {Deco}\ and\ \citenamefont
  {Jirsa}(2012)}]{deco2012ongoing}%
  \BibitemOpen
  \bibfield  {author} {\bibinfo {author} {\bibfnamefont {G.}~\bibnamefont
  {Deco}}\ and\ \bibinfo {author} {\bibfnamefont {V.~K.}\ \bibnamefont
  {Jirsa}},\ }\bibfield  {title} {\bibinfo {title} {Ongoing cortical activity
  at rest: criticality, multistability, and ghost attractors},\ }\href@noop {}
  {\bibfield  {journal} {\bibinfo  {journal} {Journal of Neuroscience}\
  }\textbf {\bibinfo {volume} {32}},\ \bibinfo {pages} {3366} (\bibinfo {year}
  {2012})}\BibitemShut {NoStop}%
\bibitem [{\citenamefont {Deco}\ \emph {et~al.}(2013)\citenamefont {Deco},
  \citenamefont {Ponce-Alvarez}, \citenamefont {Mantini}, \citenamefont
  {Romani}, \citenamefont {Hagmann},\ and\ \citenamefont
  {Corbetta}}]{deco2013resting}%
  \BibitemOpen
  \bibfield  {author} {\bibinfo {author} {\bibfnamefont {G.}~\bibnamefont
  {Deco}}, \bibinfo {author} {\bibfnamefont {A.}~\bibnamefont {Ponce-Alvarez}},
  \bibinfo {author} {\bibfnamefont {D.}~\bibnamefont {Mantini}}, \bibinfo
  {author} {\bibfnamefont {G.~L.}\ \bibnamefont {Romani}}, \bibinfo {author}
  {\bibfnamefont {P.}~\bibnamefont {Hagmann}},\ and\ \bibinfo {author}
  {\bibfnamefont {M.}~\bibnamefont {Corbetta}},\ }\bibfield  {title} {\bibinfo
  {title} {Resting-state functional connectivity emerges from structurally and
  dynamically shaped slow linear fluctuations},\ }\href@noop {} {\bibfield
  {journal} {\bibinfo  {journal} {Journal of Neuroscience}\ }\textbf {\bibinfo
  {volume} {33}},\ \bibinfo {pages} {11239} (\bibinfo {year}
  {2013})}\BibitemShut {NoStop}%
\bibitem [{\citenamefont {Demirta{\c{s}}}\ \emph {et~al.}(2019)\citenamefont
  {Demirta{\c{s}}}, \citenamefont {Burt}, \citenamefont {Helmer}, \citenamefont
  {Ji}, \citenamefont {Adkinson}, \citenamefont {Glasser}, \citenamefont
  {Van~Essen}, \citenamefont {Sotiropoulos}, \citenamefont {Anticevic},\ and\
  \citenamefont {Murray}}]{demirtacs2019hierarchical}%
  \BibitemOpen
  \bibfield  {author} {\bibinfo {author} {\bibfnamefont {M.}~\bibnamefont
  {Demirta{\c{s}}}}, \bibinfo {author} {\bibfnamefont {J.~B.}\ \bibnamefont
  {Burt}}, \bibinfo {author} {\bibfnamefont {M.}~\bibnamefont {Helmer}},
  \bibinfo {author} {\bibfnamefont {J.~L.}\ \bibnamefont {Ji}}, \bibinfo
  {author} {\bibfnamefont {B.~D.}\ \bibnamefont {Adkinson}}, \bibinfo {author}
  {\bibfnamefont {M.~F.}\ \bibnamefont {Glasser}}, \bibinfo {author}
  {\bibfnamefont {D.~C.}\ \bibnamefont {Van~Essen}}, \bibinfo {author}
  {\bibfnamefont {S.~N.}\ \bibnamefont {Sotiropoulos}}, \bibinfo {author}
  {\bibfnamefont {A.}~\bibnamefont {Anticevic}},\ and\ \bibinfo {author}
  {\bibfnamefont {J.~D.}\ \bibnamefont {Murray}},\ }\bibfield  {title}
  {\bibinfo {title} {Hierarchical heterogeneity across human cortex shapes
  large-scale neural dynamics},\ }\href@noop {} {\bibfield  {journal} {\bibinfo
   {journal} {Neuron}\ }\textbf {\bibinfo {volume} {101}},\ \bibinfo {pages}
  {1181} (\bibinfo {year} {2019})}\BibitemShut {NoStop}%
\bibitem [{\citenamefont {Cakan}\ \emph {et~al.}(2020)\citenamefont {Cakan},
  \citenamefont {Dimulescu}, \citenamefont {Khakimova}, \citenamefont {Obst},
  \citenamefont {Fl{\"o}el},\ and\ \citenamefont {Obermayer}}]{cakan2020deep}%
  \BibitemOpen
  \bibfield  {author} {\bibinfo {author} {\bibfnamefont {C.}~\bibnamefont
  {Cakan}}, \bibinfo {author} {\bibfnamefont {C.}~\bibnamefont {Dimulescu}},
  \bibinfo {author} {\bibfnamefont {L.}~\bibnamefont {Khakimova}}, \bibinfo
  {author} {\bibfnamefont {D.}~\bibnamefont {Obst}}, \bibinfo {author}
  {\bibfnamefont {A.}~\bibnamefont {Fl{\"o}el}},\ and\ \bibinfo {author}
  {\bibfnamefont {K.}~\bibnamefont {Obermayer}},\ }\bibfield  {title} {\bibinfo
  {title} {A deep sleep model of the human brain: how slow waves emerge due to
  adaptation and are guided by the connectome},\ }\href@noop {} {\bibfield
  {journal} {\bibinfo  {journal} {arXiv preprint arXiv:2011.14731}\ } (\bibinfo
  {year} {2020})}\BibitemShut {NoStop}%
\bibitem [{\citenamefont {Tang}\ and\ \citenamefont
  {Bassett}(2018)}]{tang2018colloquium}%
  \BibitemOpen
  \bibfield  {author} {\bibinfo {author} {\bibfnamefont {E.}~\bibnamefont
  {Tang}}\ and\ \bibinfo {author} {\bibfnamefont {D.~S.}\ \bibnamefont
  {Bassett}},\ }\bibfield  {title} {\bibinfo {title} {Colloquium: Control of
  dynamics in brain networks},\ }\href@noop {} {\bibfield  {journal} {\bibinfo
  {journal} {Reviews of Modern Physics}\ }\textbf {\bibinfo {volume} {90}},\
  \bibinfo {pages} {031003} (\bibinfo {year} {2018})}\BibitemShut {NoStop}%
\bibitem [{\citenamefont {Gu}\ \emph {et~al.}(2017)\citenamefont {Gu},
  \citenamefont {Betzel}, \citenamefont {Mattar}, \citenamefont {Cieslak},
  \citenamefont {Delio}, \citenamefont {Grafton}, \citenamefont {Pasqualetti},\
  and\ \citenamefont {Bassett}}]{gu2017optimal}%
  \BibitemOpen
  \bibfield  {author} {\bibinfo {author} {\bibfnamefont {S.}~\bibnamefont
  {Gu}}, \bibinfo {author} {\bibfnamefont {R.~F.}\ \bibnamefont {Betzel}},
  \bibinfo {author} {\bibfnamefont {M.~G.}\ \bibnamefont {Mattar}}, \bibinfo
  {author} {\bibfnamefont {M.}~\bibnamefont {Cieslak}}, \bibinfo {author}
  {\bibfnamefont {P.~R.}\ \bibnamefont {Delio}}, \bibinfo {author}
  {\bibfnamefont {S.~T.}\ \bibnamefont {Grafton}}, \bibinfo {author}
  {\bibfnamefont {F.}~\bibnamefont {Pasqualetti}},\ and\ \bibinfo {author}
  {\bibfnamefont {D.~S.}\ \bibnamefont {Bassett}},\ }\bibfield  {title}
  {\bibinfo {title} {Optimal trajectories of brain state transitions},\
  }\href@noop {} {\bibfield  {journal} {\bibinfo  {journal} {Neuroimage}\
  }\textbf {\bibinfo {volume} {148}},\ \bibinfo {pages} {305} (\bibinfo {year}
  {2017})}\BibitemShut {NoStop}%
\bibitem [{\citenamefont {Gu}\ \emph {et~al.}(2015)\citenamefont {Gu},
  \citenamefont {Pasqualetti}, \citenamefont {Cieslak}, \citenamefont
  {Telesford}, \citenamefont {Alfred}, \citenamefont {Kahn}, \citenamefont
  {Medaglia}, \citenamefont {Vettel}, \citenamefont {Miller}, \citenamefont
  {Grafton} \emph {et~al.}}]{gu2015controllability}%
  \BibitemOpen
  \bibfield  {author} {\bibinfo {author} {\bibfnamefont {S.}~\bibnamefont
  {Gu}}, \bibinfo {author} {\bibfnamefont {F.}~\bibnamefont {Pasqualetti}},
  \bibinfo {author} {\bibfnamefont {M.}~\bibnamefont {Cieslak}}, \bibinfo
  {author} {\bibfnamefont {Q.~K.}\ \bibnamefont {Telesford}}, \bibinfo {author}
  {\bibfnamefont {B.~Y.}\ \bibnamefont {Alfred}}, \bibinfo {author}
  {\bibfnamefont {A.~E.}\ \bibnamefont {Kahn}}, \bibinfo {author}
  {\bibfnamefont {J.~D.}\ \bibnamefont {Medaglia}}, \bibinfo {author}
  {\bibfnamefont {J.~M.}\ \bibnamefont {Vettel}}, \bibinfo {author}
  {\bibfnamefont {M.~B.}\ \bibnamefont {Miller}}, \bibinfo {author}
  {\bibfnamefont {S.~T.}\ \bibnamefont {Grafton}}, \emph {et~al.},\ }\bibfield
  {title} {\bibinfo {title} {Controllability of structural brain networks},\
  }\href@noop {} {\bibfield  {journal} {\bibinfo  {journal} {Nature
  Communications}\ }\textbf {\bibinfo {volume} {6}},\ \bibinfo {pages} {1}
  (\bibinfo {year} {2015})}\BibitemShut {NoStop}%
\bibitem [{\citenamefont {Golos}\ \emph {et~al.}(2015)\citenamefont {Golos},
  \citenamefont {Jirsa},\ and\ \citenamefont
  {Dauc{\'e}}}]{golos2015multistability}%
  \BibitemOpen
  \bibfield  {author} {\bibinfo {author} {\bibfnamefont {M.}~\bibnamefont
  {Golos}}, \bibinfo {author} {\bibfnamefont {V.}~\bibnamefont {Jirsa}},\ and\
  \bibinfo {author} {\bibfnamefont {E.}~\bibnamefont {Dauc{\'e}}},\ }\bibfield
  {title} {\bibinfo {title} {Multistability in large scale models of brain
  activity},\ }\href@noop {} {\bibfield  {journal} {\bibinfo  {journal} {PLoS
  Computational Biology}\ }\textbf {\bibinfo {volume} {11}},\ \bibinfo {pages}
  {e1004644} (\bibinfo {year} {2015})}\BibitemShut {NoStop}%
\bibitem [{\citenamefont {Muldoon}\ \emph {et~al.}(2016)\citenamefont
  {Muldoon}, \citenamefont {Pasqualetti}, \citenamefont {Gu}, \citenamefont
  {Cieslak}, \citenamefont {Grafton}, \citenamefont {Vettel},\ and\
  \citenamefont {Bassett}}]{muldoon2016stimulation}%
  \BibitemOpen
  \bibfield  {author} {\bibinfo {author} {\bibfnamefont {S.~F.}\ \bibnamefont
  {Muldoon}}, \bibinfo {author} {\bibfnamefont {F.}~\bibnamefont
  {Pasqualetti}}, \bibinfo {author} {\bibfnamefont {S.}~\bibnamefont {Gu}},
  \bibinfo {author} {\bibfnamefont {M.}~\bibnamefont {Cieslak}}, \bibinfo
  {author} {\bibfnamefont {S.~T.}\ \bibnamefont {Grafton}}, \bibinfo {author}
  {\bibfnamefont {J.~M.}\ \bibnamefont {Vettel}},\ and\ \bibinfo {author}
  {\bibfnamefont {D.~S.}\ \bibnamefont {Bassett}},\ }\bibfield  {title}
  {\bibinfo {title} {Stimulation-based control of dynamic brain networks},\
  }\href@noop {} {\bibfield  {journal} {\bibinfo  {journal} {PLoS Computational
  Biology}\ }\textbf {\bibinfo {volume} {12}},\ \bibinfo {pages} {e1005076}
  (\bibinfo {year} {2016})}\BibitemShut {NoStop}%
\bibitem [{\citenamefont {Berkovitz}\ and\ \citenamefont
  {Medhin}(2012)}]{berkovitz2012nonlinear}%
  \BibitemOpen
  \bibfield  {author} {\bibinfo {author} {\bibfnamefont {L.~D.}\ \bibnamefont
  {Berkovitz}}\ and\ \bibinfo {author} {\bibfnamefont {N.~G.}\ \bibnamefont
  {Medhin}},\ }\href@noop {} {\emph {\bibinfo {title} {Nonlinear optimal
  control theory}}}\ (\bibinfo  {publisher} {CRC press},\ \bibinfo {year}
  {2012})\BibitemShut {NoStop}%
\bibitem [{\citenamefont {Herzog}\ \emph {et~al.}(2012)\citenamefont {Herzog},
  \citenamefont {Stadler},\ and\ \citenamefont
  {Wachsmuth}}]{herzog2012directional}%
  \BibitemOpen
  \bibfield  {author} {\bibinfo {author} {\bibfnamefont {R.}~\bibnamefont
  {Herzog}}, \bibinfo {author} {\bibfnamefont {G.}~\bibnamefont {Stadler}},\
  and\ \bibinfo {author} {\bibfnamefont {G.}~\bibnamefont {Wachsmuth}},\
  }\bibfield  {title} {\bibinfo {title} {Directional sparsity in optimal
  control of partial differential equations},\ }\href@noop {} {\bibfield
  {journal} {\bibinfo  {journal} {SIAM Journal on Control and Optimization}\
  }\textbf {\bibinfo {volume} {50}},\ \bibinfo {pages} {943} (\bibinfo {year}
  {2012})}\BibitemShut {NoStop}%
\bibitem [{\citenamefont {Casas}\ \emph {et~al.}(2017)\citenamefont {Casas},
  \citenamefont {Herzog},\ and\ \citenamefont {Wachsmuth}}]{casas2017analysis}%
  \BibitemOpen
  \bibfield  {author} {\bibinfo {author} {\bibfnamefont {E.}~\bibnamefont
  {Casas}}, \bibinfo {author} {\bibfnamefont {R.}~\bibnamefont {Herzog}},\ and\
  \bibinfo {author} {\bibfnamefont {G.}~\bibnamefont {Wachsmuth}},\ }\bibfield
  {title} {\bibinfo {title} {Analysis of spatio-temporally sparse optimal
  control problems of semilinear parabolic equations},\ }\href@noop {}
  {\bibfield  {journal} {\bibinfo  {journal} {ESAIM: Control, Optimisation and
  Calculus of Variations}\ }\textbf {\bibinfo {volume} {23}},\ \bibinfo {pages}
  {263} (\bibinfo {year} {2017})}\BibitemShut {NoStop}%
\bibitem [{\citenamefont {Tr{\"o}ltzsch}(2010)}]{troltzsch2010optimal}%
  \BibitemOpen
  \bibfield  {author} {\bibinfo {author} {\bibfnamefont {F.}~\bibnamefont
  {Tr{\"o}ltzsch}},\ }\href@noop {} {\emph {\bibinfo {title} {Optimal control
  of partial differential equations: theory, methods, and applications}}},\
  Vol.\ \bibinfo {volume} {112}\ (\bibinfo  {publisher} {American Mathematical
  Soc.},\ \bibinfo {year} {2010})\BibitemShut {NoStop}%
\bibitem [{\citenamefont {Stannat}\ and\ \citenamefont
  {Wessels}(2020)}]{stannat2020deterministic}%
  \BibitemOpen
  \bibfield  {author} {\bibinfo {author} {\bibfnamefont {W.}~\bibnamefont
  {Stannat}}\ and\ \bibinfo {author} {\bibfnamefont {L.}~\bibnamefont
  {Wessels}},\ }\bibfield  {title} {\bibinfo {title} {Deterministic control of
  stochastic reaction-diffusion equations},\ }\href@noop {} {\bibfield
  {journal} {\bibinfo  {journal} {Evolution Equations \& Control Theory}\ }
  (\bibinfo {year} {2020})}\BibitemShut {NoStop}%
\bibitem [{\citenamefont {Casas}\ \emph {et~al.}(2013)\citenamefont {Casas},
  \citenamefont {Ryll},\ and\ \citenamefont {Tr{\"o}ltzsch}}]{casas2013sparse}%
  \BibitemOpen
  \bibfield  {author} {\bibinfo {author} {\bibfnamefont {E.}~\bibnamefont
  {Casas}}, \bibinfo {author} {\bibfnamefont {C.}~\bibnamefont {Ryll}},\ and\
  \bibinfo {author} {\bibfnamefont {F.}~\bibnamefont {Tr{\"o}ltzsch}},\
  }\bibfield  {title} {\bibinfo {title} {Sparse optimal control of the
  schl{\"o}gl and {FitzHugh--Nagumo} systems},\ }\href@noop {} {\bibfield
  {journal} {\bibinfo  {journal} {Computational Methods in Applied
  Mathematics}\ }\textbf {\bibinfo {volume} {13}},\ \bibinfo {pages} {415}
  (\bibinfo {year} {2013})}\BibitemShut {NoStop}%
\bibitem [{\citenamefont {FitzHugh}(1961)}]{FitzHugh1961}%
  \BibitemOpen
  \bibfield  {author} {\bibinfo {author} {\bibfnamefont {R.}~\bibnamefont
  {FitzHugh}},\ }\bibfield  {title} {\bibinfo {title} {Impulses and
  physiological states in theoretical models of nerve membrane},\ }\href
  {https://doi.org/https://doi.org/10.1016/S0006-3495(61)86902-6} {\bibfield
  {journal} {\bibinfo  {journal} {Biophysical Journal}\ }\textbf {\bibinfo
  {volume} {1}},\ \bibinfo {pages} {445 } (\bibinfo {year} {1961})}\BibitemShut
  {NoStop}%
\bibitem [{\citenamefont {{Nagumo}}\ \emph {et~al.}(1962)\citenamefont
  {{Nagumo}}, \citenamefont {{Arimoto}},\ and\ \citenamefont
  {{Yoshizawa}}}]{Nagumo1962}%
  \BibitemOpen
  \bibfield  {author} {\bibinfo {author} {\bibfnamefont {J.}~\bibnamefont
  {{Nagumo}}}, \bibinfo {author} {\bibfnamefont {S.}~\bibnamefont
  {{Arimoto}}},\ and\ \bibinfo {author} {\bibfnamefont {S.}~\bibnamefont
  {{Yoshizawa}}},\ }\bibfield  {title} {\bibinfo {title} {An active pulse
  transmission line simulating nerve axon},\ }\href@noop {} {\bibfield
  {journal} {\bibinfo  {journal} {Proceedings of the IRE}\ }\textbf {\bibinfo
  {volume} {50}},\ \bibinfo {pages} {2061} (\bibinfo {year}
  {1962})}\BibitemShut {NoStop}%
\bibitem [{\citenamefont {Kostova}\ \emph {et~al.}(2004)\citenamefont
  {Kostova}, \citenamefont {Ravindran},\ and\ \citenamefont
  {Schonbek}}]{kostova2004fitzhugh}%
  \BibitemOpen
  \bibfield  {author} {\bibinfo {author} {\bibfnamefont {T.}~\bibnamefont
  {Kostova}}, \bibinfo {author} {\bibfnamefont {R.}~\bibnamefont {Ravindran}},\
  and\ \bibinfo {author} {\bibfnamefont {M.}~\bibnamefont {Schonbek}},\
  }\bibfield  {title} {\bibinfo {title} {{FitzHugh--Nagumo} revisited: Types of
  bifurcations, periodical forcing and stability regions by a lyapunov
  functional},\ }\href@noop {} {\bibfield  {journal} {\bibinfo  {journal}
  {International Journal of Bifurcation and Chaos}\ }\textbf {\bibinfo {volume}
  {14}},\ \bibinfo {pages} {913} (\bibinfo {year} {2004})}\BibitemShut
  {NoStop}%
\bibitem [{\citenamefont {Sch{\"o}ll}\ \emph {et~al.}(2009)\citenamefont
  {Sch{\"o}ll}, \citenamefont {Hiller}, \citenamefont {H{\"o}vel},\ and\
  \citenamefont {Dahlem}}]{scholl2009time}%
  \BibitemOpen
  \bibfield  {author} {\bibinfo {author} {\bibfnamefont {E.}~\bibnamefont
  {Sch{\"o}ll}}, \bibinfo {author} {\bibfnamefont {G.}~\bibnamefont {Hiller}},
  \bibinfo {author} {\bibfnamefont {P.}~\bibnamefont {H{\"o}vel}},\ and\
  \bibinfo {author} {\bibfnamefont {M.~A.}\ \bibnamefont {Dahlem}},\ }\bibfield
   {title} {\bibinfo {title} {Time-delayed feedback in neurosystems},\
  }\href@noop {} {\bibfield  {journal} {\bibinfo  {journal} {Philosophical
  Transactions of the Royal Society A: Mathematical, Physical and Engineering
  Sciences}\ }\textbf {\bibinfo {volume} {367}},\ \bibinfo {pages} {1079}
  (\bibinfo {year} {2009})}\BibitemShut {NoStop}%
\bibitem [{\citenamefont {Eydam}\ \emph {et~al.}(2019)\citenamefont {Eydam},
  \citenamefont {Franovi{\'c}},\ and\ \citenamefont {Wolfrum}}]{eydam2019FHN}%
  \BibitemOpen
  \bibfield  {author} {\bibinfo {author} {\bibfnamefont {S.}~\bibnamefont
  {Eydam}}, \bibinfo {author} {\bibfnamefont {I.}~\bibnamefont
  {Franovi{\'c}}},\ and\ \bibinfo {author} {\bibfnamefont {M.}~\bibnamefont
  {Wolfrum}},\ }\bibfield  {title} {\bibinfo {title} {Leap-frog patterns in
  systems of two coupled {FitzHugh--Nagumo} units},\ }\href@noop {} {\bibfield
  {journal} {\bibinfo  {journal} {Physical Review E}\ }\textbf {\bibinfo
  {volume} {99}},\ \bibinfo {pages} {042207} (\bibinfo {year}
  {2019})}\BibitemShut {NoStop}%
\bibitem [{\citenamefont {Shepelev}\ and\ \citenamefont
  {Vadivasova}(2019)}]{shepelev2019FHN}%
  \BibitemOpen
  \bibfield  {author} {\bibinfo {author} {\bibfnamefont {I.}~\bibnamefont
  {Shepelev}}\ and\ \bibinfo {author} {\bibfnamefont {T.}~\bibnamefont
  {Vadivasova}},\ }\bibfield  {title} {\bibinfo {title} {Variety of
  spatio-temporal regimes in a 2d lattice of coupled bistable
  {FitzHugh--Nagumo} oscillators. {F}ormation mechanisms of spiral and
  double-well chimeras},\ }\href@noop {} {\bibfield  {journal} {\bibinfo
  {journal} {Communications in Nonlinear Science and Numerical Simulation}\
  }\textbf {\bibinfo {volume} {79}},\ \bibinfo {pages} {104925} (\bibinfo
  {year} {2019})}\BibitemShut {NoStop}%
\bibitem [{\citenamefont {Plotnikov}\ \emph {et~al.}(2016)\citenamefont
  {Plotnikov}, \citenamefont {Lehnert}, \citenamefont {Fradkov},\ and\
  \citenamefont {Sch\"oll}}]{plotnikov2016synchronization}%
  \BibitemOpen
  \bibfield  {author} {\bibinfo {author} {\bibfnamefont {S.~A.}\ \bibnamefont
  {Plotnikov}}, \bibinfo {author} {\bibfnamefont {J.}~\bibnamefont {Lehnert}},
  \bibinfo {author} {\bibfnamefont {A.~L.}\ \bibnamefont {Fradkov}},\ and\
  \bibinfo {author} {\bibfnamefont {E.}~\bibnamefont {Sch\"oll}},\ }\bibfield
  {title} {\bibinfo {title} {Synchronization in heterogeneous
  {FitzHugh--Nagumo} networks with hierarchical architecture},\ }\href
  {https://doi.org/10.1103/PhysRevE.94.012203} {\bibfield  {journal} {\bibinfo
  {journal} {Physical Review E}\ }\textbf {\bibinfo {volume} {94}},\ \bibinfo
  {pages} {012203} (\bibinfo {year} {2016})}\BibitemShut {NoStop}%
\bibitem [{\citenamefont {Lehnert}\ \emph {et~al.}(2011)\citenamefont
  {Lehnert}, \citenamefont {Dahms}, \citenamefont {H{\"o}vel},\ and\
  \citenamefont {Sch{\"o}ll}}]{lehnert2011loss}%
  \BibitemOpen
  \bibfield  {author} {\bibinfo {author} {\bibfnamefont {J.}~\bibnamefont
  {Lehnert}}, \bibinfo {author} {\bibfnamefont {T.}~\bibnamefont {Dahms}},
  \bibinfo {author} {\bibfnamefont {P.}~\bibnamefont {H{\"o}vel}},\ and\
  \bibinfo {author} {\bibfnamefont {E.}~\bibnamefont {Sch{\"o}ll}},\ }\bibfield
   {title} {\bibinfo {title} {Loss of synchronization in complex neuronal
  networks with delay},\ }\href@noop {} {\bibfield  {journal} {\bibinfo
  {journal} {EPL (Europhysics Letters)}\ }\textbf {\bibinfo {volume} {96}},\
  \bibinfo {pages} {60013} (\bibinfo {year} {2011})}\BibitemShut {NoStop}%
\bibitem [{\citenamefont {Cakan}\ \emph {et~al.}(2014)\citenamefont {Cakan},
  \citenamefont {Lehnert},\ and\ \citenamefont
  {Sch{\"o}ll}}]{cakan2014heterogeneous}%
  \BibitemOpen
  \bibfield  {author} {\bibinfo {author} {\bibfnamefont {C.}~\bibnamefont
  {Cakan}}, \bibinfo {author} {\bibfnamefont {J.}~\bibnamefont {Lehnert}},\
  and\ \bibinfo {author} {\bibfnamefont {E.}~\bibnamefont {Sch{\"o}ll}},\
  }\bibfield  {title} {\bibinfo {title} {Heterogeneous delays in neural
  networks},\ }\href@noop {} {\bibfield  {journal} {\bibinfo  {journal} {The
  European Physical Journal B}\ }\textbf {\bibinfo {volume} {87}},\ \bibinfo
  {pages} {1} (\bibinfo {year} {2014})}\BibitemShut {NoStop}%
\bibitem [{\citenamefont {Nikitin}\ \emph {et~al.}(2019)\citenamefont
  {Nikitin}, \citenamefont {Omelchenko}, \citenamefont {Zakharova},
  \citenamefont {Avetyan}, \citenamefont {Fradkov},\ and\ \citenamefont
  {Sch{\"o}ll}}]{nikitin2019complex}%
  \BibitemOpen
  \bibfield  {author} {\bibinfo {author} {\bibfnamefont {D.}~\bibnamefont
  {Nikitin}}, \bibinfo {author} {\bibfnamefont {I.}~\bibnamefont {Omelchenko}},
  \bibinfo {author} {\bibfnamefont {A.}~\bibnamefont {Zakharova}}, \bibinfo
  {author} {\bibfnamefont {M.}~\bibnamefont {Avetyan}}, \bibinfo {author}
  {\bibfnamefont {A.~L.}\ \bibnamefont {Fradkov}},\ and\ \bibinfo {author}
  {\bibfnamefont {E.}~\bibnamefont {Sch{\"o}ll}},\ }\bibfield  {title}
  {\bibinfo {title} {Complex partial synchronization patterns in networks of
  delay-coupled neurons},\ }\href@noop {} {\bibfield  {journal} {\bibinfo
  {journal} {Philosophical Transactions of the Royal Society A}\ }\textbf
  {\bibinfo {volume} {377}},\ \bibinfo {pages} {20180128} (\bibinfo {year}
  {2019})}\BibitemShut {NoStop}%
\bibitem [{\citenamefont {Omelchenko}\ \emph {et~al.}(2019)\citenamefont
  {Omelchenko}, \citenamefont {H{\"u}lser}, \citenamefont {Zakharova},\ and\
  \citenamefont {Sch{\"o}ll}}]{omelchenko2019control}%
  \BibitemOpen
  \bibfield  {author} {\bibinfo {author} {\bibfnamefont {I.}~\bibnamefont
  {Omelchenko}}, \bibinfo {author} {\bibfnamefont {T.}~\bibnamefont
  {H{\"u}lser}}, \bibinfo {author} {\bibfnamefont {A.}~\bibnamefont
  {Zakharova}},\ and\ \bibinfo {author} {\bibfnamefont {E.}~\bibnamefont
  {Sch{\"o}ll}},\ }\bibfield  {title} {\bibinfo {title} {Control of chimera
  states in multilayer networks},\ }\href@noop {} {\bibfield  {journal}
  {\bibinfo  {journal} {Frontiers in Applied Mathematics and Statistics}\
  }\textbf {\bibinfo {volume} {4}},\ \bibinfo {pages} {67} (\bibinfo {year}
  {2019})}\BibitemShut {NoStop}%
\bibitem [{\citenamefont {Ruzzene}\ \emph {et~al.}(2020)\citenamefont
  {Ruzzene}, \citenamefont {Omelchenko}, \citenamefont {Sawicki}, \citenamefont
  {Zakharova}, \citenamefont {Sch{\"o}ll},\ and\ \citenamefont
  {Andrzejak}}]{ruzzene2020remote}%
  \BibitemOpen
  \bibfield  {author} {\bibinfo {author} {\bibfnamefont {G.}~\bibnamefont
  {Ruzzene}}, \bibinfo {author} {\bibfnamefont {I.}~\bibnamefont {Omelchenko}},
  \bibinfo {author} {\bibfnamefont {J.}~\bibnamefont {Sawicki}}, \bibinfo
  {author} {\bibfnamefont {A.}~\bibnamefont {Zakharova}}, \bibinfo {author}
  {\bibfnamefont {E.}~\bibnamefont {Sch{\"o}ll}},\ and\ \bibinfo {author}
  {\bibfnamefont {R.~G.}\ \bibnamefont {Andrzejak}},\ }\bibfield  {title}
  {\bibinfo {title} {Remote pacemaker control of chimera states in multilayer
  networks of neurons},\ }\href@noop {} {\bibfield  {journal} {\bibinfo
  {journal} {Physical Review E}\ }\textbf {\bibinfo {volume} {102}},\ \bibinfo
  {pages} {052216} (\bibinfo {year} {2020})}\BibitemShut {NoStop}%
\bibitem [{\citenamefont {Chouzouris}\ \emph {et~al.}(2018)\citenamefont
  {Chouzouris}, \citenamefont {Omelchenko}, \citenamefont {Zakharova},
  \citenamefont {Hlinka}, \citenamefont {Jiruska},\ and\ \citenamefont
  {Sch{\"o}ll}}]{chouzouris2018chimera}%
  \BibitemOpen
  \bibfield  {author} {\bibinfo {author} {\bibfnamefont {T.}~\bibnamefont
  {Chouzouris}}, \bibinfo {author} {\bibfnamefont {I.}~\bibnamefont
  {Omelchenko}}, \bibinfo {author} {\bibfnamefont {A.}~\bibnamefont
  {Zakharova}}, \bibinfo {author} {\bibfnamefont {J.}~\bibnamefont {Hlinka}},
  \bibinfo {author} {\bibfnamefont {P.}~\bibnamefont {Jiruska}},\ and\ \bibinfo
  {author} {\bibfnamefont {E.}~\bibnamefont {Sch{\"o}ll}},\ }\bibfield  {title}
  {\bibinfo {title} {Chimera states in brain networks: Empirical neural vs.
  modular fractal connectivity},\ }\href@noop {} {\bibfield  {journal}
  {\bibinfo  {journal} {Chaos: An Interdisciplinary Journal of Nonlinear
  Science}\ }\textbf {\bibinfo {volume} {28}},\ \bibinfo {pages} {045112}
  (\bibinfo {year} {2018})}\BibitemShut {NoStop}%
\bibitem [{\citenamefont {Gerster}\ \emph {et~al.}(2020)\citenamefont
  {Gerster}, \citenamefont {Berner}, \citenamefont {Sawicki}, \citenamefont
  {Zakharova}, \citenamefont {{\v{S}}koch}, \citenamefont {Hlinka},
  \citenamefont {Lehnertz},\ and\ \citenamefont
  {Sch{\"o}ll}}]{gerster2020fitzhugh}%
  \BibitemOpen
  \bibfield  {author} {\bibinfo {author} {\bibfnamefont {M.}~\bibnamefont
  {Gerster}}, \bibinfo {author} {\bibfnamefont {R.}~\bibnamefont {Berner}},
  \bibinfo {author} {\bibfnamefont {J.}~\bibnamefont {Sawicki}}, \bibinfo
  {author} {\bibfnamefont {A.}~\bibnamefont {Zakharova}}, \bibinfo {author}
  {\bibfnamefont {A.}~\bibnamefont {{\v{S}}koch}}, \bibinfo {author}
  {\bibfnamefont {J.}~\bibnamefont {Hlinka}}, \bibinfo {author} {\bibfnamefont
  {K.}~\bibnamefont {Lehnertz}},\ and\ \bibinfo {author} {\bibfnamefont
  {E.}~\bibnamefont {Sch{\"o}ll}},\ }\bibfield  {title} {\bibinfo {title}
  {{FitzHugh--Nagumo} oscillators on complex networks mimic
  epileptic-seizure-related synchronization phenomena},\ }\href@noop {}
  {\bibfield  {journal} {\bibinfo  {journal} {Chaos: An Interdisciplinary
  Journal of Nonlinear Science}\ }\textbf {\bibinfo {volume} {30}},\ \bibinfo
  {pages} {123130} (\bibinfo {year} {2020})}\BibitemShut {NoStop}%
\bibitem [{\citenamefont {Ramlow}\ \emph {et~al.}(2019)\citenamefont {Ramlow},
  \citenamefont {Sawicki}, \citenamefont {Zakharova}, \citenamefont {Hlinka},
  \citenamefont {Claussen},\ and\ \citenamefont
  {Sch{\"o}ll}}]{ramlow2019partial}%
  \BibitemOpen
  \bibfield  {author} {\bibinfo {author} {\bibfnamefont {L.}~\bibnamefont
  {Ramlow}}, \bibinfo {author} {\bibfnamefont {J.}~\bibnamefont {Sawicki}},
  \bibinfo {author} {\bibfnamefont {A.}~\bibnamefont {Zakharova}}, \bibinfo
  {author} {\bibfnamefont {J.}~\bibnamefont {Hlinka}}, \bibinfo {author}
  {\bibfnamefont {J.~C.}\ \bibnamefont {Claussen}},\ and\ \bibinfo {author}
  {\bibfnamefont {E.}~\bibnamefont {Sch{\"o}ll}},\ }\bibfield  {title}
  {\bibinfo {title} {Partial synchronization in empirical brain networks as a
  model for unihemispheric sleep},\ }\href@noop {} {\bibfield  {journal}
  {\bibinfo  {journal} {EPL (Europhysics Letters)}\ }\textbf {\bibinfo {volume}
  {126}},\ \bibinfo {pages} {50007} (\bibinfo {year} {2019})}\BibitemShut
  {NoStop}%
\bibitem [{\citenamefont {Ghosh}\ \emph {et~al.}(2008)\citenamefont {Ghosh},
  \citenamefont {Rho}, \citenamefont {McIntosh}, \citenamefont {K{\"o}tter},\
  and\ \citenamefont {Jirsa}}]{ghosh2008cortical}%
  \BibitemOpen
  \bibfield  {author} {\bibinfo {author} {\bibfnamefont {A.}~\bibnamefont
  {Ghosh}}, \bibinfo {author} {\bibfnamefont {Y.}~\bibnamefont {Rho}}, \bibinfo
  {author} {\bibfnamefont {A.}~\bibnamefont {McIntosh}}, \bibinfo {author}
  {\bibfnamefont {R.}~\bibnamefont {K{\"o}tter}},\ and\ \bibinfo {author}
  {\bibfnamefont {V.}~\bibnamefont {Jirsa}},\ }\bibfield  {title} {\bibinfo
  {title} {Cortical network dynamics with time delays reveals functional
  connectivity in the resting brain},\ }\href@noop {} {\bibfield  {journal}
  {\bibinfo  {journal} {Cognitive Neurodynamics}\ }\textbf {\bibinfo {volume}
  {2}},\ \bibinfo {pages} {115} (\bibinfo {year} {2008})}\BibitemShut {NoStop}%
\bibitem [{\citenamefont {Vuksanovi{\'c}}\ and\ \citenamefont
  {H{\"o}vel}(2015)}]{vuksanovic2015dynamic}%
  \BibitemOpen
  \bibfield  {author} {\bibinfo {author} {\bibfnamefont {V.}~\bibnamefont
  {Vuksanovi{\'c}}}\ and\ \bibinfo {author} {\bibfnamefont {P.}~\bibnamefont
  {H{\"o}vel}},\ }\bibfield  {title} {\bibinfo {title} {Dynamic changes in
  network synchrony reveal resting-state functional networks},\ }\href@noop {}
  {\bibfield  {journal} {\bibinfo  {journal} {Chaos: An Interdisciplinary
  Journal of Nonlinear Science}\ }\textbf {\bibinfo {volume} {25}},\ \bibinfo
  {pages} {023116} (\bibinfo {year} {2015})}\BibitemShut {NoStop}%
\bibitem [{\citenamefont {Mess{\'e}}\ \emph {et~al.}(2015)\citenamefont
  {Mess{\'e}}, \citenamefont {H{\"u}tt}, \citenamefont {K{\"o}nig},\ and\
  \citenamefont {Hilgetag}}]{messe2015closer}%
  \BibitemOpen
  \bibfield  {author} {\bibinfo {author} {\bibfnamefont {A.}~\bibnamefont
  {Mess{\'e}}}, \bibinfo {author} {\bibfnamefont {M.-T.}\ \bibnamefont
  {H{\"u}tt}}, \bibinfo {author} {\bibfnamefont {P.}~\bibnamefont
  {K{\"o}nig}},\ and\ \bibinfo {author} {\bibfnamefont {C.~C.}\ \bibnamefont
  {Hilgetag}},\ }\bibfield  {title} {\bibinfo {title} {A closer look at the
  apparent correlation of structural and functional connectivity in excitable
  neural networks},\ }\href@noop {} {\bibfield  {journal} {\bibinfo  {journal}
  {Scientific Reports}\ }\textbf {\bibinfo {volume} {5}},\ \bibinfo {pages}
  {7870} (\bibinfo {year} {2015})}\BibitemShut {NoStop}%
\bibitem [{\citenamefont {Buchholz}\ \emph {et~al.}(2013)\citenamefont
  {Buchholz}, \citenamefont {Engel}, \citenamefont {Kammann},\ and\
  \citenamefont {Tr{\"o}ltzsch}}]{buchholz2013optimal}%
  \BibitemOpen
  \bibfield  {author} {\bibinfo {author} {\bibfnamefont {R.}~\bibnamefont
  {Buchholz}}, \bibinfo {author} {\bibfnamefont {H.}~\bibnamefont {Engel}},
  \bibinfo {author} {\bibfnamefont {E.}~\bibnamefont {Kammann}},\ and\ \bibinfo
  {author} {\bibfnamefont {F.}~\bibnamefont {Tr{\"o}ltzsch}},\ }\bibfield
  {title} {\bibinfo {title} {On the optimal control of the schl{\"o}gl-model},\
  }\href@noop {} {\bibfield  {journal} {\bibinfo  {journal} {Computational
  Optimization and Applications}\ }\textbf {\bibinfo {volume} {56}},\ \bibinfo
  {pages} {153} (\bibinfo {year} {2013})}\BibitemShut {NoStop}%
\bibitem [{\citenamefont {Polak}\ and\ \citenamefont
  {Ribiere}(1969)}]{polak1969note}%
  \BibitemOpen
  \bibfield  {author} {\bibinfo {author} {\bibfnamefont {E.}~\bibnamefont
  {Polak}}\ and\ \bibinfo {author} {\bibfnamefont {G.}~\bibnamefont
  {Ribiere}},\ }\bibfield  {title} {\bibinfo {title} {Note sur la convergence
  de m{\'e}thodes de directions conjugu{\'e}es},\ }\href@noop {} {\bibfield
  {journal} {\bibinfo  {journal} {ESAIM: Mathematical Modelling and Numerical
  Analysis-Mod{\'e}lisation Math{\'e}matique et Analyse Num{\'e}rique}\
  }\textbf {\bibinfo {volume} {3}},\ \bibinfo {pages} {35} (\bibinfo {year}
  {1969})}\BibitemShut {NoStop}%
\bibitem [{\citenamefont {Nocedal}\ and\ \citenamefont
  {Wright}(2006)}]{nocedal2006numerical}%
  \BibitemOpen
  \bibfield  {author} {\bibinfo {author} {\bibfnamefont {J.}~\bibnamefont
  {Nocedal}}\ and\ \bibinfo {author} {\bibfnamefont {S.}~\bibnamefont
  {Wright}},\ }\href@noop {} {\emph {\bibinfo {title} {Numerical
  optimization}}}\ (\bibinfo  {publisher} {Springer Science \& Business
  Media},\ \bibinfo {year} {2006})\BibitemShut {NoStop}%
\bibitem [{\citenamefont {Fletcher}\ and\ \citenamefont
  {Reeves}(1964)}]{fletcher1964function}%
  \BibitemOpen
  \bibfield  {author} {\bibinfo {author} {\bibfnamefont {R.}~\bibnamefont
  {Fletcher}}\ and\ \bibinfo {author} {\bibfnamefont {C.~M.}\ \bibnamefont
  {Reeves}},\ }\bibfield  {title} {\bibinfo {title} {Function minimization by
  conjugate gradients},\ }\href@noop {} {\bibfield  {journal} {\bibinfo
  {journal} {The Computer Journal}\ }\textbf {\bibinfo {volume} {7}},\ \bibinfo
  {pages} {149} (\bibinfo {year} {1964})}\BibitemShut {NoStop}%
\bibitem [{\citenamefont {Izhikevich}(2007)}]{izhikevich2007dynamical}%
  \BibitemOpen
  \bibfield  {author} {\bibinfo {author} {\bibfnamefont {E.~M.}\ \bibnamefont
  {Izhikevich}},\ }\href@noop {} {\emph {\bibinfo {title} {Dynamical systems in
  neuroscience}}}\ (\bibinfo  {publisher} {MIT press},\ \bibinfo {year}
  {2007})\BibitemShut {NoStop}%
\bibitem [{\citenamefont {Cakan}\ and\ \citenamefont
  {Obermayer}(2020)}]{cakan2019}%
  \BibitemOpen
  \bibfield  {author} {\bibinfo {author} {\bibfnamefont {C.}~\bibnamefont
  {Cakan}}\ and\ \bibinfo {author} {\bibfnamefont {K.}~\bibnamefont
  {Obermayer}},\ }\bibfield  {title} {\bibinfo {title} {Biophysically grounded
  mean-field models of neural populations under electrical stimulation},\
  }\href {https://doi.org/10.1371/journal.pcbi.1007822} {\bibfield  {journal}
  {\bibinfo  {journal} {PLoS Computational Biology}\ }\textbf {\bibinfo
  {volume} {16}},\ \bibinfo {pages} {1} (\bibinfo {year} {2020})}\BibitemShut
  {NoStop}%
\bibitem [{\citenamefont {Van~Essen}\ \emph {et~al.}(2013)\citenamefont
  {Van~Essen}, \citenamefont {Smith}, \citenamefont {Barch}, \citenamefont
  {Behrens}, \citenamefont {Yacoub}, \citenamefont {Ugurbil}, \citenamefont
  {Consortium} \emph {et~al.}}]{van2013hcp}%
  \BibitemOpen
  \bibfield  {author} {\bibinfo {author} {\bibfnamefont {D.~C.}\ \bibnamefont
  {Van~Essen}}, \bibinfo {author} {\bibfnamefont {S.~M.}\ \bibnamefont
  {Smith}}, \bibinfo {author} {\bibfnamefont {D.~M.}\ \bibnamefont {Barch}},
  \bibinfo {author} {\bibfnamefont {T.~E.}\ \bibnamefont {Behrens}}, \bibinfo
  {author} {\bibfnamefont {E.}~\bibnamefont {Yacoub}}, \bibinfo {author}
  {\bibfnamefont {K.}~\bibnamefont {Ugurbil}}, \bibinfo {author} {\bibfnamefont
  {W.-M.~H.}\ \bibnamefont {Consortium}}, \emph {et~al.},\ }\bibfield  {title}
  {\bibinfo {title} {The {WU-Minn} human connectome project: an overview},\
  }\href@noop {} {\bibfield  {journal} {\bibinfo  {journal} {Neuroimage}\
  }\textbf {\bibinfo {volume} {80}},\ \bibinfo {pages} {62} (\bibinfo {year}
  {2013})}\BibitemShut {NoStop}%
\bibitem [{\citenamefont {Jenkinson}\ \emph {et~al.}(2012)\citenamefont
  {Jenkinson}, \citenamefont {Beckmann}, \citenamefont {Behrens}, \citenamefont
  {Woolrich},\ and\ \citenamefont {Smith}}]{jenkinson2012fsl}%
  \BibitemOpen
  \bibfield  {author} {\bibinfo {author} {\bibfnamefont {M.}~\bibnamefont
  {Jenkinson}}, \bibinfo {author} {\bibfnamefont {C.~F.}\ \bibnamefont
  {Beckmann}}, \bibinfo {author} {\bibfnamefont {T.~E.}\ \bibnamefont
  {Behrens}}, \bibinfo {author} {\bibfnamefont {M.~W.}\ \bibnamefont
  {Woolrich}},\ and\ \bibinfo {author} {\bibfnamefont {S.~M.}\ \bibnamefont
  {Smith}},\ }\bibfield  {title} {\bibinfo {title} {{FSL}},\ }\href@noop {}
  {\bibfield  {journal} {\bibinfo  {journal} {Neuroimage}\ }\textbf {\bibinfo
  {volume} {62}},\ \bibinfo {pages} {782} (\bibinfo {year} {2012})}\BibitemShut
  {NoStop}%
\bibitem [{hor(1984)}]{horsthemke1984noise}%
  \BibitemOpen
  \bibinfo {title} {Noise-induced transitions in physics, chemistry, and
  biology},\ in\ \href {https://doi.org/10.1007/3-540-36852-3_7} {\emph
  {\bibinfo {booktitle} {Noise-Induced Transitions: Theory and Applications in
  Physics, Chemistry, and Biology}}}\ (\bibinfo  {publisher} {Springer Berlin
  Heidelberg},\ \bibinfo {address} {Berlin, Heidelberg},\ \bibinfo {year}
  {1984})\ pp.\ \bibinfo {pages} {164--200}\BibitemShut {NoStop}%
\bibitem [{\citenamefont {Schmidt}\ \emph {et~al.}(2013)\citenamefont
  {Schmidt}, \citenamefont {Scholz}, \citenamefont {Obermayer},\ and\
  \citenamefont {Brandt}}]{schmidt2013micha}%
  \BibitemOpen
  \bibfield  {author} {\bibinfo {author} {\bibfnamefont {S.}~\bibnamefont
  {Schmidt}}, \bibinfo {author} {\bibfnamefont {M.}~\bibnamefont {Scholz}},
  \bibinfo {author} {\bibfnamefont {K.}~\bibnamefont {Obermayer}},\ and\
  \bibinfo {author} {\bibfnamefont {S.~A.}\ \bibnamefont {Brandt}},\ }\bibfield
   {title} {\bibinfo {title} {Patterned brain stimulation, what a framework
  with rhythmic and noisy components might tell us about recovery
  maximization},\ }\href@noop {} {\bibfield  {journal} {\bibinfo  {journal}
  {Frontiers in Human Neuroscience}\ }\textbf {\bibinfo {volume} {7}},\
  \bibinfo {pages} {325} (\bibinfo {year} {2013})}\BibitemShut {NoStop}%
\bibitem [{\citenamefont {Honey}\ \emph {et~al.}(2009)\citenamefont {Honey},
  \citenamefont {Sporns}, \citenamefont {Cammoun}, \citenamefont {Gigandet},
  \citenamefont {Thiran}, \citenamefont {Meuli},\ and\ \citenamefont
  {Hagmann}}]{honey2009predicting}%
  \BibitemOpen
  \bibfield  {author} {\bibinfo {author} {\bibfnamefont {C.}~\bibnamefont
  {Honey}}, \bibinfo {author} {\bibfnamefont {O.}~\bibnamefont {Sporns}},
  \bibinfo {author} {\bibfnamefont {L.}~\bibnamefont {Cammoun}}, \bibinfo
  {author} {\bibfnamefont {X.}~\bibnamefont {Gigandet}}, \bibinfo {author}
  {\bibfnamefont {J.-P.}\ \bibnamefont {Thiran}}, \bibinfo {author}
  {\bibfnamefont {R.}~\bibnamefont {Meuli}},\ and\ \bibinfo {author}
  {\bibfnamefont {P.}~\bibnamefont {Hagmann}},\ }\bibfield  {title} {\bibinfo
  {title} {Predicting human resting-state functional connectivity from
  structural connectivity},\ }\href@noop {} {\bibfield  {journal} {\bibinfo
  {journal} {Proceedings of the National Academy of Sciences}\ }\textbf
  {\bibinfo {volume} {106}},\ \bibinfo {pages} {2035} (\bibinfo {year}
  {2009})}\BibitemShut {NoStop}%
\bibitem [{\citenamefont {Gal{\'a}n}(2008)}]{galan2008network}%
  \BibitemOpen
  \bibfield  {author} {\bibinfo {author} {\bibfnamefont {R.~F.}\ \bibnamefont
  {Gal{\'a}n}},\ }\bibfield  {title} {\bibinfo {title} {On how network
  architecture determines the dominant patterns of spontaneous neural
  activity},\ }\href@noop {} {\bibfield  {journal} {\bibinfo  {journal} {PLoS
  One}\ }\textbf {\bibinfo {volume} {3}},\ \bibinfo {pages} {e2148} (\bibinfo
  {year} {2008})}\BibitemShut {NoStop}%
\bibitem [{\citenamefont {Hamdan}\ and\ \citenamefont
  {Nayfeh}(1989)}]{hamdan1989measures}%
  \BibitemOpen
  \bibfield  {author} {\bibinfo {author} {\bibfnamefont {A.}~\bibnamefont
  {Hamdan}}\ and\ \bibinfo {author} {\bibfnamefont {A.}~\bibnamefont
  {Nayfeh}},\ }\bibfield  {title} {\bibinfo {title} {Measures of modal
  controllability and observability for first-and second-order linear
  systems},\ }\href@noop {} {\bibfield  {journal} {\bibinfo  {journal} {Journal
  of Gidance, Control, and Dynamics}\ }\textbf {\bibinfo {volume} {12}},\
  \bibinfo {pages} {421} (\bibinfo {year} {1989})}\BibitemShut {NoStop}%
\bibitem [{\citenamefont {Tang}\ \emph {et~al.}(2020)\citenamefont {Tang},
  \citenamefont {Ju}, \citenamefont {Baum}, \citenamefont {Roalf},
  \citenamefont {Satterthwaite}, \citenamefont {Pasqualetti},\ and\
  \citenamefont {Bassett}}]{tang2020control}%
  \BibitemOpen
  \bibfield  {author} {\bibinfo {author} {\bibfnamefont {E.}~\bibnamefont
  {Tang}}, \bibinfo {author} {\bibfnamefont {H.}~\bibnamefont {Ju}}, \bibinfo
  {author} {\bibfnamefont {G.~L.}\ \bibnamefont {Baum}}, \bibinfo {author}
  {\bibfnamefont {D.~R.}\ \bibnamefont {Roalf}}, \bibinfo {author}
  {\bibfnamefont {T.~D.}\ \bibnamefont {Satterthwaite}}, \bibinfo {author}
  {\bibfnamefont {F.}~\bibnamefont {Pasqualetti}},\ and\ \bibinfo {author}
  {\bibfnamefont {D.~S.}\ \bibnamefont {Bassett}},\ }\bibfield  {title}
  {\bibinfo {title} {Control of brain network dynamics across diverse scales of
  space and time},\ }\href@noop {} {\bibfield  {journal} {\bibinfo  {journal}
  {Physical Review E}\ }\textbf {\bibinfo {volume} {101}},\ \bibinfo {pages}
  {062301} (\bibinfo {year} {2020})}\BibitemShut {NoStop}%
\bibitem [{\citenamefont {Mazzoni}\ \emph {et~al.}(2015)\citenamefont
  {Mazzoni}, \citenamefont {Lind{\'e}n}, \citenamefont {Cuntz}, \citenamefont
  {Lansner}, \citenamefont {Panzeri},\ and\ \citenamefont
  {Einevoll}}]{mazzoni2015computing}%
  \BibitemOpen
  \bibfield  {author} {\bibinfo {author} {\bibfnamefont {A.}~\bibnamefont
  {Mazzoni}}, \bibinfo {author} {\bibfnamefont {H.}~\bibnamefont {Lind{\'e}n}},
  \bibinfo {author} {\bibfnamefont {H.}~\bibnamefont {Cuntz}}, \bibinfo
  {author} {\bibfnamefont {A.}~\bibnamefont {Lansner}}, \bibinfo {author}
  {\bibfnamefont {S.}~\bibnamefont {Panzeri}},\ and\ \bibinfo {author}
  {\bibfnamefont {G.~T.}\ \bibnamefont {Einevoll}},\ }\bibfield  {title}
  {\bibinfo {title} {Computing the local field potential ({LFP}) from
  integrate-and-fire network models},\ }\href@noop {} {\bibfield  {journal}
  {\bibinfo  {journal} {PLoS Computational Biology}\ }\textbf {\bibinfo
  {volume} {11}},\ \bibinfo {pages} {e1004584} (\bibinfo {year}
  {2015})}\BibitemShut {NoStop}%
\bibitem [{\citenamefont {Caminiti}\ \emph {et~al.}(2013)\citenamefont
  {Caminiti}, \citenamefont {Carducci}, \citenamefont {Piervincenzi},
  \citenamefont {Battaglia-Mayer}, \citenamefont {Confalone}, \citenamefont
  {Visco-Comandini}, \citenamefont {Pantano},\ and\ \citenamefont
  {Innocenti}}]{caminiti2013diameter}%
  \BibitemOpen
  \bibfield  {author} {\bibinfo {author} {\bibfnamefont {R.}~\bibnamefont
  {Caminiti}}, \bibinfo {author} {\bibfnamefont {F.}~\bibnamefont {Carducci}},
  \bibinfo {author} {\bibfnamefont {C.}~\bibnamefont {Piervincenzi}}, \bibinfo
  {author} {\bibfnamefont {A.}~\bibnamefont {Battaglia-Mayer}}, \bibinfo
  {author} {\bibfnamefont {G.}~\bibnamefont {Confalone}}, \bibinfo {author}
  {\bibfnamefont {F.}~\bibnamefont {Visco-Comandini}}, \bibinfo {author}
  {\bibfnamefont {P.}~\bibnamefont {Pantano}},\ and\ \bibinfo {author}
  {\bibfnamefont {G.~M.}\ \bibnamefont {Innocenti}},\ }\bibfield  {title}
  {\bibinfo {title} {Diameter, length, speed, and conduction delay of callosal
  axons in macaque monkeys and humans: comparing data from histology and
  magnetic resonance imaging diffusion tractography},\ }\href@noop {}
  {\bibfield  {journal} {\bibinfo  {journal} {Journal of Neuroscience}\
  }\textbf {\bibinfo {volume} {33}},\ \bibinfo {pages} {14501} (\bibinfo {year}
  {2013})}\BibitemShut {NoStop}%
\bibitem [{\citenamefont {Marshall}\ \emph {et~al.}(2004)\citenamefont
  {Marshall}, \citenamefont {M{\"o}lle}, \citenamefont {Hallschmid},\ and\
  \citenamefont {Born}}]{marshall2004transcranial}%
  \BibitemOpen
  \bibfield  {author} {\bibinfo {author} {\bibfnamefont {L.}~\bibnamefont
  {Marshall}}, \bibinfo {author} {\bibfnamefont {M.}~\bibnamefont {M{\"o}lle}},
  \bibinfo {author} {\bibfnamefont {M.}~\bibnamefont {Hallschmid}},\ and\
  \bibinfo {author} {\bibfnamefont {J.}~\bibnamefont {Born}},\ }\bibfield
  {title} {\bibinfo {title} {Transcranial direct current stimulation during
  sleep improves declarative memory},\ }\href@noop {} {\bibfield  {journal}
  {\bibinfo  {journal} {Journal of Neuroscience}\ }\textbf {\bibinfo {volume}
  {24}},\ \bibinfo {pages} {9985} (\bibinfo {year} {2004})}\BibitemShut
  {NoStop}%
\bibitem [{\citenamefont {Ladenbauer}\ \emph {et~al.}(2016)\citenamefont
  {Ladenbauer}, \citenamefont {K{\"u}lzow}, \citenamefont {Passmann},
  \citenamefont {Antonenko}, \citenamefont {Grittner}, \citenamefont {Tamm},\
  and\ \citenamefont {Fl{\"o}el}}]{Ladenbauer2016}%
  \BibitemOpen
  \bibfield  {author} {\bibinfo {author} {\bibfnamefont {J.}~\bibnamefont
  {Ladenbauer}}, \bibinfo {author} {\bibfnamefont {N.}~\bibnamefont
  {K{\"u}lzow}}, \bibinfo {author} {\bibfnamefont {S.}~\bibnamefont
  {Passmann}}, \bibinfo {author} {\bibfnamefont {D.}~\bibnamefont {Antonenko}},
  \bibinfo {author} {\bibfnamefont {U.}~\bibnamefont {Grittner}}, \bibinfo
  {author} {\bibfnamefont {S.}~\bibnamefont {Tamm}},\ and\ \bibinfo {author}
  {\bibfnamefont {A.}~\bibnamefont {Fl{\"o}el}},\ }\bibfield  {title} {\bibinfo
  {title} {Brain stimulation during an afternoon nap boosts slow oscillatory
  activity and memory consolidation in older adults},\ }\href@noop {}
  {\bibfield  {journal} {\bibinfo  {journal} {Neuroimage}\ }\textbf {\bibinfo
  {volume} {142}},\ \bibinfo {pages} {311} (\bibinfo {year}
  {2016})}\BibitemShut {NoStop}%
\bibitem [{\citenamefont {Nitsche}\ and\ \citenamefont
  {Paulus}(2011)}]{nitsche2011transcranial}%
  \BibitemOpen
  \bibfield  {author} {\bibinfo {author} {\bibfnamefont {M.~A.}\ \bibnamefont
  {Nitsche}}\ and\ \bibinfo {author} {\bibfnamefont {W.}~\bibnamefont
  {Paulus}},\ }\bibfield  {title} {\bibinfo {title} {Transcranial direct
  current stimulation--update 2011},\ }\href@noop {} {\bibfield  {journal}
  {\bibinfo  {journal} {Restorative Neurology and Neuroscience}\ }\textbf
  {\bibinfo {volume} {29}},\ \bibinfo {pages} {463} (\bibinfo {year}
  {2011})}\BibitemShut {NoStop}%
\bibitem [{\citenamefont {Antal}\ \emph {et~al.}(2008)\citenamefont {Antal},
  \citenamefont {Boros}, \citenamefont {Poreisz}, \citenamefont {Chaieb},
  \citenamefont {Terney},\ and\ \citenamefont
  {Paulus}}]{antal2008comparatively}%
  \BibitemOpen
  \bibfield  {author} {\bibinfo {author} {\bibfnamefont {A.}~\bibnamefont
  {Antal}}, \bibinfo {author} {\bibfnamefont {K.}~\bibnamefont {Boros}},
  \bibinfo {author} {\bibfnamefont {C.}~\bibnamefont {Poreisz}}, \bibinfo
  {author} {\bibfnamefont {L.}~\bibnamefont {Chaieb}}, \bibinfo {author}
  {\bibfnamefont {D.}~\bibnamefont {Terney}},\ and\ \bibinfo {author}
  {\bibfnamefont {W.}~\bibnamefont {Paulus}},\ }\bibfield  {title} {\bibinfo
  {title} {Comparatively weak after-effects of transcranial alternating current
  stimulation ({tACS}) on cortical excitability in humans},\ }\href@noop {}
  {\bibfield  {journal} {\bibinfo  {journal} {Brain Stimulation}\ }\textbf
  {\bibinfo {volume} {1}},\ \bibinfo {pages} {97} (\bibinfo {year}
  {2008})}\BibitemShut {NoStop}%
\bibitem [{\citenamefont {Terney}\ \emph {et~al.}(2008)\citenamefont {Terney},
  \citenamefont {Chaieb}, \citenamefont {Moliadze}, \citenamefont {Antal},\
  and\ \citenamefont {Paulus}}]{terney2008increasing}%
  \BibitemOpen
  \bibfield  {author} {\bibinfo {author} {\bibfnamefont {D.}~\bibnamefont
  {Terney}}, \bibinfo {author} {\bibfnamefont {L.}~\bibnamefont {Chaieb}},
  \bibinfo {author} {\bibfnamefont {V.}~\bibnamefont {Moliadze}}, \bibinfo
  {author} {\bibfnamefont {A.}~\bibnamefont {Antal}},\ and\ \bibinfo {author}
  {\bibfnamefont {W.}~\bibnamefont {Paulus}},\ }\bibfield  {title} {\bibinfo
  {title} {Increasing human brain excitability by transcranial high-frequency
  random noise stimulation},\ }\href@noop {} {\bibfield  {journal} {\bibinfo
  {journal} {Journal of Neuroscience}\ }\textbf {\bibinfo {volume} {28}},\
  \bibinfo {pages} {14147} (\bibinfo {year} {2008})}\BibitemShut {NoStop}%
\bibitem [{\citenamefont {Behrens}\ \emph {et~al.}(2017)\citenamefont
  {Behrens}, \citenamefont {Kraft}, \citenamefont {Irlbacher}, \citenamefont
  {Gerhardt}, \citenamefont {Olma},\ and\ \citenamefont
  {Brandt}}]{behrens2017long}%
  \BibitemOpen
  \bibfield  {author} {\bibinfo {author} {\bibfnamefont {J.~R.}\ \bibnamefont
  {Behrens}}, \bibinfo {author} {\bibfnamefont {A.}~\bibnamefont {Kraft}},
  \bibinfo {author} {\bibfnamefont {K.}~\bibnamefont {Irlbacher}}, \bibinfo
  {author} {\bibfnamefont {H.}~\bibnamefont {Gerhardt}}, \bibinfo {author}
  {\bibfnamefont {M.~C.}\ \bibnamefont {Olma}},\ and\ \bibinfo {author}
  {\bibfnamefont {S.~A.}\ \bibnamefont {Brandt}},\ }\bibfield  {title}
  {\bibinfo {title} {Long-lasting enhancement of visual perception with
  repetitive noninvasive transcranial direct current stimulation},\ }\href@noop
  {} {\bibfield  {journal} {\bibinfo  {journal} {Frontiers in Cellular
  Neuroscience}\ }\textbf {\bibinfo {volume} {11}},\ \bibinfo {pages} {238}
  (\bibinfo {year} {2017})}\BibitemShut {NoStop}%
\bibitem [{\citenamefont {Moisa}\ \emph {et~al.}(2016)\citenamefont {Moisa},
  \citenamefont {Polania}, \citenamefont {Grueschow},\ and\ \citenamefont
  {Ruff}}]{moisa2016brain}%
  \BibitemOpen
  \bibfield  {author} {\bibinfo {author} {\bibfnamefont {M.}~\bibnamefont
  {Moisa}}, \bibinfo {author} {\bibfnamefont {R.}~\bibnamefont {Polania}},
  \bibinfo {author} {\bibfnamefont {M.}~\bibnamefont {Grueschow}},\ and\
  \bibinfo {author} {\bibfnamefont {C.~C.}\ \bibnamefont {Ruff}},\ }\bibfield
  {title} {\bibinfo {title} {Brain network mechanisms underlying motor
  enhancement by transcranial entrainment of gamma oscillations},\ }\href@noop
  {} {\bibfield  {journal} {\bibinfo  {journal} {Journal of Neuroscience}\
  }\textbf {\bibinfo {volume} {36}},\ \bibinfo {pages} {12053} (\bibinfo {year}
  {2016})}\BibitemShut {NoStop}%
\bibitem [{\citenamefont {Alagapan}\ \emph {et~al.}(2016)\citenamefont
  {Alagapan}, \citenamefont {Schmidt}, \citenamefont {Lefebvre}, \citenamefont
  {Hadar}, \citenamefont {Shin},\ and\ \citenamefont
  {Fr{\"o}hlich}}]{alagapan2016modulation}%
  \BibitemOpen
  \bibfield  {author} {\bibinfo {author} {\bibfnamefont {S.}~\bibnamefont
  {Alagapan}}, \bibinfo {author} {\bibfnamefont {S.~L.}\ \bibnamefont
  {Schmidt}}, \bibinfo {author} {\bibfnamefont {J.}~\bibnamefont {Lefebvre}},
  \bibinfo {author} {\bibfnamefont {E.}~\bibnamefont {Hadar}}, \bibinfo
  {author} {\bibfnamefont {H.~W.}\ \bibnamefont {Shin}},\ and\ \bibinfo
  {author} {\bibfnamefont {F.}~\bibnamefont {Fr{\"o}hlich}},\ }\bibfield
  {title} {\bibinfo {title} {Modulation of cortical oscillations by
  low-frequency direct cortical stimulation is state-dependent},\ }\href@noop
  {} {\bibfield  {journal} {\bibinfo  {journal} {PLoS Biology}\ }\textbf
  {\bibinfo {volume} {14}},\ \bibinfo {pages} {e1002424} (\bibinfo {year}
  {2016})}\BibitemShut {NoStop}%
\bibitem [{\citenamefont {Thut}\ \emph {et~al.}(2017)\citenamefont {Thut},
  \citenamefont {Bergmann}, \citenamefont {Fr{\"o}hlich}, \citenamefont
  {Soekadar}, \citenamefont {Brittain}, \citenamefont {Valero-Cabr{\'e}},
  \citenamefont {Sack}, \citenamefont {Miniussi}, \citenamefont {Antal},
  \citenamefont {Siebner} \emph {et~al.}}]{thut2017guiding}%
  \BibitemOpen
  \bibfield  {author} {\bibinfo {author} {\bibfnamefont {G.}~\bibnamefont
  {Thut}}, \bibinfo {author} {\bibfnamefont {T.~O.}\ \bibnamefont {Bergmann}},
  \bibinfo {author} {\bibfnamefont {F.}~\bibnamefont {Fr{\"o}hlich}}, \bibinfo
  {author} {\bibfnamefont {S.~R.}\ \bibnamefont {Soekadar}}, \bibinfo {author}
  {\bibfnamefont {J.-S.}\ \bibnamefont {Brittain}}, \bibinfo {author}
  {\bibfnamefont {A.}~\bibnamefont {Valero-Cabr{\'e}}}, \bibinfo {author}
  {\bibfnamefont {A.~T.}\ \bibnamefont {Sack}}, \bibinfo {author}
  {\bibfnamefont {C.}~\bibnamefont {Miniussi}}, \bibinfo {author}
  {\bibfnamefont {A.}~\bibnamefont {Antal}}, \bibinfo {author} {\bibfnamefont
  {H.~R.}\ \bibnamefont {Siebner}}, \emph {et~al.},\ }\bibfield  {title}
  {\bibinfo {title} {Guiding transcranial brain stimulation by {EEG}/{MEG} to
  interact with ongoing brain activity and associated functions: a position
  paper},\ }\href@noop {} {\bibfield  {journal} {\bibinfo  {journal} {Clinical
  Neurophysiology}\ }\textbf {\bibinfo {volume} {128}},\ \bibinfo {pages} {843}
  (\bibinfo {year} {2017})}\BibitemShut {NoStop}%
\bibitem [{\citenamefont {Bergmann}(2018)}]{bergmann2018brain}%
  \BibitemOpen
  \bibfield  {author} {\bibinfo {author} {\bibfnamefont {T.~O.}\ \bibnamefont
  {Bergmann}},\ }\bibfield  {title} {\bibinfo {title} {Brain state-dependent
  brain stimulation},\ }\href@noop {} {\bibfield  {journal} {\bibinfo
  {journal} {Frontiers in Psychology}\ }\textbf {\bibinfo {volume} {9}},\
  \bibinfo {pages} {2108} (\bibinfo {year} {2018})}\BibitemShut {NoStop}%
\bibitem [{\citenamefont {Deco}\ \emph {et~al.}(2017)\citenamefont {Deco},
  \citenamefont {Cabral}, \citenamefont {Woolrich}, \citenamefont {Stevner},
  \citenamefont {Van~Hartevelt},\ and\ \citenamefont
  {Kringelbach}}]{deco2017single}%
  \BibitemOpen
  \bibfield  {author} {\bibinfo {author} {\bibfnamefont {G.}~\bibnamefont
  {Deco}}, \bibinfo {author} {\bibfnamefont {J.}~\bibnamefont {Cabral}},
  \bibinfo {author} {\bibfnamefont {M.~W.}\ \bibnamefont {Woolrich}}, \bibinfo
  {author} {\bibfnamefont {A.~B.}\ \bibnamefont {Stevner}}, \bibinfo {author}
  {\bibfnamefont {T.~J.}\ \bibnamefont {Van~Hartevelt}},\ and\ \bibinfo
  {author} {\bibfnamefont {M.~L.}\ \bibnamefont {Kringelbach}},\ }\bibfield
  {title} {\bibinfo {title} {Single or multiple frequency generators in
  on-going brain activity: A mechanistic whole-brain model of empirical meg
  data},\ }\href@noop {} {\bibfield  {journal} {\bibinfo  {journal}
  {Neuroimage}\ }\textbf {\bibinfo {volume} {152}},\ \bibinfo {pages} {538}
  (\bibinfo {year} {2017})}\BibitemShut {NoStop}%
\bibitem [{\citenamefont {Gong}\ \emph {et~al.}(2009)\citenamefont {Gong},
  \citenamefont {Rosa-Neto}, \citenamefont {Carbonell}, \citenamefont {Chen},
  \citenamefont {He},\ and\ \citenamefont {Evans}}]{gong2009age}%
  \BibitemOpen
  \bibfield  {author} {\bibinfo {author} {\bibfnamefont {G.}~\bibnamefont
  {Gong}}, \bibinfo {author} {\bibfnamefont {P.}~\bibnamefont {Rosa-Neto}},
  \bibinfo {author} {\bibfnamefont {F.}~\bibnamefont {Carbonell}}, \bibinfo
  {author} {\bibfnamefont {Z.~J.}\ \bibnamefont {Chen}}, \bibinfo {author}
  {\bibfnamefont {Y.}~\bibnamefont {He}},\ and\ \bibinfo {author}
  {\bibfnamefont {A.~C.}\ \bibnamefont {Evans}},\ }\bibfield  {title} {\bibinfo
  {title} {Age-and gender-related differences in the cortical anatomical
  network},\ }\href@noop {} {\bibfield  {journal} {\bibinfo  {journal} {Journal
  of Neuroscience}\ }\textbf {\bibinfo {volume} {29}},\ \bibinfo {pages}
  {15684} (\bibinfo {year} {2009})}\BibitemShut {NoStop}%
\end{thebibliography}%
\end{document}


\maketitle
 


\section{Whole-brain network -- state space exploration}

\subsection{Limit cycle criterion}

\begin{figure}
    \centering 
    \includegraphics[width=\textwidth]{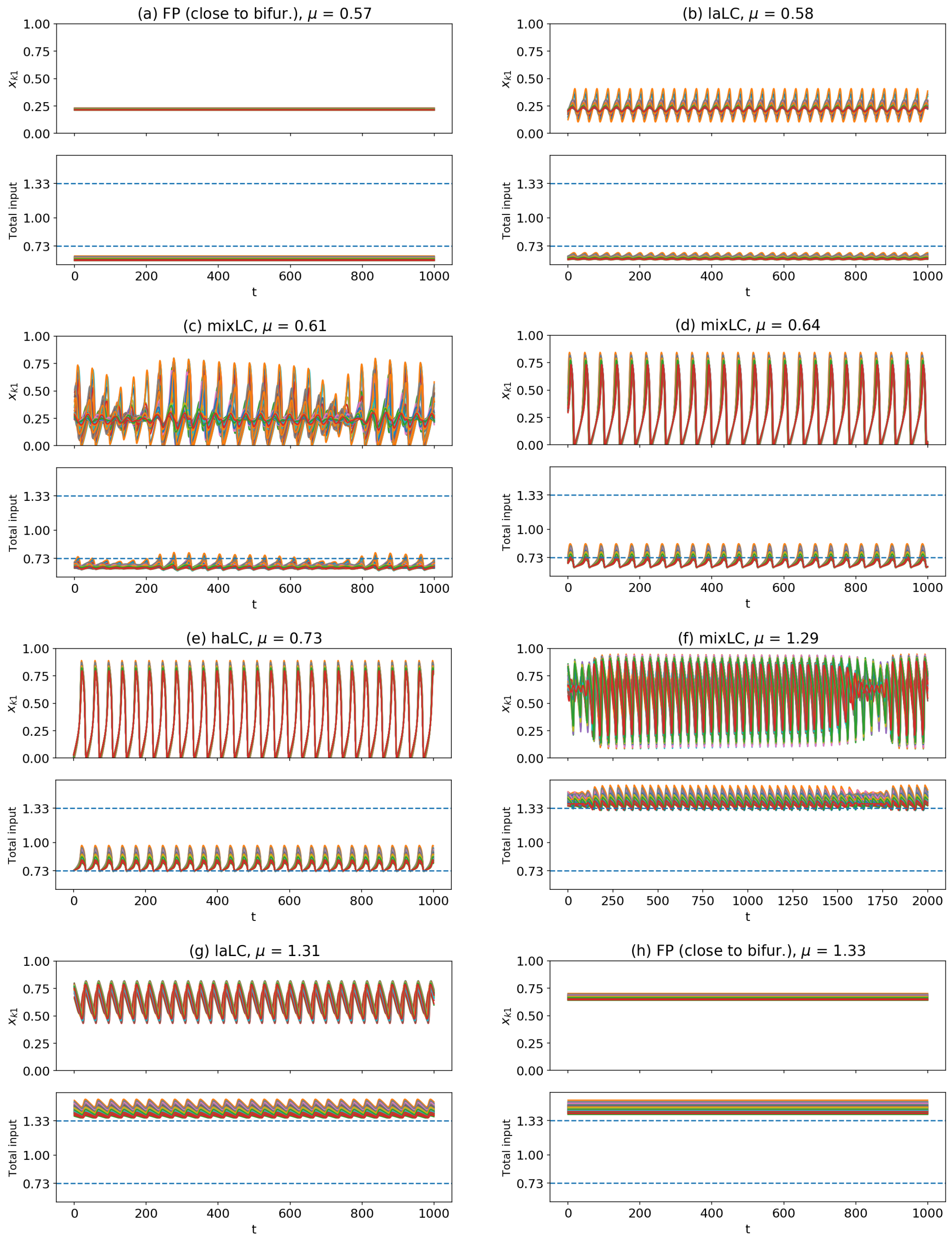}
    \caption{Activity traces over time of all nodes in the network (different colors denote different nodes) for different external input $\mu$, as indicated in the respective title. Parameter values of (a)-(h) correspond to Figs.~3c(1-8) of the main paper with $\sigma=0.08$ and $\eta=0$ (noise-free case). 
    Top panels show the activity variables $x_{k1}$ as a function of time for random initial conditions after a sufficiently long transient ($\bar{t}=10000$).
    Bottom panels show the respective time series of the total input [Eq.~\eqref{totalinput}] to the network nodes. 
    Blue dashed lines indicate the input values for which a single FHN oscillator transitions into its individual LC [$0.73$ and $1.33$, see Eq.~(17) of the main paper].}
    \label{fig:nonoisetraces}
\end{figure}

Figure \ref{fig:nonoisetraces} shows traces of the activity variables $x_{k1}$ and the respective total inputs for each node over time for different background inputs $\mu$ (see bifurcation diagram shown in Fig.~3b of the main paper). The total input 
\begin{align}
\label{totalinput}
    \mu + \sigma \sum^N_{i} A_{ki}x_{i1}(t) + \eta \, \xi_k(t)
\end{align}
to each node consists of the sum of the background input $\mu$, the input $\sigma \sum^N_{i} A_{ki}x_{i1}(t)$ from other nodes in the network, and the input $\eta \, \xi_k(t)$ from the noise process. We expect qualitatively different dynamics if the total input of a node lies inside vs.\ outside the interval given by Eq.~(17) of the main paper (dashed blue lines in Fig.~\ref{fig:nonoisetraces}), because total inputs inside this interval correspond to an individual FHN oscillator being in its limit cycle.

For small $\mu$, as shown in Fig.~\ref{fig:nonoisetraces}a, the total input is below the lower threshold, and the network is in a stable fixed point. If $\mu$ is larger with the total input still below but closer to the lower threshold, a low amplitude limit cycle (laLC) emerges (Fig.~\ref{fig:nonoisetraces}b). These oscillations cannot be explained by the dynamics of the individual FHN oscillator but emerge as a result of network effects.
If the total inputs of some of the nodes enter the described interval, as it happens for some (Fig.~\ref{fig:nonoisetraces}c) or all (Fig.~\ref{fig:nonoisetraces}d) of the maxima of activity, the network dynamics exhibits high amplitude oscillations. The elevated amplitudes for nodes with total input within the interval increase the coupling inputs to other nodes in the network and this creates oscillations in nodes that are not in their individual LC. 
We call this regime the mixed limit cycle (mixLC) state.
If $\mu$ is high enough so that all nodes satisfy the condition in Eq.~(17), as shown in Fig.~\ref{fig:nonoisetraces}e, the network is in the high amplitude limit cycle (haLC) state. 
For higher $\mu$, the total input of some nodes is exceeds the upper threshold and we observe mixLC states, as in Fig.~\ref{fig:nonoisetraces}f, that show similar aperiodic behavior as mixLC states at the low bifurcation (cf.\ Fig.~\ref{fig:nonoisetraces}c).
We also observe a laLC at the high bifurcation (Fig.~\ref{fig:nonoisetraces}g) and a stable fixed point (Fig.~\ref{fig:nonoisetraces}h) for even higher values for the background input $\mu$.

The interval criterion for the node input alone is not sufficient to explain the asynchronous or aperiodic network dynamics observed in Figs.~\ref{fig:nonoisetraces}c and~f, but in these traces we see that the network dynamics change qualitatively, depending on whether some of the node inputs fall into the considered interval. Consequently, small changes in the input to individual nodes can result in large effects on the network dynamics. 

\subsection{Aperiodic behavior in mixed states}

\begin{figure}
    \centering
    \includegraphics[width=\textwidth]{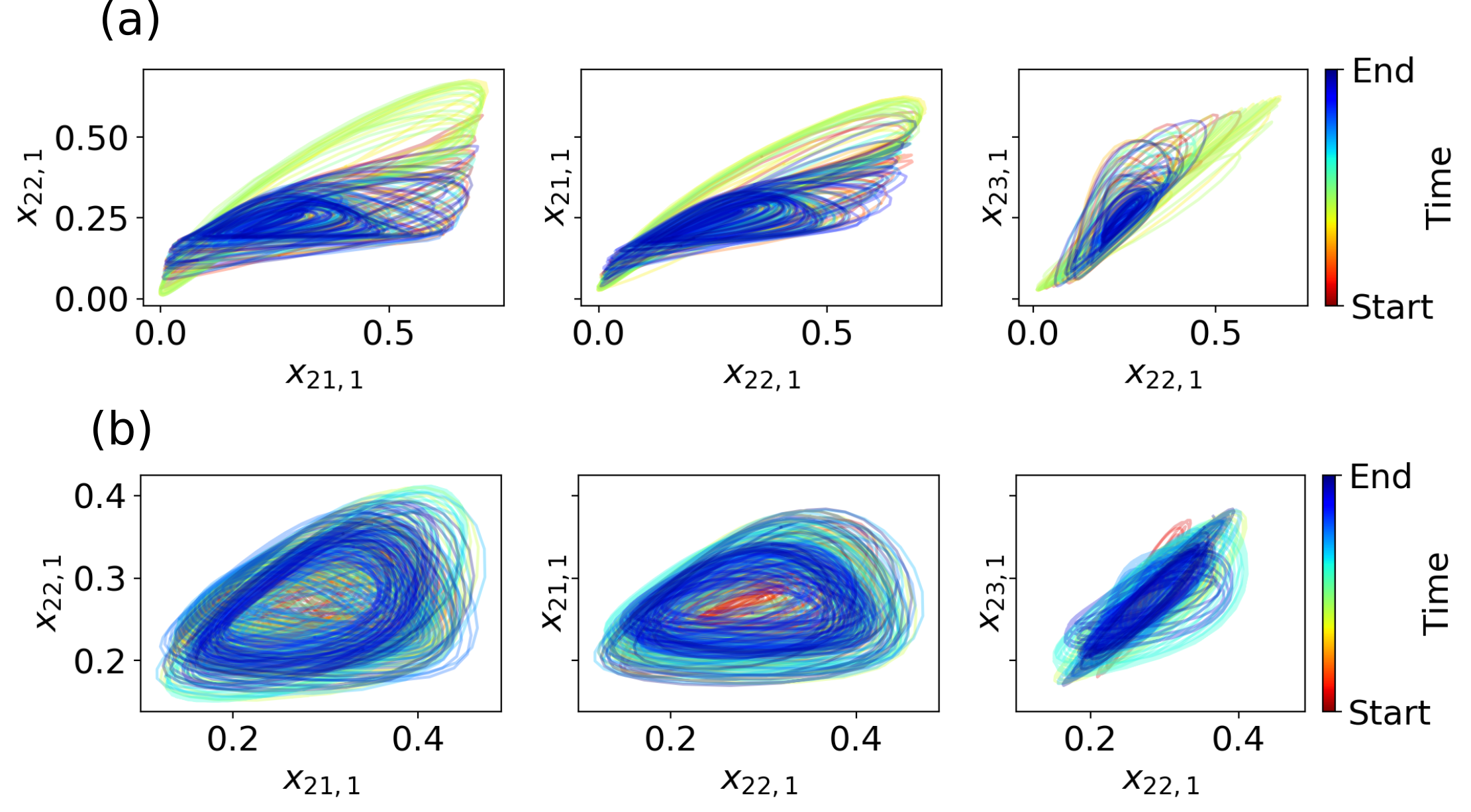}
    \caption{Trajectories of activity variables $x_{k1}$, for nodes $k\in \{21,22,23\}$ for different node combinations, in the mixLC state without noise ($\eta=0$). Trajectories are shown after $\Bar{t}=10000$ (to aviod transient effects) for a time interval of length 10000 arb.\ units. The open trajectories indicate aperiodic network dynamics. (a) Network parameters are $\sigma = 0.08, \mu=0.61$ (same as Fig.~\ref{fig:nonoisetraces}c or Fig.~3c(3) of the main paper). (b) Parameters are $\sigma = 0.025, \mu=0.7$ (same as point S1, cf.\ Fig.~3a of the main paper).}
    \label{fig:chaotic}
\end{figure}

Figure~\ref{fig:chaotic}a shows the activity variables $x_{k1}$ for nodes $k\in \{21,22,23\}$ plotted against each other for the asynchronous mixed state at the low bifurcation in Fig.~\ref{fig:nonoisetraces}c (equivalent to Fig.~3c(3) of the main paper). We observe 
open trajectories, which differ between different oscillation periods.
This is a strong indication that the analyzed network state is not only asynchronous but also aperiodic.
Figure~\ref{fig:chaotic}b shows the equivalent phenomenon for the trajectories of the investigated asynchronous states at the low bifurcation S1 (cf.\ Fig.~3a of the main paper).

\begin{figure}
    \centering 
    \includegraphics[width=\textwidth]{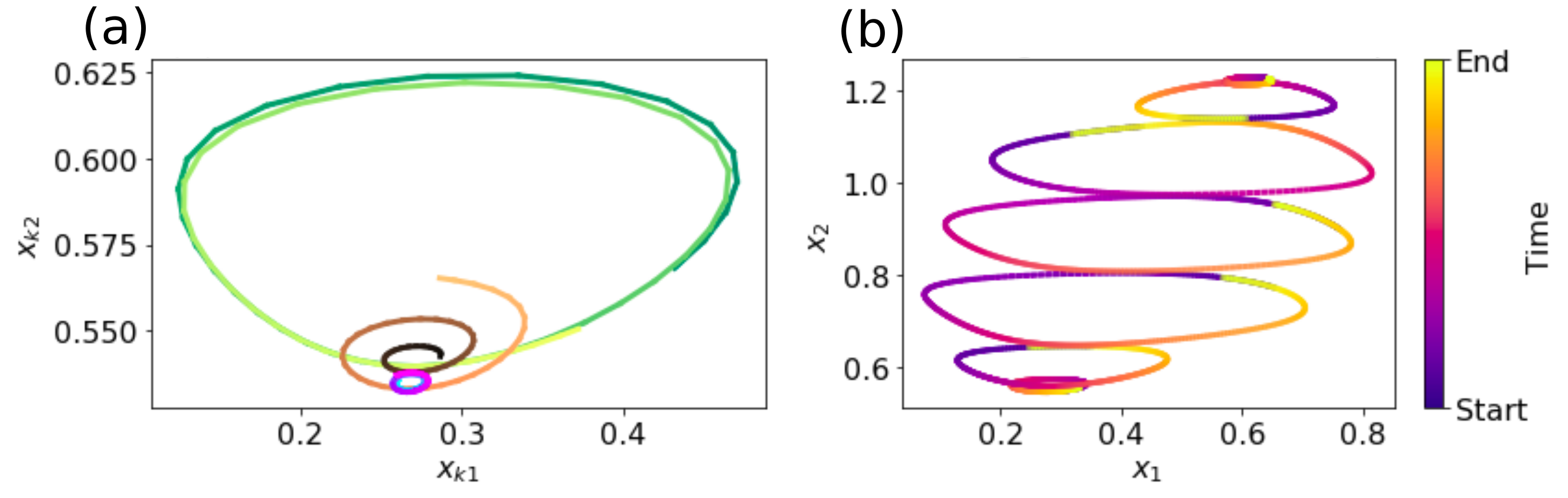}
    \caption{(a) Plot of the trajectories in the ($x_{k1}, x_{k2}$) phase plane for 3 of the 94 nodes in the network at location S1 ($\sigma=0.025, \mu=0.7$, cf. Fig.~3a of the main paper). All nodes have the same start and end time and are plotted for 64 time units, which is equivalent to roughly 2 cycles (after transient of $\Bar{t}=5000$). Cyan to pink corresponds to node 71 which has the lowest weighted degree $d_k$ in the network. Black to orange corresponds to node 33 with an intermediate value of $d_k$, and green to yellow corresponds to node 72 which has the highest value of $d_k$ in the network. Color changes along the trajectory in discrete time. (b) Shape of the LC of a single, uncoupled node for different background inputs $\mu = \{0.73,0.75,0.85,1.03,1.21,1.31,1.33\}$ (bottom left to top right).}
    \label{fig:lcshape}
\end{figure}

Figure~\ref{fig:lcshape}a shows the phase space oscillations in the ($x_{k1}, x_{k2}$) plane of three other network nodes, $k\in \{71, 33, 72\}$ (lowest, intermediate, and highest weighted degree $d_k$ in the network), for point S1 (cf.\ Fig.~3a of the main paper). 
The large differences in ranges of the trajectories are caused by total inputs [Eq.~\eqref{totalinput}] of different strength due to their different weighted in-degree. 
Even small changes in the input to a node can have a large effect on the shape of the limit cycle, especially close to the bifurcations (see also Fig.~\ref{fig:lcshape}b).  
The starting and end points of these trajectories (all plotted for $t\in[5000,5064]$) show that the oscillations of these nodes are not in phase and that the phase difference between them does not remain constant.

\subsection{Influence of noise on the network dynamics}

\begin{figure}
    \centering 
    \includegraphics[width=0.9\textwidth]{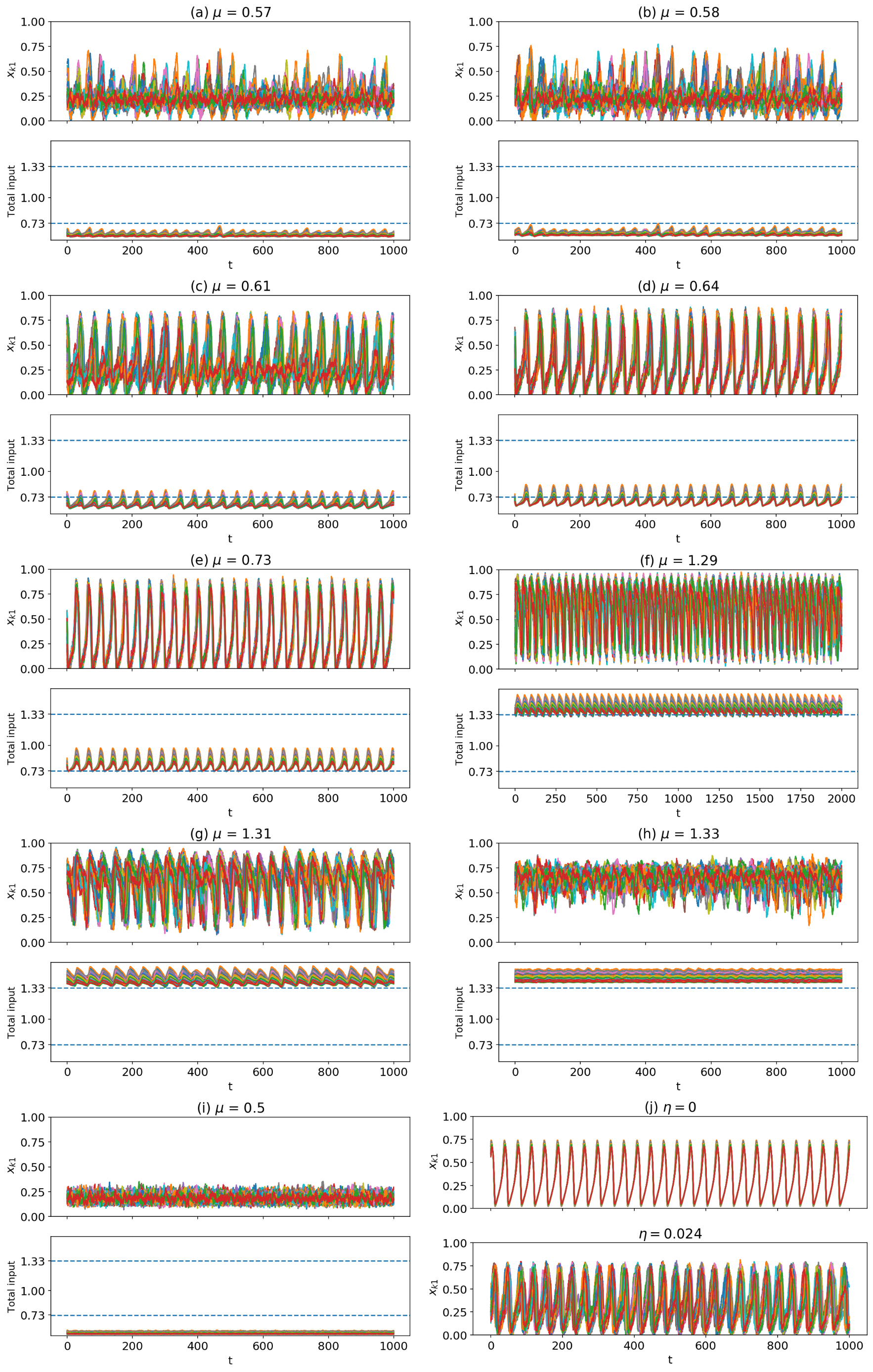}
    \caption{
    (a)-(i) Activity traces over time of all nodes in the network (different colors denote different nodes) for different external input $\mu$, as indicated in the respective title. Parameters for (a)-(h) are the same as in Figs.~\ref{fig:nonoisetraces}a-h, except for added Gaussian white noise with variance $\eta=0.024$. 
    Top panels show the activity variables $x_{k1}$ as a function of time for random initial conditions after a sufficiently long transient ($\bar{t}=10000$).
    Bottom panels show the respective time series of the total input [Eq.~\eqref{totalinput}] to the network nodes. 
    Blue dashed lines indicate the input values for which a single FHN oscillator transitions into its individual LC ($0.73$ and $1.33$, see Eq.~(17) of the main paper). (j) Activity traces over time of all nodes for $\mu=0.75, \sigma=0.025$ and different noise levels, top: noise-free case ($\eta=0$, resulting in $R=0.99$); bottom: noisy network dynamics with reduced average cross-correlation ($\eta=0.024$, resulting in $R=0.75$).}
    \label{fig:noisetraces}
\thisfloatpagestyle{empty}
\end{figure}

Figures \ref{fig:noisetraces}a--h show the activity variables $x_{k1}$ and the total inputs [Eq.~\eqref{totalinput}] for each node for the same parametrizations of the network dynamics as Figs.~\ref{fig:nonoisetraces}a--h, with additional Gaussian white noise ($\eta=0.024$).
If the network without noise is in the stable down state close the the bifurcation, as in Fig.~\ref{fig:nonoisetraces}a, small fluctuations in the node input can lead to large changes in the activity variable. Such oscillations of individual nodes are then sustained by the coupling and drive the network oscillation.
This effect blurs the distinction between FPs and the laLC for increasing noise levels. 
The oscillation amplitudes of nodes in the laLC fluctuate due to the noise, which makes it even harder to distinguish the states based on the time-series of activity variables. 
Interestingly, the mean network cross-correlation of the noise-free asynchronous state in Fig.~\ref{fig:nonoisetraces}c is increased from $R=0.47$ ($\eta=0$) to  $R=0.58$ ($\eta=0.024$) in Fig.~\ref{fig:noisetraces}c. A potential explanation for the higher synchrony with additional Gaussian noise is the increased number of nodes that fulfill the criterion in Eq.~(17) of the main paper (36 for $\eta=0.024$, compared to 15 for $\eta=0$). If a random fluctuation drives a node into the respective individual LC, this increases the coupling input to the other nodes in the network. With more nodes oscillating in the individual LC, the driving input to the remaining nodes is increased, which can then be sufficient to entrain a synchronous network oscillation.
Close to the high bifurcation in Fig.~\ref{fig:noisetraces}f, a similar effect counteracts the aperiodic collapsing of the network oscillation (cf. Fig.~\ref{fig:nonoisetraces}f).
The periodic and synchronous states in Figs.~\ref{fig:noisetraces}d,~e are robust against the influence of noise, showing no qualitative change in their dynamics compared to the noise-free case in Figs.~\ref{fig:nonoisetraces}d,~e. Figure~\ref{fig:noisetraces}h shows, similar to Fig.~\ref{fig:noisetraces}a at the low bifurcation, a self-sustained noise induced network oscillation at the high bifurcation. 
Figure~\ref{fig:noisetraces}i shows that if the network dynamics are in a fixed point far away from a bifurcation, we observe uncorrelated fluctuations in the activity variable and the combined input, instead of the emerging oscillations in Figs.~\ref{fig:noisetraces}a and~h.

When evaluating the results of the optimal control method for the synchronization task in the case of finite noise, it is unreasonable to expect that the optimal control algorithm will achieve the target mean network cross-correlation of $R_T=1$. 
In the synchronization task in Section IV B of the main paper, the optimal control method increases the average network cross-correlation from $R=0.25$ in the uncontrolled case to $R = 0.83$ (parameters $\mu=0.7, \sigma=0.025, \eta=0.024$, location S1, Fig.~10a of the main paper).
We compare the deviation of this result from the target $R_T=1$ with the effect of noise on $R$ of a previously synchronous network state.
Figure \ref{fig:noisetraces}j shows the effect of noise on an uncontrolled, synchronous state with low global coupling strength as S1 ($\sigma=0.025$, used for synchronization task of the main paper). The mean network cross-correlation of $R = 0.99$ in the noise-free ($\eta=0$) case is substantially reduced to $R=0.75$ when noise is introduced ($\eta=0.024$, same as for the synchronization task of the main paper).

\section{Optimal control of the brain network dynamics}

\subsection{Recovery of laLC target state}

\begin{figure}
    \centering
    \includegraphics[width=\textwidth]{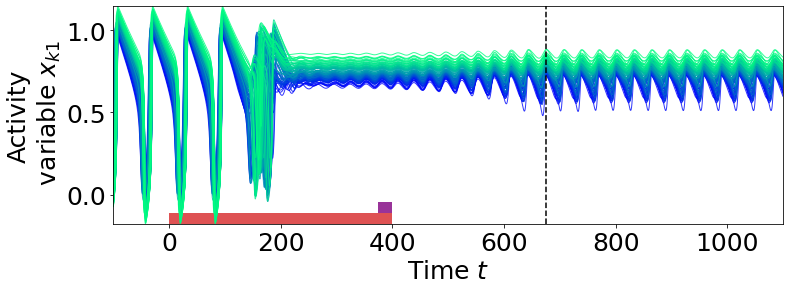}
    \caption{Same as Fig.6b of the main paper but shown for an extended time interval. The purple bar indicates the time interval in which the deviation of the target state is penalized ($\tau=25$) and the red bar indicates the time interval the control is active. 
    The dashed line at $t=675$ indicates the first point in time when the cross-correlation with the laLC target state is larger than $0.99$.}
    \label{fig:cont6b}
\end{figure}

The network dynamics as shown in Fig.~6b of the main paper does not achieve the target state (cf. initial state in Fig.~6a of the main paper) within the time interval in which the control is active. The biphasic pulse of the optimal control method occurs around $t=150$, after which the network oscillation almost comes to rest. Since, however, no stable fixed point exists in the absence of control input, the oscillation amplitude grows until it reaches the laLC target state. Figure~\ref{fig:cont6b} shows that the laLC target state is stable and reached about 550 time units after the control pulse. Although the control target is eventually reached, this trajectory results in a nonzero precision cost, which is evaluated in the last $\tau=25$ time units of the control interval. Due to energy constraints, however, this is still the optimal control solution.

\subsection{Computation of critical times $t_c$}
The critical time $t_c$ is defined as the first point in time when all nodes are ``in synchrony''. The cross-correlation measure for synchrony [Eq.~(7) of the main paper] is defined for time intervals and does not allow to determine synchrony for a given point in time. We therefore compute the  Kuramoto order parameter~$r$,
\begin{align}
    r(t) = \dfrac{1}{N} \left|\sum^N_{k=1} e^{i\theta_k(t)}\right|,
\end{align}
for $N$ nodes of the network with their respective phases $\theta_k$. We calculate the phases for node $k$ by assigning $\theta_k(\hat{t})=0$ to all times $\hat{t}$ that correspond to its oscillation maxima. We then linearly interpolate the phases $\theta_k \in [0,2\pi)$ between subsequent maxima. 
The critical time $t_c$ is then defined as the first point in time in the control period, for which $r(t)\ge 0.999$ holds.

\subsection{Controlled nodes in case of sparse control}

\begin{figure}
     \centering
    \includegraphics[width=0.9\textwidth]{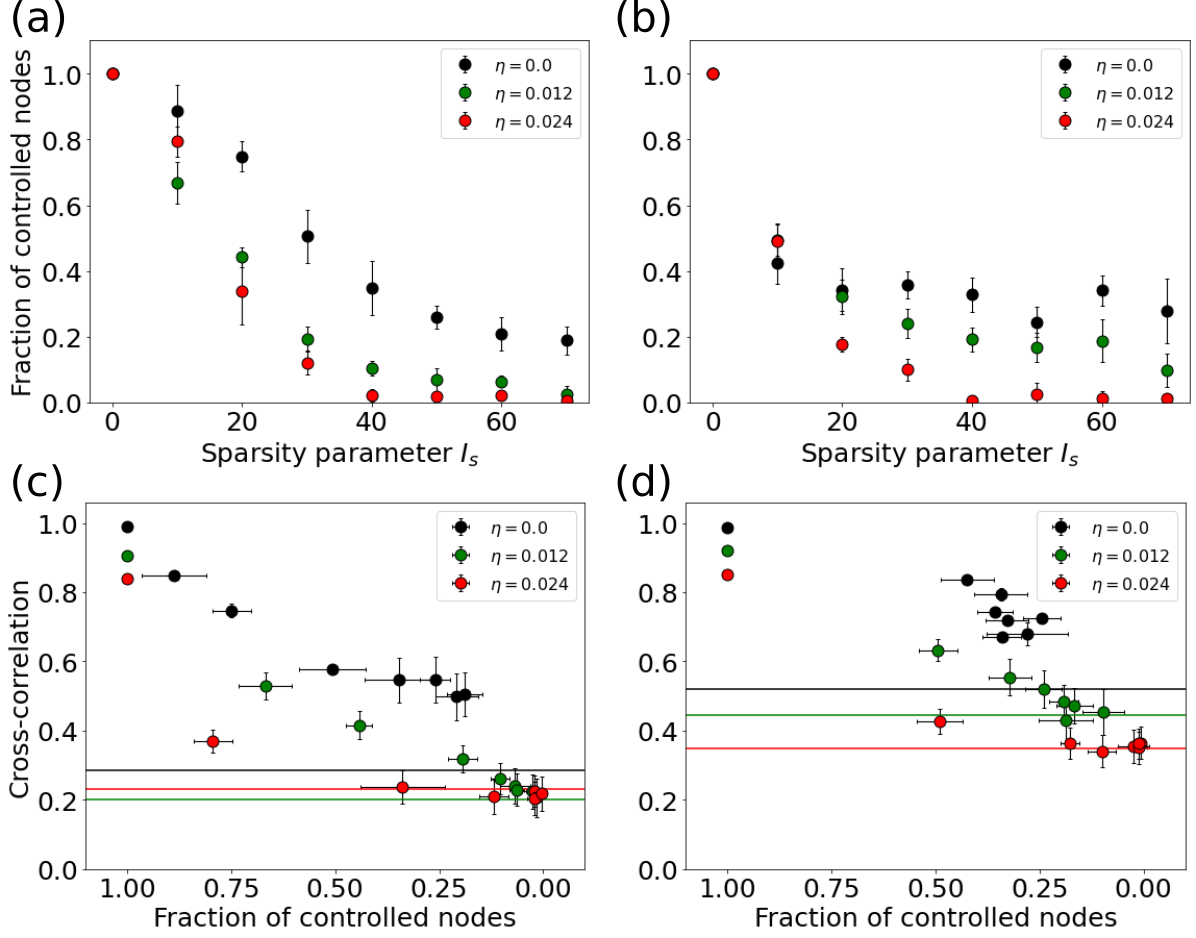}
    \caption{Synchronization with sparse optimal control for different noise levels $\eta$, equivalent to Fig.~12 of the main paper. (a) Fraction of nodes in the network that receive finite control input as a function of the sparsity parameter $I_s$ at point S1 (low bifurcation). (b) Same for S2 (high bifurcation). (c) Achieved synchonization measured by the average cross-correlation  $R$ as a function of the fraction of controlled nodes at point S1. The mean and standard deviation are shown for 5 different initial conditions of the network dynamics, each with 20 independent realizations of the noise. Horizontal lines denote the average cross-correlations $R$ for the uncontrolled case. (d) Same for S2. Parameters as in Fig.~12 of the main paper.}
    \label{fig:sparsity}
\end{figure}

Figure~\ref{fig:sparsity} provides additional information to Fig.~12 of the main paper by showing the fraction of controlled nodes as a function of the sparsity parameter $I_s$ and the resulting average cross-correlation $R$.
As expected, the fraction of controlled nodes decreases with increasing penalty on sparsity in space. 
Plotting the average cross-correlations $R$ against the fraction of controlled nodes intuitively shows how many nodes have to receive a finite control input in order to achieve a certain synchrony for different noise levels. 
It is not possible to achieve a significantly higher synchrony compared to the uncontrolled dynamics by controlling less than 15\% of the network nodes.

\subsection{Videos of the node dynamics in phase space with optimal control}
Videos can be viewed at \href{https://github.com/rederoth/nonlinearControlFHN-videos}{https://github.com/rederoth/nonlinearControlFHN-videos}. The states ($x_{k1}, x_{k2}$) for all network nodes $k$ are plotted as dots in the ($x_{k1}, x_{k2}$) phase plane. 
The color of each node corresponds to its weighted degree (green to blue indicates high to low values).
The optimal control inputs $\Bar{u}_k$ are indicated by black arrows whose direction denotes the sign and whose length denote absolute strength (see top right corner for scale). The time $t$ and the duration of the control interval is indicated in the bottom right corner. If applicable, the limit cycle of a single node (no coupling, $\sigma = 0$) is indicated in light gray. 
\begin{itemize}
    \item Video 1: \href{https://github.com/rederoth/nonlinearControlFHN-videos/blob/main/video_1_switch_low.avi}{Attractor switching at point A1 (low bifurcation) from down-state to oscillatory state}  
    \item Video 2: \href{https://github.com/rederoth/nonlinearControlFHN-videos/blob/main/video_2_switch_low_tolowamp.avi}{Attractor switching at point A1 (low bifurcation) from oscillatory state to down-state}  
    \item Video 3: \href{https://github.com/rederoth/nonlinearControlFHN-videos/blob/main/video_3_switch_high.avi}{Attractor switching at point B2 (high bifurcation) from low-amplitude oscillation to high-amplitude oscillation} 
    \item Video 4: \href{https://github.com/rederoth/nonlinearControlFHN-videos/blob/main/video_4_switch_high_tolowamp.avi}{Attractor switching at point B2 (high bifurcation) from high-amplitude oscillation to low-amplitude oscillation} 
    \item Video 5: \href{https://github.com/rederoth/nonlinearControlFHN-videos/blob/main/video_5_sync_low.avi}{Synchronization at point S1 (low bifurcation) with $\eta=0$ (noise-free network)} 
    \item Video 6: \href{https://github.com/rederoth/nonlinearControlFHN-videos/blob/main/video_6_sync_high.avi}{Synchronization at point S2 (high bifurcation) with $\eta=0$ (noise-free network)} 
    \item Video 7: \href{https://github.com/rederoth/nonlinearControlFHN-videos/blob/main/video_7_sync_low_noise.avi}{Synchronization at point S1 (low bifurcation) with $\eta=0.024$ (noisy network)} 
    \item Video 8: \href{https://github.com/rederoth/nonlinearControlFHN-videos/blob/main/video_8_sync_high_noise.avi}{Synchronization at point S2 (high bifurcation) with $\eta=0.024$ (noisy network)} 
\end{itemize}

\section{Linear network control theory}

\subsection{Correlation of average and modal controllability with weighted degree}

\begin{figure}
    \centering
    \includegraphics[width=\textwidth]{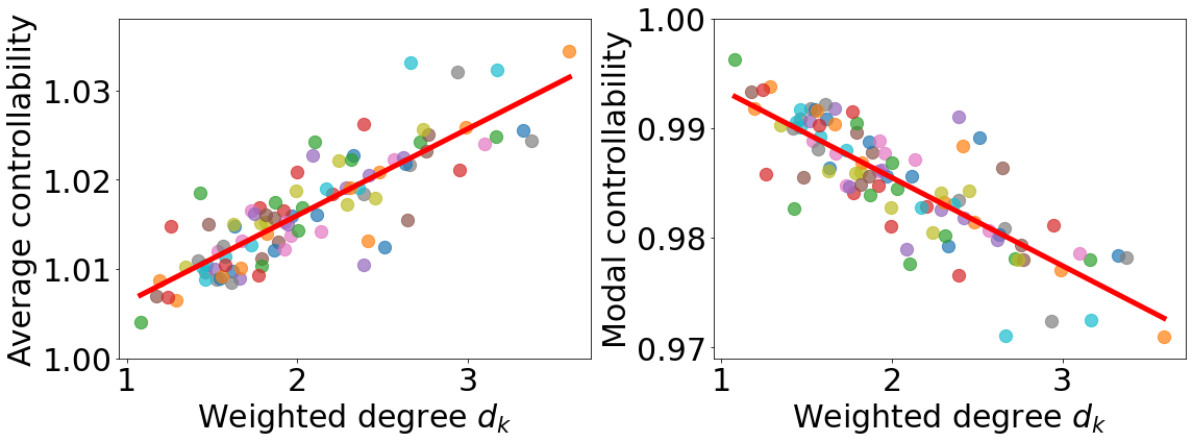}
    \caption{Linear controllability measures plotted against the weighted node degrees $d_k$. Red solid lines denote the linear regression results. Left: Average controllability with ($r=0.85, p<10^{-26}$). Right: Modal controllability with ($r=-0.82, p<10^{-23}$).}
    \label{fig:degcont}
\end{figure}

The considered linear controllability measures are known to be closely related to the weighted node degree $d_k$ [23, 25]. 
Figure~\ref{fig:degcont} shows the average and modal controllability measures as function of $d_k$ for each of the $N=94$ network nodes, where the average controllability shows a positive, and the modal controllability a negative correlation with $d_k$. Nodes with high average controllability tend to have a small modal controllability. 

\subsection{Modal controllability evaluated for the synchronisazion task (noise-free case)}

\begin{figure}
    \centering
    \includegraphics[width=\textwidth]{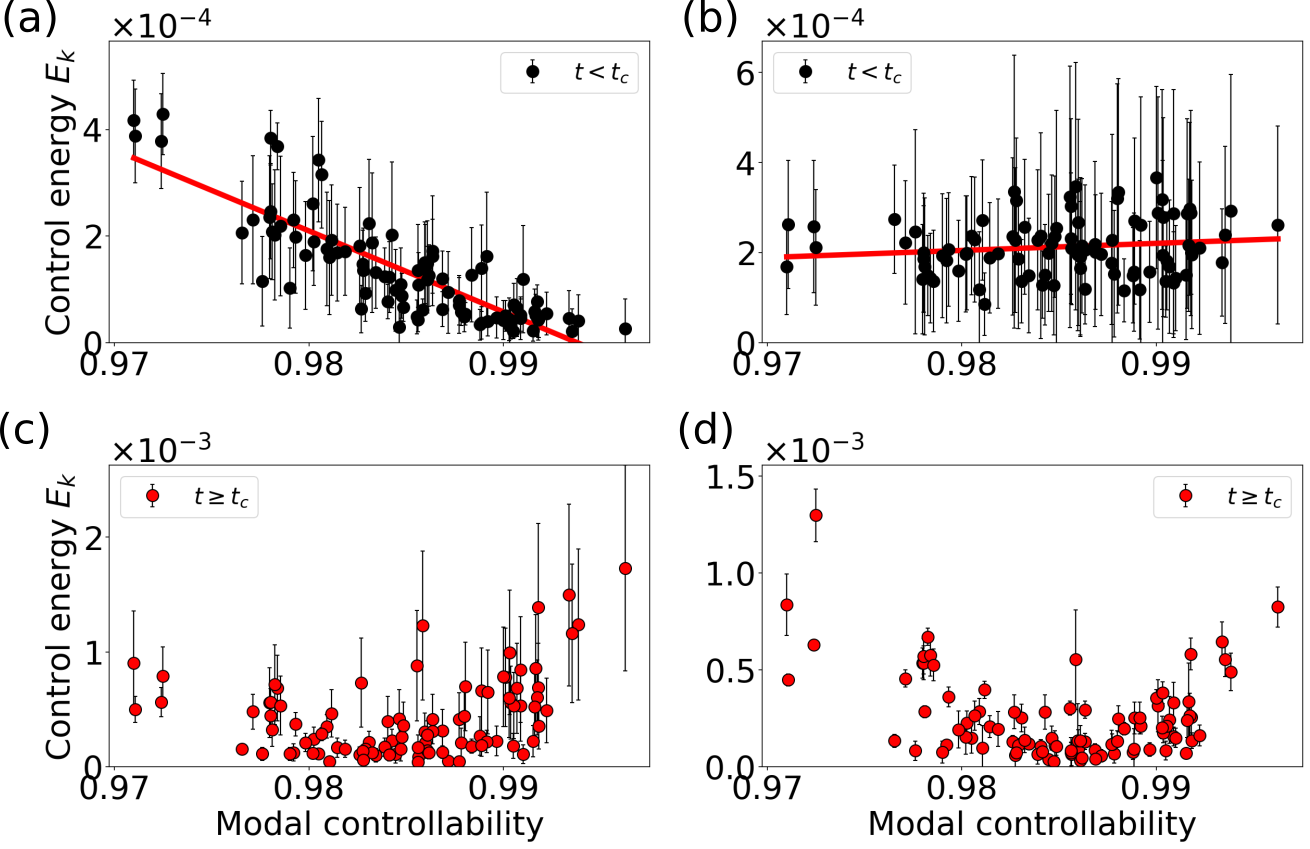}
	\caption{Mean of the optimal node-wise control energy $E_k$ plotted against the modal controllability for the synchronization task (equivalent to Fig.~14 of the main paper). 	
		(a, b) The control signal from time $t=0$ until time $t=t_c$ is considered. Solid red lines denote the linear regression results with the following obtained coefficients:
		(a) $r=-0.84, p<10^{-25}$, (b) $r=0.14, p=0.18$ (not significant).
		(c, d) The control signal from time $t=t_c$ until time $t=T$ is considered. (a, c) Point S1 (low bifurcation). (b, d) Point S2 (high bifurcation). 
		All parameters as in Fig.~14 of the main paper.}    
		\label{fig:modalsyn}
\end{figure}

Figure~\ref{fig:modalsyn} shows the correlation between the optimal node-wise control energies $E_k$ of the control inputs with the modal controllability for the synchronization task discussed in Fig.~14 of the main paper. Since nodes with high average controllability show a low modal controllability, the positive correlation at the low bifurcation for $t<t_c$ (Fig.~14a of the main paper) is reversed, while there is again no significant correlation at the high bifurcation (cf. Fig.~\ref{fig:modalsyn}b).
In the second phase of the control, with $t\ge t_c$, we again observe nodes with an intermediate modal controllability receiving smaller control energies $E_k$ (analogous to Figs.~9c,~d and 14c,~d of the main paper).

\subsection{Controllability measures for the synchronization task (noisy network dynamics)}

\begin{figure}
    \centering
	\includegraphics[width=\textwidth]{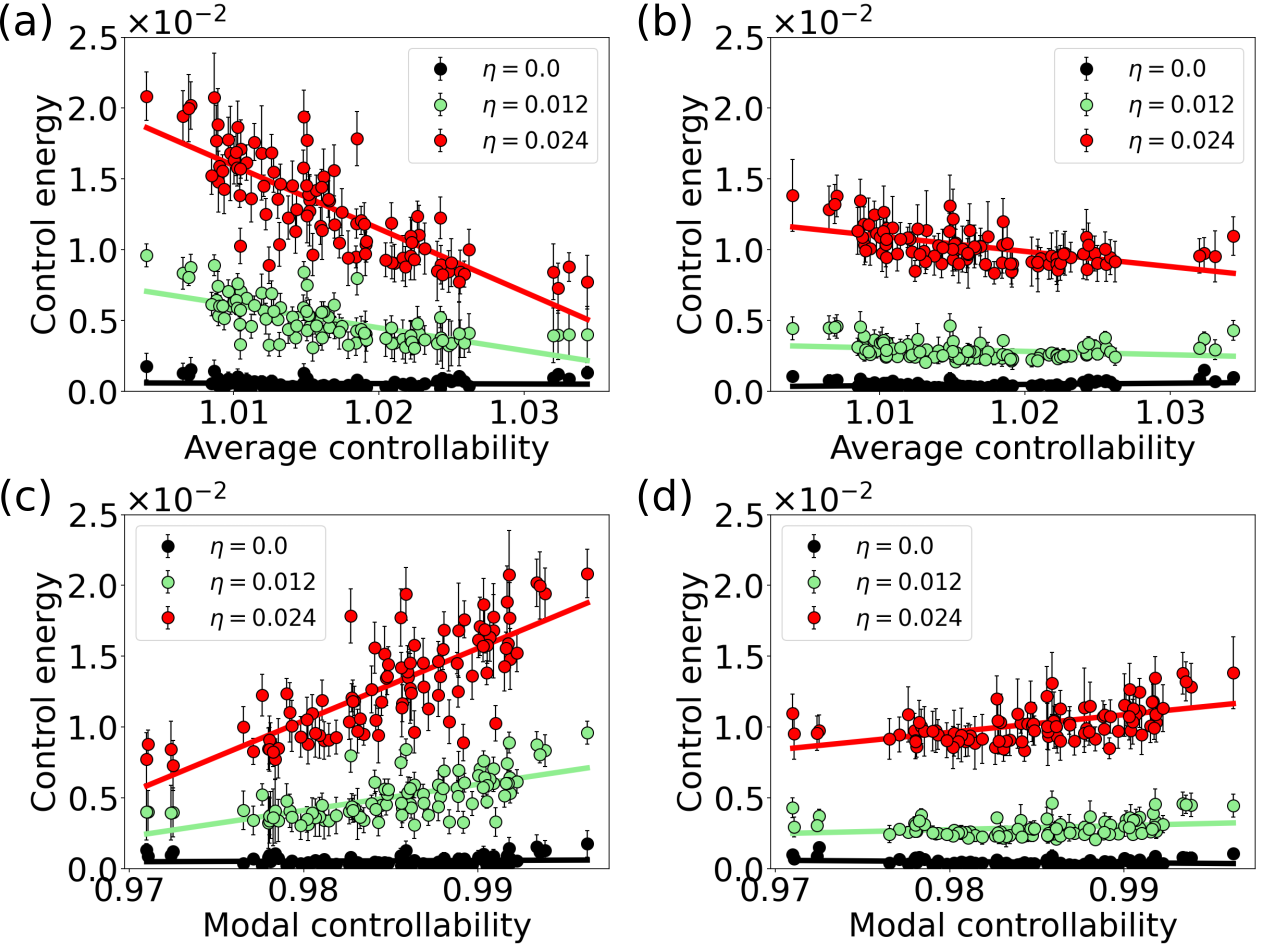}
    \caption{Mean of the optimal node-wise control energy $E_k$ plotted against average (a, b) and modal controllability (c, d) for the synchronization task with different noise levels $\eta$. (a, c) Results at the low bifurcation (point S1). (b, d) Results at the high bifurcation (point S2).
    Solid lines denote the linear regression results, where black corresponds to the noise-free case ( $\eta=0.0$), green to small ($\eta=0.012$), and red to high noise levels ($\eta=0.024$). Regression coefficients are: 
    (a) for 
    $\eta=0.0$ $(r=-0.05, p=0.63)$ not significant,
    $\eta=0.012$ $(r=-0.66, p<10^{-12})$,
    $\eta=0.024$ $(r=-0.80, p<10^{-21})$.
    (b) 
    $(r=0.24, p=0.02)$, $(r=-0.24, p=0.02)$, $(r=-0.54, p<10^{-7})$, 
    (c)
    $(r=0.07, p=0.49)$ n.s., $(r=0.65, p<10^{-11})$, $(r=0.79, p<10^{-20})$, 
    (d)
    $(r=-0.2, p=0.04)$, $(r=0.25, p=0.02)$, $(r=0.54, p<10^{-7})$.
    All parameters as in Fig.~11 of the main paper.}
    \label{fig:ln_syn}
\end{figure}

For the synchronization task under finite noise, we observe a change in the control strategy with increasing noise levels (cf. Section II B of the main paper). 
Figure~\ref{fig:ln_syn} shows that the influence of noise also affects the correlation of the node-wise control energy $E_k$ with the introduced measures from linear control theory. 
At the low bifurcation (Fig.~\ref{fig:ln_syn}a), $E_k$ is not correlated with the average controllability over the whole time interval of the control for $\eta=0$ (combination of Figs.~14a and c of the main paper). With increasing noise levels, however,  we also observe an increased negative correlation. At the high bifurcation (Fig.~\ref{fig:ln_syn}b) the correlation of $E_k$ with the average controllability is even inverted from positive for $\eta=0$ to increasingly negative for $\eta=0.012$ and $\eta=0.024$. Similar effects are observed for the changes in the changes in the correlation between the node-wise control energy $E_k$ and the modal controllability (Fig.~\ref{fig:ln_syn}c,~d) when noise is increased.

